\documentclass[nofootinbib,prd,superscriptaddress,twocolumn,showpacs]{revtex4}

\usepackage{amsmath}
\usepackage{amsfonts}
\usepackage{amssymb}
\usepackage{graphicx}
\usepackage{subcaption}
\usepackage[titletoc]{appendix}
\usepackage{color}
\usepackage{hyperref}
\usepackage{cleveref}
\usepackage[rightcaption]{sidecap}
\usepackage{comment}
\usepackage{soul}
\usepackage{cancel}
\usepackage{caption}
\usepackage{dcolumn}

\usepackage{floatrow}
\usepackage{array}
\usepackage{ctable}
\usepackage{multirow}
\usepackage{siunitx}
\usepackage{tabularx}
\usepackage{booktabs}
\usepackage{supertabular}

\graphicspath{{Graphics/}}

\def\be{\begin{equation}}
\def\ee{\end{equation}}
\def\bea{\begin{eqnarray}}
\def\eea{\end{eqnarray}}

\definecolor{vividviolet}{rgb}{0.62, 0.0, 1.0}
\definecolor{amaranth}{rgb}{0.9, 0.17, 0.31}
\definecolor{palatinateblue}{rgb}{0.15, 0.23, 0.89}
\definecolor{brightpink}{rgb}{1.0, 0.0, 0.5}
\definecolor{cornflowerblue}{rgb}{0.39, 0.58, 0.93}
\definecolor{deepcarminepink}{rgb}{0.94, 0.19, 0.22}
\definecolor{radicalred}{rgb}{1.0, 0.21, 0.37}

\hypersetup{ linktoc=all,
    colorlinks, linkcolor={palatinateblue},
    citecolor={brightpink}, urlcolor={amaranth}
}

\DeclareUnicodeCharacter{2212}{-}

\begin{document}

\title{Redshift dependence and dipolar velocity corrections in cosmographic reconstructions through Type Ia supernova samples}

\author{Anna Chiara Alfano}
\email{a.alfano@ssmeridionale.it}
\affiliation{Scuola Superiore Meridionale, Largo S. Marcellino 10, 80138 Napoli, Italy.}
\affiliation{Istituto Nazionale di Fisica Nucleare (INFN), Via Cinthia 9, 80138 Napoli, Italy.}

\author{Youri Carloni}
\email{youri.carloni@unicam.it}
\affiliation{Universit\`a di Camerino, Divisione di Fisica, Via Madonna delle carceri 9, 62032 Camerino, Italy.}
\affiliation{INAF, Osservatorio Astronomico di Brera, Milano, Italy.}
\affiliation{INFN, Sezione di Perugia, Perugia, 06123, Italy}

\author{Orlando Luongo}
\email{orlando.luongo@unicam.it}
\affiliation{Universit\`a di Camerino, Divisione di Fisica, Via Madonna delle carceri 9, 62032 Camerino, Italy.}
\affiliation{INAF, Osservatorio Astronomico di Brera, Milano, Italy.}
\affiliation{INFN, Sezione di Perugia, Perugia, 06123, Italy}
\affiliation{Department of Nanoscale Science and Engineering, University at Albany SUNY, Albany, NY 12222, USA.}
\affiliation{Al-Farabi Kazakh National University, Al-Farabi av. 71, 050040 Almaty, Kazakhstan.}

\author{Francesco Pace}
\email{francesco.pace@unito.it}
\affiliation{Dipartimento di Fisica, Universit\`a degli Studi di Torino, Via P. Giuria 1, I-10125 Torino, Italy.}
\affiliation{INFN, Sezione di Torino, Via P. Giuria 1, I-10125 Torino, Italy.}
\affiliation{INAF, Osservatorio Astrofisico di Torino, strada Osservatorio 20, 10025, Pino Torinese, Italy.}

\begin{abstract}
We perform a cosmographic analysis of Type Ia supernovae using the Pantheon+\&SH0ES, DES-SNY5 and Union3 compilations. We consider different redshift intervals, analyzed through a third-order Taylor expansion and a Pad\'e $(1,2)$ approximation of the luminosity distance. First, we first infer the cosmographic parameters directly from the observed supernova redshifts. Afterwards, we extend the analysis by including a dipole correction associated with the local peculiar velocity field, constraining both its amplitude and direction. For Pantheon+\&SH0ES, the Cepheid calibration allows a direct determination of the Hubble constant, whereas for DES-SNY5 and Union3 we fix $H_0$ to set the absolute distance scale. By progressively increasing the maximum redshift of the sample, we study how the inferred cosmographic parameters depend on the adopted redshift interval. We find that the Hubble constant obtained from Pantheon+\&SH0ES remains consistent with previous determinations for both cosmographic parameterizations. Instead, the agreement of the deceleration $q_0$ and jerk $j_0$ parameters with the $\Lambda$CDM values depends on the adopted compilation and improves mainly for Pantheon+\&SH0ES as the redshift interval is enlarged. Moreover, the reconstructed dipole parameters remain stable across the redshift intervals considered, with velocity amplitudes of order $300\,\mathrm{km\,s^{-1}}$, for all three supernova samples. Finally, fixing the dipole parameters with the cosmic microwave background values leads to a slight shift from the fiducial $\Lambda$CDM values of the deceleration $q_0$ and jerk $j_0$ parameters.
\end{abstract}



\maketitle
\tableofcontents

\section{Introduction}

The current cosmological scenario requires a dark energy component to account for the observed late-time dynamics \cite{Peebles:2002gy, Copeland:2006wr}. In the concordance $\Lambda$CDM model, dark energy is identified with a cosmological constant $\Lambda$ \cite{Carroll:2000fy, Planck:2018vyg}, although its physical origin remains unknown and alternative descriptions are still observationally allowed \cite{Weinberg:1988cp, Martin:2012bt,Belfiglio:2023rxb,Luongo:2015zaa,Dunsby:2023qpb,Aviles:2011sfa,Belfiglio:2022qai,Dunsby:2016lkw}. At the same time, the persistence of discrepancies between early- and late-time findings of the Hubble constant, together with the milder tension affecting measurements of the clustering amplitude commonly expressed through $S_8$, has further motivated tests of the standard cosmological framework \cite{Riess:2021jrx, Riess:2022oxy, Planck:2018vyg, DES:2021wwk, KiDS:2020suj, Luongo:2022bju}. 

Several extensions of the concordance paradigm have consequently been proposed to alleviate these discrepancies, including modifications of the pre-recombination expansion and mechanisms able to suppress the late-time growth of structures\footnote{More generally, these tensions have reinforced the motivation for testing the assumptions underlying the $\Lambda$CDM model across different cosmological epochs, including the hypothesis that the late-time accelerated expansion is driven by a strictly constant dark energy component.} \cite{Poulin:2018cxd, Carloni:2025jlk, Carloni:2026yut, Carloni:2026mrv, DiValentino:2017iww, Lucca:2021dxo, Clark:2021hlo, Capozziello:2018hly}.

Here, recent baryon acoustic oscillation measurements from the Dark Energy Spectroscopic Instrument (DESI) have provided more stringent tests on the late-time expansion history and increased the interest in possible departures from a cosmological constant \cite{DESI:2024mwx, DESI:2025zgx}.

The significance of these indications, however, remains sensitive to the adopted datasets, statistical methodology and dark energy degeneracy \cite{Alfano:2025gie, Carloni:2024zpl, Alfano:2025awf, Carloni:2026ard}.

Within this context, a careful characterization of the observational probes tracing the late-time expansion history is essential for assessing the consistency and reliability of the inferred cosmological scenario.

Among these probes, Type Ia supernovae (SNe Ia) have played a central role since the discovery of the accelerated expansion of the Universe \cite{SupernovaSearchTeam:1998fmf, SupernovaCosmologyProject:1998vns}. Owing to their well-established standardization, SNe Ia provide precise constraints on the late-time distance-redshift relation and, consequently, on the parameters governing the recent expansion history\footnote{Moreover, when anchored to the Cepheid distance scale, they yield a direct local determination of the Hubble constant $H_0$ \cite{Riess:2021jrx, Riess:2022oxy}.}.

Currently, we have at our disposal three main SNe Ia catalogs, namely Pantheon+ \cite{Scolnic:2021amr, Brout:2022vxf}, the Dark Energy Survey Supernova Program (DES-SNY5) \cite{DES:2024jxu, DES:2024upw, DES:2024hip} and Union3 \cite{Rubin:2023jdq}. The main differences between the samples lie in $1)$ the redshift range they cover, $2)$ the number of data and $3)$ improvements between the samples, in particular between Pantheon+ and DES-SNY5. Regarding point number one, the Pantheon+ catalog exhibits the largest redshift interval,  encompassing data up to $z=2.33$, while DES-SNY5 and Union3 end at $z=1.13$ and $z=2.26$, respectively. On the other hand, the data points in every compilation are 1701 for Pantheon+, 1635 for DES-SNY5 with 145 in common with Pantheon+ and 2087 for Union3 with 1363 shared with Pantheon+. Nevertheless, regarding point number three, as discussed in Ref. \cite{DES:2025tir}, the main differences and improvements between Pantheon+ and DES-SNY5 compilations are an upgrade of the model used for fitting the light-curve switching from SALT2 adopted in Pantheon+ to SALT3 adopted by the DES Collaboration, an improvement in DES-SNY5 in the determination of the free parameters describing the SNe Ia intrinsic scatter model \cite{Popovic:2021yuo} and an upgrade in the stellar masses of host-galaxies adopting deeper coadd photometry \cite{DES:2020xhr}. Last but not least, the Union3 sample  differs from the Pantheon+ catalog in, e.g. different calibrations, the use of the SALT3 model to fit the full rest-frame optical wavelength range and the selection of SNe Ia \cite{Rubin:2023jdq}.

Although all samples present differences between each other, they have in common the presence of the host-galaxy peculiar velocities in low-$z$ regions that could alter the constraints on key parameters when one does not take into account the corrections for these velocities \cite{Davis:2010jq, Huterer:2020szt, Carr:2021lcj, Peterson:2021hel, Pasten:2023rpc, Sorrenti:2024ugq}.

Motivated by the above considerations, we here investigate the impact of peculiar velocities adopting a \emph{cosmographic approach} \cite{Visser:2004bf, Visser:2003vq, Cattoen:2007sk, Cattoen:2007id,Dunsby:2015ers},  widely used in the recent literature \cite{Luongo:2024fww, Alfano:2024ukk, Alfano:2026sxg, Carloni:2024zpl, Fazzari:2025lzd, Jesus:2024nrl, Macpherson:2025qec, Rodrigues:2025tfg, Pourojaghi:2024bxa}. Cosmography requires the validity of the cosmological principle and, since it does not depend {\it a priori} on any cosmological model, appears to be model-independent \cite{Dunsby:2015ers}. Its basic demands consist of expanding the scale factor $a(t)$ up to a selected order in a Taylor series and then reconstructing the luminosity distance $d_L(z)$ and/or any other observable quantities. Clearly, this procedure leads to a truncation, jeopardizing the results, and to a severe convergence problem \cite{Gruber:2013wua}. The latter has been healed  by proposing different parameterizations of the luminosity distance. The most promising ones are given by the original first auxiliary variable proposal \cite{Cattoen:2007sk, Busti:2015xqa, Capozziello:2020ctn}, widely-criticized \cite{Aviles:2012ay,Busti:2015xqa}, and by the use of alternatives to the Taylor series, namely the Chebychev polynomials \cite{Capozziello:2017nbu} or, even better, the Pad\'e approximants \cite{Gruber:2013wua, Aviles:2014rma, Capozziello:2017ddd, Capozziello:2020ctn, Hu:2022udt,Carloni:2025dqt}.

In this context, we adopt a standard Taylor expansion of the luminosity distance and a Pad\'e series of order $(1,2)$ to constrain the deceleration $q_0$ and jerk $j_0$ parameters. When the Pantheon+\&SH0ES sample is involved, in every redshift interval we also bound the Hubble constant $H_0$. Afterwards, our strategy employs five different redshift intervals, for each sample, by performing two main analyses,  following first and, then,  extending the approach proposed in Ref.  \cite{Sorrenti:2024ugq}. Specifically, in the first analysis we define the redshift interval through the cosmic microwave background (CMB) frame redshift $z_{\text{CMB}}$,  accounting for the motion of the observer with respect to the CMB. In the second analysis, we introduce a dipolar correction considering this time the heliocentric redshift $z_{\text{HEL}}$ to define the five redshift intervals. From these two analyses, we find a compatibility at $1$-$\sigma$ with our inferred $H_0$ and $H_0=73.6\pm 1.1\ \mathrm{km\,s^{-1}\,Mpc^{-1}}$ from Ref. \cite{Riess:2021jrx},  when the Pantheon+\&SH0ES sample is adopted for both the luminosity distance parameterizations and in every redshift interval. This shows that the value of the Hubble constant remains overall consistent, independently of the precise interval in which the analysis is performed. On the other hand, we observe that the $q_0$ and $j_0$ parameters are strongly dependent on the adopted redshift interval first and on the supernova samples employed too. More precisely, in the lowest intervals the parameters show larger error bars that decrease with the enlarging of the $z$-intervals leading to more constrained parameters.

Comparing our results with the concordance paradigm fiducial values, inferred by assuming \emph{exactly} $\Omega_{\rm m}=0.3$, i.e. $q_0=-0.55$ and $j_0=1$, gives the major agreement with the Pantheon+\&SH0ES sample at the highest redshifts and only partially when DES-SNY5 is employed. On the other hand, Union3 exhibits a more persistent deviation from the fiducial values. 

Last but not least, focusing on the dipole parameters $(v_0,\, \text{ra},\, \text{dec})$ we observed that Pantheon+\&SH0ES and DES-SNY5 seem to prefer very similar directions while Union3 seems to prefer, especially for declination, different values while the amplitude $v_0$ remains of more or less the same order for all three samples.

Complementary with the two main analyses we also perform a comparison with the second analysis by fixing the amplitude of the dipole and its direction to the CMB values inferred by the Planck Collaboration, i.e. $v_0=369.82\pm 0.11~\mathrm{km\,s^{-1}}$, $\text{ra}=167.942^\circ\pm 0.007^\circ$ and $\text{dec}=-6.944^\circ\pm0.007^\circ$ \cite{Sullivan:2021yms, Planck:2013kqc, Planck:2018nkj}. Even though in this case the compatibility between the inferred Hubble constant and $H_0=73.6\pm 1.1\ \mathrm{km\,s^{-1}\,Mpc^{-1}}$ \cite{Riess:2021jrx} is still at $1$-$\sigma$ too, we observe a slight shift from the fiducial values of the cosmographic parameters in the $(q_0,\, j_0)$ plane compared with allowing the dipole parameters to vary freely.

This work is presented as follows. In Sect. \ref{sec:1} we introduce the cosmographic parameters and discuss how the luminosity distance is reconstructed adopting this approach. Sect. \ref{sec:2} explains the methodology adopted for the two main analyses of this work while Sect. \ref{sec:4} presents the three main supernova samples. Then, in Sect. \ref{sec:5} we discuss the numerical outcomes of our analyses and compare them with the CMB dipole in Sect. \ref{sec:6}. Finally, in Sect. \ref{conc} the conclusions and future work are discussed.

\section{Cosmographic set up}\label{sec:1}

The cosmographic approach aims to constrain kinematic quantities without imposing {\it a priori} cosmological model and only under the validity of homogeneity and isotropy of the Universe \cite{Visser:2004bf, Visser:2003vq, Cattoen:2007sk, Cattoen:2007id}. Each parameter describes kinematic quantities such as the acceleration (or deceleration) of the Universe or the variation of the acceleration. They are probed through the expansion in a Taylor series of the scale factor $a(t)$
\begin{equation}\label{scalefac}
 a(t)\simeq1+H_0\Delta t-\frac{1}{2}H^2_0q_0\Delta t^2+\frac{1}{6}H^3_0j_0\Delta t^3+\dots   
\end{equation}
where $\Delta t^n\equiv (t-t_0)^n$ with $n=\{1,\ 2,\ 3, \dots\}$ and $t_0$ being the present time. In Eq. \eqref{scalefac} $q_0$ is the deceleration parameter probing the current acceleration or deceleration of the Universe while $j_0$ is labeled as the jerk parameter linked with the variation of $q$ through
\begin{equation}\label{eq:j0}
    j_0=\frac{dq}{dz}\Big|_{z=0}+2q^2_0+q_0.
\end{equation}
Considering Eq. \eqref{eq:j0} and the expression for the deceleration parameter, i.e. $q(z)=-1+(1+z)H(z)^{-1}(dH/dz)$ we are able to express these cosmographic parameters for a cosmological model of choice. 

For our purposes we take into consideration the $\Lambda$CDM scenario in which the Hubble rate $H(z)=H_0\sqrt{\Omega_{\rm m}(1+z)^3+\Omega_\Lambda}$ with $\Omega_\Lambda\equiv1-\Omega_{\rm m}$ yielding
\begin{equation}
    q_0=\frac{3}{2}\Omega_{\rm m}-1,\quad j_0=1.
\end{equation}
Considering $\Omega_{\rm m}=0.3$ we end up with 
\begin{equation}
    q_0=-0.55,\quad j_0=1.
\end{equation}
Furthermore, the cosmographic parameters can be linked to the luminosity distance up to the, e.g. third order, through a Taylor expansion
\begin{equation}
    d^{(3)}_{\rm L}(z)=\frac{cz}{H_0}\left(\alpha_0+\alpha_1\frac{z}{2}+\alpha_2\frac{z^2}{6}\right),\label{eq:taylor}
\end{equation}
where $c$ is the speed of light, while the $\alpha_i$ quantities are
\begin{subequations}
\begin{align}
    &\alpha_0=1,&\\
    &\alpha_1=1-q_0,&\\
    &\alpha_2=-1+q_0(1+3q_0)-j_0.&
\end{align}
\end{subequations}
Even though cosmography is a powerful approach to constrain key parameters without the need to assume a background cosmological model, it suffers from two main issues when Eq. \eqref{eq:taylor} is adopted, specifically \cite{Gruber:2013wua}
\begin{itemize}
    \item [-] {\bf Truncation problem.} Because of the unfeasibility of considering an unlimited number of terms in the expansion for numerical analysis the series needs to be truncated at a certain order, typically to the one where the cosmographic parameters are constrained by current data. However, this gives rise to errors that can be alleviated by considering higher orders but with the shortcoming of complicating the analysis.
    \item [-] {\bf Convergence problem at high-$z$.} This problem comes from how the series diverges when it reaches redshift $z\geq 1$. Over the years, many strategies have been proposed to heal this issue, e.g. adoption of auxiliary variables \cite{Cattoen:2007sk, Aviles:2012ay, Busti:2015xqa, Capozziello:2020ctn}, Pad\'e approximants \cite{Gruber:2013wua, Aviles:2014rma, Capozziello:2017ddd, Capozziello:2020ctn, Hu:2022udt} or Chebyshev polynomials \cite{Capozziello:2017nbu}.
\end{itemize}

Even though in analyses employing Pad\'e parameterization of the luminosity distance it is common to employ the $(2,1)$ Pad\'e approximant due to its convergence at $z\geq 1$, see e.g. Refs. \cite{Aviles:2014rma, Capozziello:2020ctn}, we employ a $(1, 2)$ Pad\'e series for two main reasons, i.e., first  we work at $z\leq 1$, where the convergence problem is quite limited, second at small redshifts only, the $(1, 2)$ expansion appears well-performing \cite{Gruber:2013wua}. The luminosity distance becomes, 
\begin{align}\label{eq:pade}
d^{(1,2)}_L(z)
&= \frac{cz}{H_0}\biggl[
1-\frac{1}{2}(1-q_0)z+ \notag\\
&\qquad
+\frac{1}{12}\left(2j_0+5-(8+3q_0)q_0\right)z^2
\biggr]^{-1}\,.
\end{align}
In our subsequent analyses, our priors are taken in order to avoid divergences in the above Pad\'e approximant, avoiding poles in the denominator. This property is perfectly in line with previous literature, in which it has been shown that viable cosmological priors do not break down Pad\'e expansion validity \cite{Aviles:2014rma}.

\section{Data analysis}\label{sec:2}

We assume the background described by a spatially flat Friedmann--Lema\^itre--Robertson--Walker metric. Within this framework, cosmography allows the expansion history of the Universe to be characterized without assuming a specific cosmological model.

In a homogeneous and isotropic Universe, a single set of cosmographic parameters should describe the SNe Ia Hubble diagram over the full redshift range. 

However, the inferred distance-redshift relation may be affected by local structure, peculiar motions, calibration uncertainties and residual systematics. These effects are expected to be especially relevant at low redshift, where peculiar velocities can represent a non-negligible fraction of the total observed recession velocity \cite{Davis:2010jq, Huterer:2020szt, Carr:2021lcj, Peterson:2021hel, Pasten:2023rpc, Sorrenti:2024ugq}.

In this work, we extend the results presented in Ref. \cite{Sorrenti:2024ugq}, applying the same procedure used for Pantheon+\&SH0ES \cite{Scolnic:2021amr, Brout:2022vxf, Riess:2021jrx} also to DES-SNY5 \cite{DES:2024jxu, DES:2024upw, DES:2024hip} and Union3 \cite{Rubin:2023jdq} datasets. This allows us to investigate both the redshift dependence of the inferred cosmographic parameters and their dependence on the adopted SNe Ia catalogue. 

In particular, we perform two different analyses:
\begin{itemize}
    \item [-] in this first analysis, we assume an isotropic distance-redshift relation and use the CMB-frame redshift, $z_{\rm CMB}$, both to define the redshift intervals and to evaluate the luminosity distance. The CMB-frame redshift accounts for the motion of the observer with respect to the CMB rest frame. It is related to the heliocentric redshift through the relation \cite{Carr:2021lcj}
\begin{equation}
1+z_{\rm CMB}
=
\frac{1+z_{\rm HEL}}{1+z_{\odot}}\,,
\label{eq:zcmb_relation}
\end{equation}
where $z_{\odot}$ denotes the line-of-sight Doppler contribution due to the motion of the observer relative to the CMB frame.

The CMB-frame correction removes the contribution associated with the motion of the observer, but it does not remove the peculiar velocity of the SNe Ia host galaxies.

\item [-] in the second analysis, we introduce a free direction-dependent dipolar correction in the distance-redshift relation, starting from the heliocentric redshift, $z_{\rm HEL}$. 

Specifically, for a SN Ia observed in the direction $\hat{\boldsymbol{n}}$, the corrected redshift is defined as \cite{Carr:2021lcj}
\begin{equation}
1+z^{\rm COR}
=
\frac{1+z_{\rm HEL}}{1+\delta z}\,,
\label{eq:dipole_corrected_redshift}
\end{equation}
where
\begin{equation}
1+\delta z
=
\sqrt{
\frac{
1-\hat{\boldsymbol{n}}\cdot\boldsymbol{\beta}_{0}
}{
1+\hat{\boldsymbol{n}}\cdot\boldsymbol{\beta}_{0}
}
}, \quad \text{with}\quad \boldsymbol{\beta}_{0}
=
\frac{\boldsymbol{v}_{0}}{c}\,.
\label{eq:dipole_redshift}
\end{equation}
Here, $\boldsymbol{v}_0$ denotes the free velocity vector entering the dipolar Doppler correction, whose amplitude and direction are constrained simultaneously with the cosmographic parameters in each redshift interval.

It is important to stress that $\boldsymbol{v}_0$ is not directly identified with the bulk velocity of the SNe Ia hosts. 

Equations \eqref{eq:dipole_corrected_redshift} and \eqref{eq:dipole_redshift}
describe a single Doppler boost of the heliocentric redshift and therefore
define an effective rest frame for the SN sample\footnote{To give this quantity a physical interpretation, let
$\boldsymbol{v}_{\rm obs}$ be the observer velocity with respect to the
cosmological rest frame and assume that the SNe Ia contained within a given
redshift ball share a coherent component of their peculiar velocity field,
described by $\boldsymbol{v}_{\rm bulk}$. The dipolar velocity entering
Eq.~\eqref{eq:dipole_redshift} is then the relative velocity $
\boldsymbol{v}_0
=
\boldsymbol{v}_{\rm obs}
-
\boldsymbol{v}_{\rm bulk}$. Consequently, identifying $\boldsymbol{v}_{\rm obs}$ with the observer
velocity inferred from the CMB dipole gives $
\boldsymbol{v}_{\rm bulk}
=
\boldsymbol{v}_{\rm CMB}
-
\boldsymbol{v}_0$.}\,. 

\end{itemize}

Thus, in the first analysis we do not introduce an explicit model for the peculiar velocities of the SNe Ia hosts, whereas in the second analysis we include a coherent dipolar correction in the theoretical redshift. 

Both the isotropic and dipolar analyses are performed using the standard cosmographic Taylor expansion and a Padé rational approximation given in Eqs. \eqref{eq:taylor}-\eqref{eq:pade}, respectively.

\subsection{Redshift intervals}

Motivated by the redshift-layer approach of Ref. 
~\cite{Pasten:2023rpc}, we divide each supernova compilation into five redshift intervals, i.e.
\begin{equation}
\begin{split}
\mathcal{I}_{1} &: \qquad
0 \leq z \leq 0.15,
\\
\mathcal{I}_{2} &: \qquad
0 \leq z  \leq 0.30,
\\
\mathcal{I}_{3} &: \qquad
0 \leq z  \leq 0.45,
\\
\mathcal{I}_{4} &: \qquad
0 \leq z  \leq 0.60,
\\
\mathcal{I}_{5} &: \qquad
0 \leq z \leq 0.75.
\end{split}
\label{eq:redshift_int}
\end{equation}
Here, we assign the SNe Ia to the different redshift cuts using the CMB-frame redshift, $z_{\rm CMB}$, in the first analysis and the heliocentric redshift, $z_{\rm HEL}$, in the second, as provided by the corresponding catalogue.

The same prescription is applied separately to
Pantheon+\&SH0ES, DES-SNY5 and Union3.

Hence, for the $i$-th interval, we constrain a separate set of cosmographic parameters,
\begin{equation}
\boldsymbol{\theta}_{i}
=
\left\{q_{0,i},j_{0,i}
\right\},
\qquad
i=1,\ldots,5.
\label{eq:int_parameters}
\end{equation}
In the case of the Pantheon+\&SH0ES sample, the Cepheid calibration breaks the degeneracy between the absolute supernova magnitude $M$ and the Hubble constant, allowing $H_0$ to be directly constrained. By contrast, DES-SNY5 and Union3 do not provide an independent absolute calibration.
Therefore, for these two samples we fix $H_0 = 70~\mathrm{km\,s^{-1}\,Mpc^{-1}}$, which is fixed only to set the absolute normalization and should not be interpreted as a measurement of the Hubble constant.

\section{Supernova samples}\label{sec:4}

We employ the three main SNe Ia samples, namely Pantheon+ \cite{Scolnic:2021amr, Brout:2022vxf} calibrated with Cepheids from SH0ES \cite{Riess:2021jrx} to constrain the Hubble constant $H_0$, DES-SNY5 \cite{DES:2024jxu, DES:2024upw, DES:2024hip} and Union3 \cite{Rubin:2023jdq} catalogs to constrain the cosmographic parameters $(q_0,\, j_0)$ and  the dipole parameters $(v_0,\, \text{ra},\, \text{dec})$. In Tab. \ref{tab:numberSN} the number of data points, employed in every redshift interval $\mathcal{I}_i$, considering all three supernova samples, is presented. For all the SNe Ia samples considered in this work the likelihood $\ln\mathcal{L}$ takes the following form
\begin{equation}\label{eq:loglike}
    \ln\mathcal{L}=-\frac{1}{2}\Delta\mu^{\text{T}}\mathcal{C}^{-1}\Delta\mu\,,
\end{equation}
where the covariance matrix $\mathcal{C}$ varies according to the adopted catalog and the distance modulus $\mu$ being
\begin{equation}
    \mu(z)=25+5\log\left[\frac{d_L(z)}{\text{Mpc}}\right]\,,
\end{equation}
where the luminosity distance $d_L(z)$ is Eq. \eqref{eq:taylor} if we are using Taylor or Eq. \eqref{eq:pade} if we are using Pad\'e.

Now, we briefly review the three samples employed in our analyses.

\begin{itemize}

    \item[-] {\bf Pantheon+\&SH0ES.} Consisting of $N_P=1701$ data points\footnote{The complete data set can be found in the \texttt{GitHub} repository \url{https://github.com/PantheonPlusSH0ES/DataRelease}.} in the redshift interval $z\in[0.01,\ 2.33]$ the Pantheon+ is the largest SNe Ia data catalog \cite{Scolnic:2021amr}. When employing this sample we also complement our analysis with the Cepheids-calibrated distances of the galaxy hosting the SNe Ia through the SH0ES program \cite{Riess:2021jrx} providing constraints also on $H_0$. Thus, for $\Delta\mu$ in Eq. \eqref{eq:loglike} we employ the same expression used in Ref. \cite{Sorrenti:2024ugq}
    \begin{equation}
     \Delta\mu_i=\begin{cases}
     \mu^{\text{obs}}_i-\mu_i^{\text{Ceph}}+dM\quad i\in\text{Cepheid hosts},\\
     \mu^{\text{obs}}_i-\mu_i^{\text{theo}}+dM\quad \text{otherwise},
     \end{cases}
    \end{equation}
    where $\mu_i^{\text{obs}}$ is the observed SN Ia distance modulus, $\mu_i^{\text{Ceph}}$ the Cepheid-calibrated distance modulus, $\mu_i^{\text{theo}}$ the theoretical distance modulus and $dM$ represents a nuisance parameter.
    
    \item[-] {\bf DES-SNY5.} This sample encompasses $N_D=1635$ data points\footnote{The complete data set can be found in the \texttt{GitHub} repository \url{https://github.com/des-science/DES-SNY5}.} in the redshift interval $z\in[0.025,\ 1.13]$ \cite{DES:2024jxu, DES:2024upw, DES:2024hip} where 145 SNe Ia found in $z\in [0.025,\ 0.1]$ are in common with the Pantheon+ sample \cite{Scolnic:2021amr}. When using the DES-SNY5 sample we employ the likelihood as implemented in Ref.  \cite{Herold:2025hkb}\footnote{The complete likelihood can also be found in the \texttt{GitHub} repository \url{https://github.com/tkarwal/cosmo_likelihoods/tree/main/DESY5_mp_lkl_pre_Dovekie}.}. Specifically, $\Delta\mu$ is considered as
    \begin{equation}
        \Delta\mu_i=\mu^{\text{obs}}_i-\mu^{\text{theo}}_i,
    \end{equation}
    where $\mu^{\text{obs}}_i=m^{\text{corr}}_{B,i}-M$ with $m^{\text{corr}}_{B,i}$ the corrected magnitude, $M$ is a nuisance parameter and $\mu^{\text{theo}}_i$ is, as always, the theoretical distance moduli.
    
    \item[-] {\bf Union3.} Our last employed sample consists of $N_U=2087$ data points\footnote{The complete data set can be found in the \texttt{GitHub} repository \url{https://github.com/rubind/union3_release}.} in the redshift interval $z\in [0.009,\ 2.26]$ \cite{Rubin:2023jdq} where $N_U=1363$ are also in common with Pantheon+ \cite{Scolnic:2021amr}.
    In this case, although a reduced Union3 data set with a corresponding \texttt{MontePython} likelihood is available\footnote{The reduced  \texttt{MontePython} likelihood is given in the following \texttt{GitHub} repository \url{https://github.com/GreenPlanck/montepython_likelihood.git}}, it is not suitable for the analysis performed in this work. In light of this, we construct our own likelihood following the standard SALT2/Tripp standardization approach commonly adopted in SNe Ia analyses \cite{Tripp:1997wt,SDSS:2014iwm}.  Specifically, we use the released light-curve parameters $(m_B,x_1,c)$ and their corresponding $3\times 3$ covariance matrices. Here, the standardized distance modulus is computed as
\begin{equation}
\mu_i^{\rm obs}=m_{B,i}+\alpha x_{1,i}-\beta c_i-M,
\end{equation}
where the nuisance parameters $\alpha=0.14$ and $\beta=3.1$ are fixed to the values reported in Ref. \cite{SupernovaCosmologyProject:2015zlj}, while the absolute magnitude $M$ is treated as a free nuisance parameter.

In addition, we account for the residual scatter of standardized SNe Ia by fixing the intrinsic dispersion to $\sigma_{\rm int}=0.12\,{\rm mag}$, a representative value applied in cosmological analyses based on SALT2 \cite{SNLS:2011lii,SDSS:2014iwm,Brout:2022vxf}. 

This contribution is added in quadrature to the propagated uncertainty of the distance modulus for each SN Ia.

\end{itemize}

The constraints on the parameters are found by modifying the Cosmic Linear Anisotropy Solving System (\texttt{CLASS}) \cite{Lesgourgues:2011re, Blas:2011rf, Lesgourgues:2011rg} and using \texttt{MontePython} \cite{2013JCAP...02..001A, Brinckmann:2018cvx} to perform the MCMC analysis. On the other hand, the contour plots are generated using the \texttt{GetDist} package \cite{Lewis:2019xzd}. In Tab. \ref{tab:numberSN} we show the number of data points appearing in each $z$ interval. Furthermore, it is important to stress that, as highlighted above, the DES-SNY5 and Union3 samples have in common at low-$z$ data points with the Pantheon+ sample. We decided not to remove them, since we would have an impoverishment of the sample in the low redshift intervals.

\begin{table}
    \centering
    \setlength{\tabcolsep}{2.em}
    \renewcommand{\arraystretch}{1.2}
    \begin{tabular}{cccc}
    \hline
        $z$ & $N_{\text{PP+SH0ES}}$ & 
        $N_{\text{DES}}$ & $N_{\text{UN}}$ \\
        \hline
        0.15 & $825$ & $214$ & $819$\\
        0.30 & $1207$ & $403$ & $1275$\\
        0.45 & $1459$ & $808$ & $1640$\\
        0.60 & $1573$ & $1272$ & $1817$\\
        0.75 & $1653$ & $1641$ & $1945$\\
        \hline
    \end{tabular}
    \caption{Number of data points considered for every redshift interval and SNe Ia catalog, i.e. Pantheon+\&SH0ES (PP+SH0ES), DES-SNY5 (DES) and Union3 (UN).}
    \label{tab:numberSN}
\end{table}

\section{Numerical outcomes}\label{sec:5}

In this section, we present our main findings based on two main analyses and, for each investigation, we describe in detail the procedures adopted to pursue them and the main physical consequences. The tables with the results inferred from the MCMC together with the complete contour plots in Tabs. \ref{tab:cosmozCMB}-\ref{tab:cosmoDIPOLE} and Figs. \ref{fig:TaylorzCMB}-\ref{fig:PadeCMBDipole} in Appendix \ref{appendix}.

\subsection{First analysis results}

\begin{figure}[htbp]
\centering

\begin{subfigure}{0.49\textwidth}
\centering
\includegraphics[width=\linewidth]{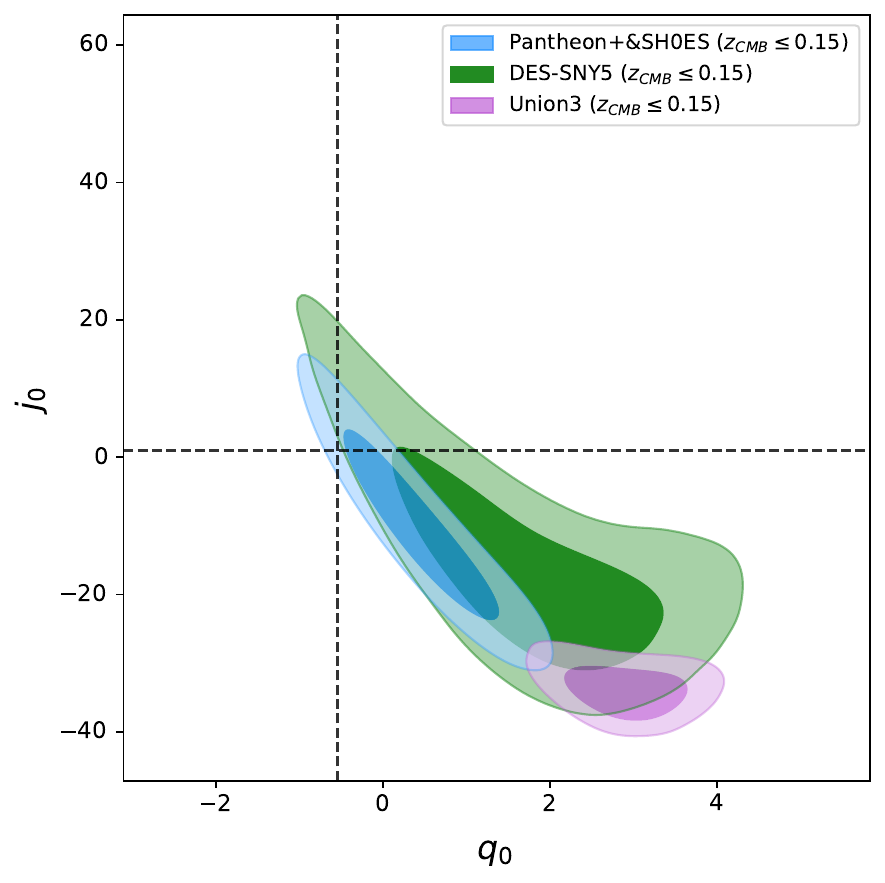}
\end{subfigure}
\begin{subfigure}{0.49\textwidth}
\centering
\includegraphics[width=\linewidth]{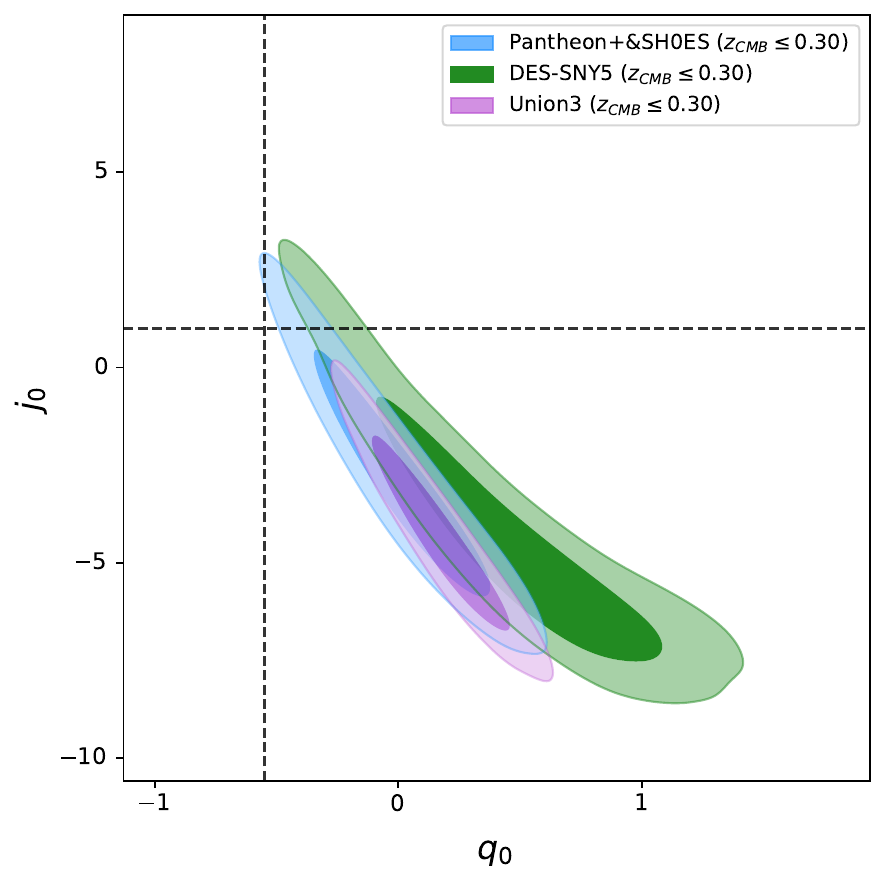}
\end{subfigure}
\begin{subfigure}{0.49\textwidth}
\centering
\includegraphics[width=\linewidth]{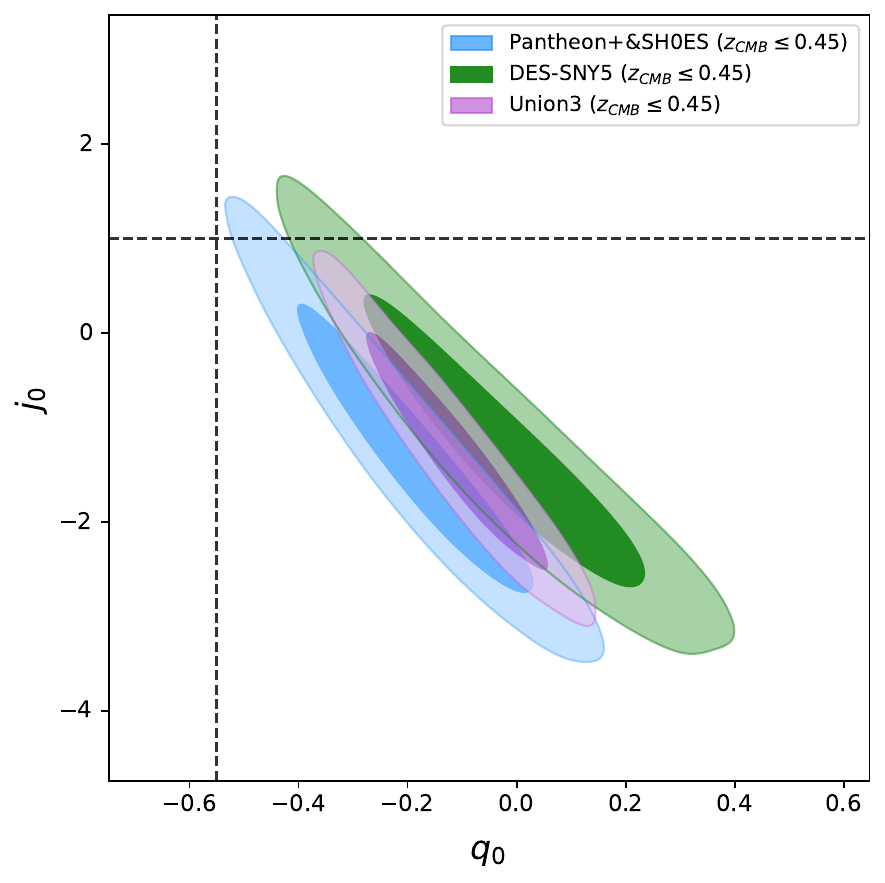}
\end{subfigure}
\begin{subfigure}{0.49\textwidth}
\centering
\includegraphics[width=\linewidth]{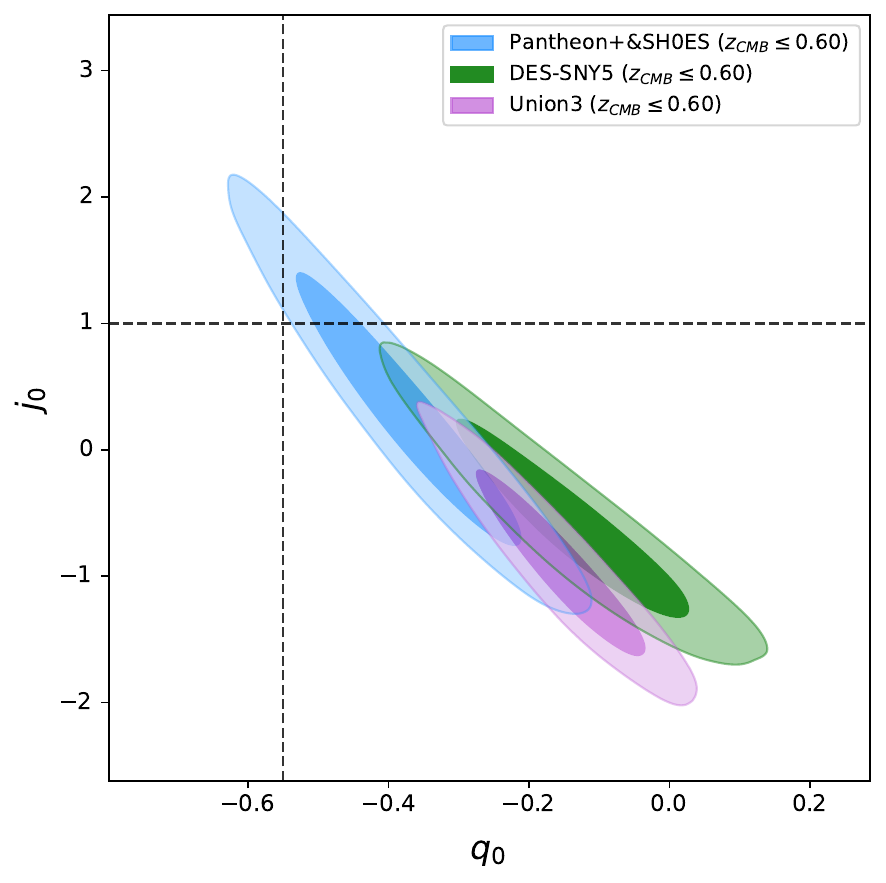}
\end{subfigure}
\begin{subfigure}{0.49\textwidth}
\centering
\includegraphics[width=\linewidth]{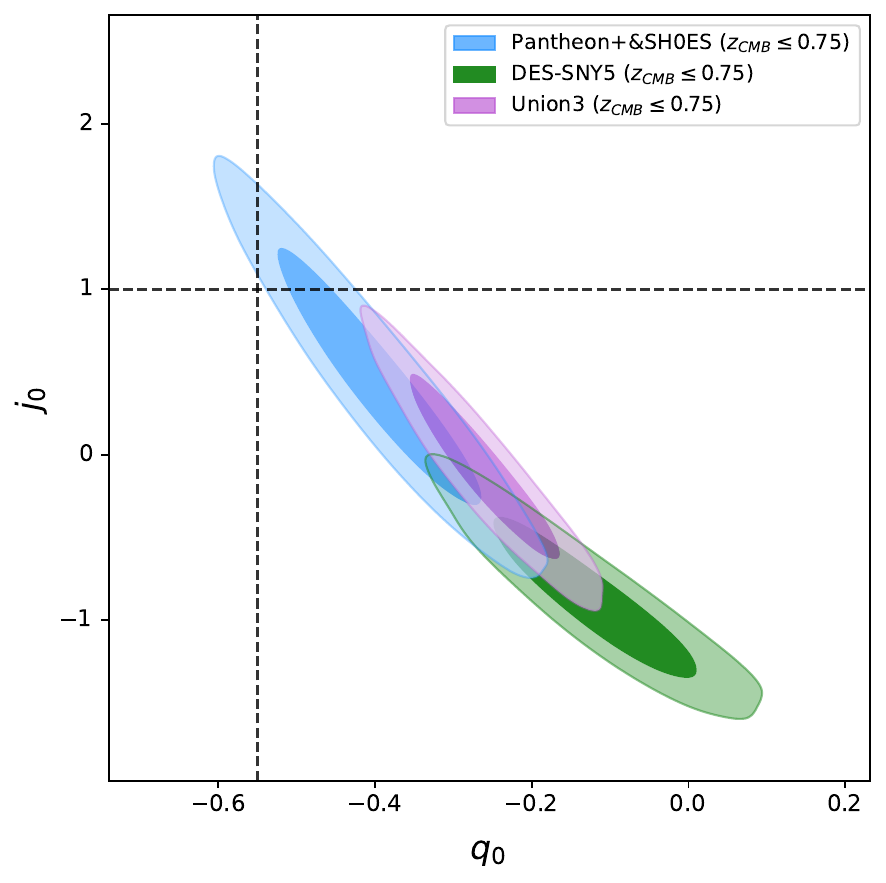}
\end{subfigure}

\caption{Confidence contours in the $(q_0,\, j_0)$ plane when the corrections due to the peculiar velocities are not taken into consideration and with the luminosity distance parameterized using the Taylor series. Blue, green and purple refer to the Pantheon+\&SH0ES, DES-SNY5 and Union3 catalogs, respectively. The dashed lines refer to the fiducial values of $q_0=-0.55$ and $j_0=1$ in the context of the $\Lambda$CDM scenario.}
\label{fig:q0j0Taylor}
\end{figure}

\begin{figure}[htbp]
\centering

\begin{subfigure}{0.49\textwidth}
\centering
\includegraphics[width=\textwidth]{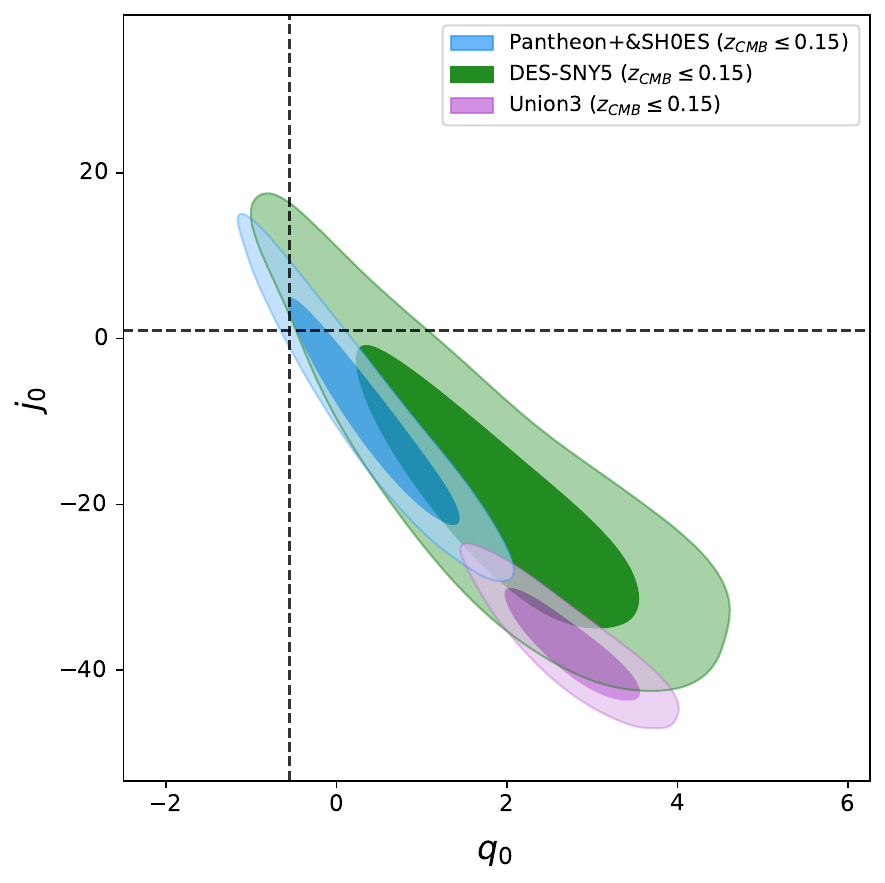}
\end{subfigure}
\begin{subfigure}{0.49\textwidth}
\centering
\includegraphics[width=\textwidth]{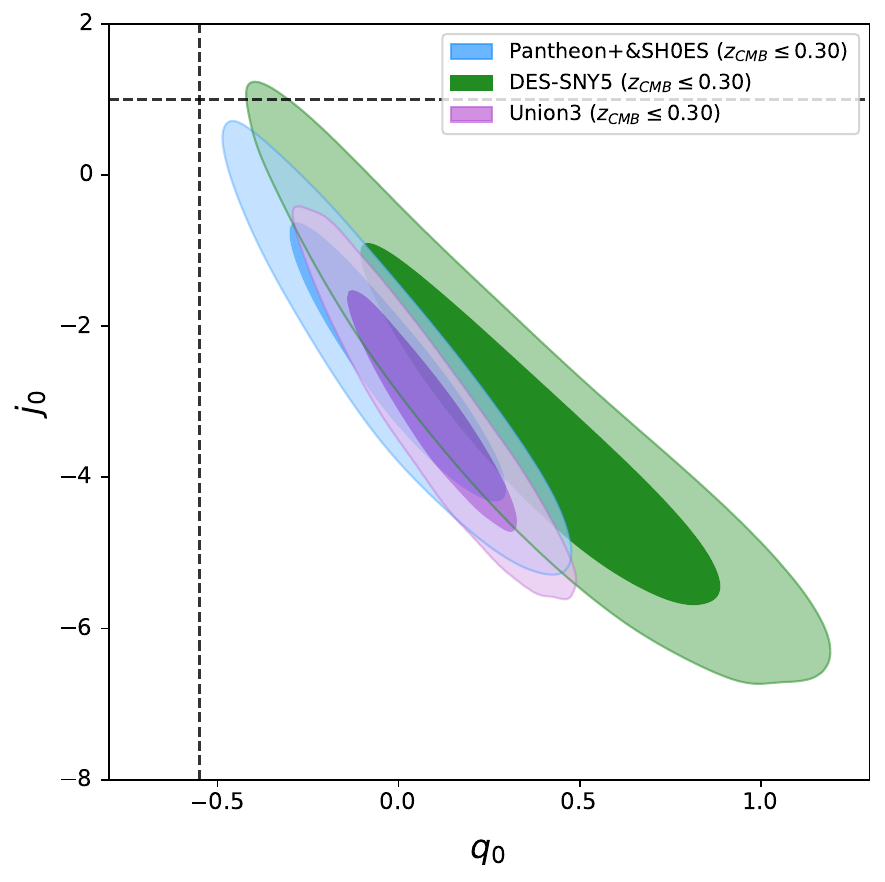}
\end{subfigure}
\begin{subfigure}{0.49\textwidth}
\centering
\includegraphics[width=\textwidth]{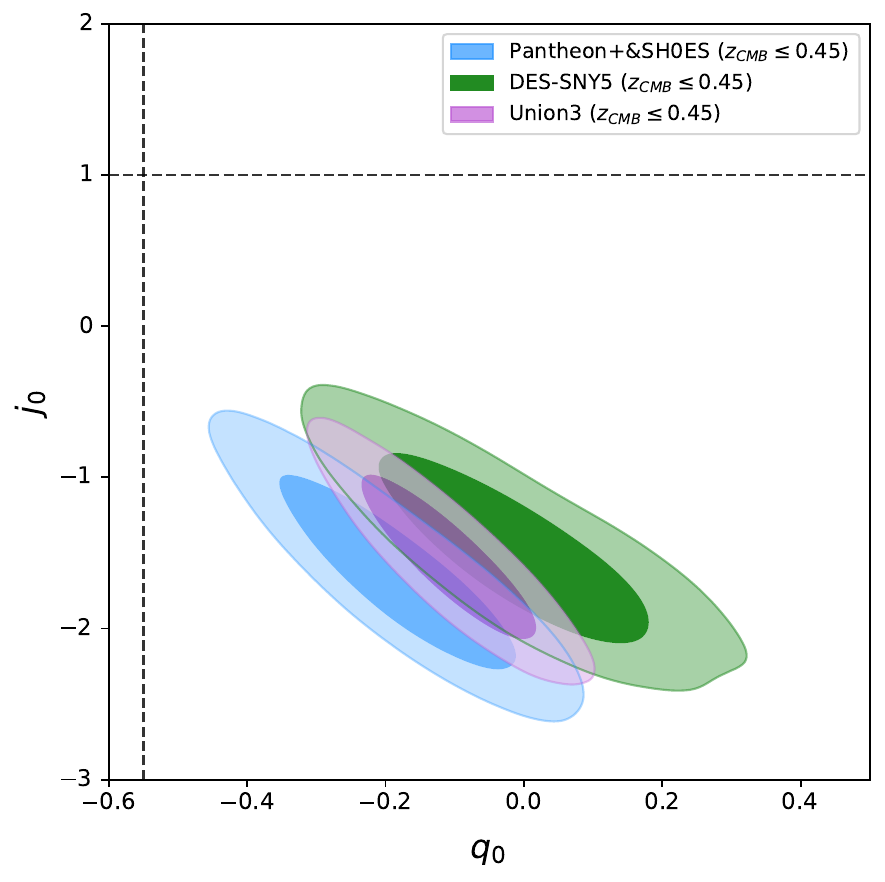}
\end{subfigure}
\begin{subfigure}{0.49\textwidth}
\centering
\includegraphics[width=\textwidth]{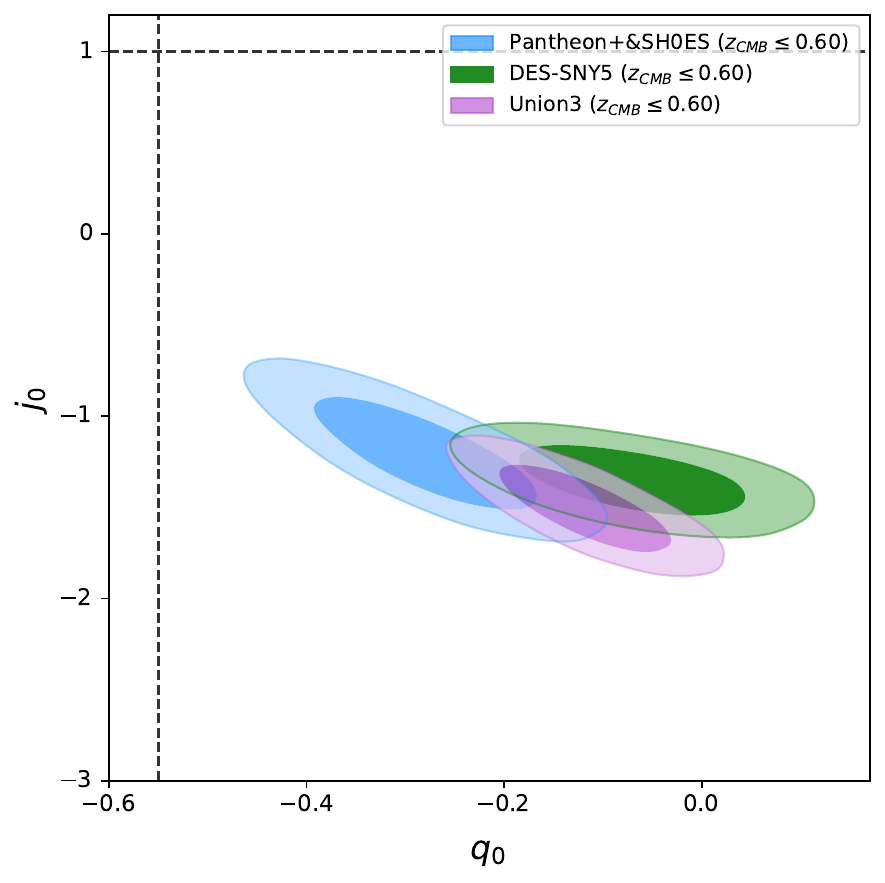}
\end{subfigure}
\begin{subfigure}{0.49\textwidth}
\centering
\includegraphics[width=\textwidth]{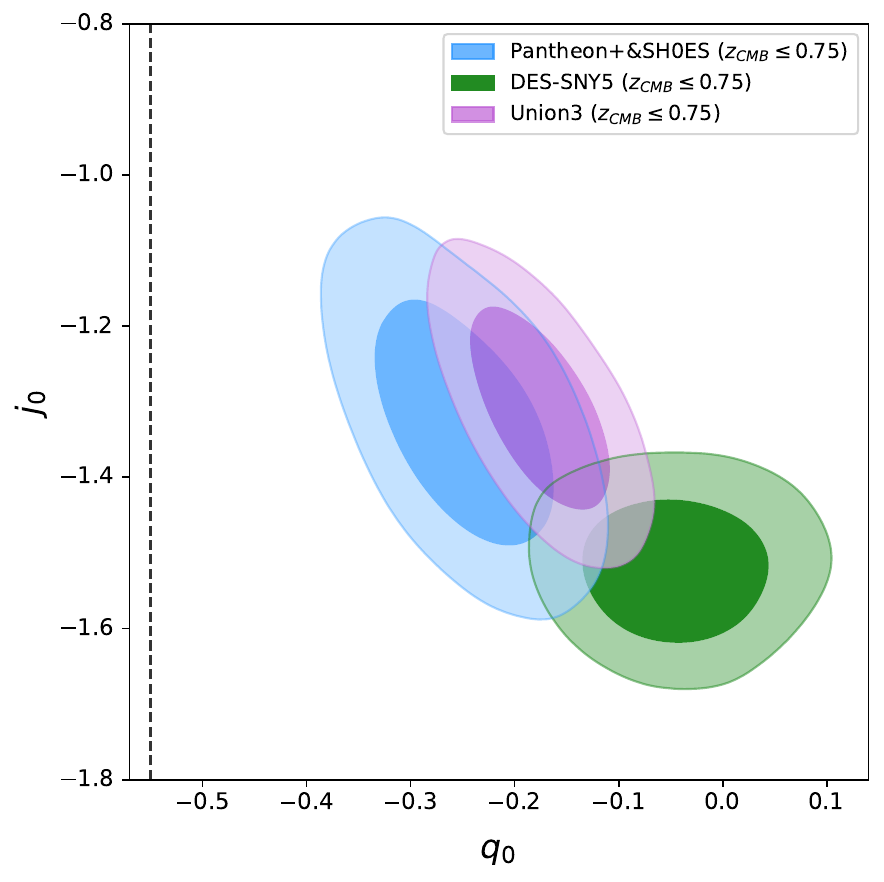}
\end{subfigure}

\caption{Same as Fig. \ref{fig:q0j0Taylor} but with the luminosity distance written adopting a $(1, 2)$ Pad\'e approximant.}
\label{fig:q0j0Pade}
\end{figure}

The first analysis deals with analyzing the cosmographic parameters without taking into account the corrections due to the peculiar velocities in five different redshift intervals considering the CMB-frame redshift $z_{\rm CMB}$. The results of this analysis are compared with the value of the cosmographic parameters as inferred from the $\Lambda$CDM scenario, i. e. $q_0=-0.55$ and $j_0=1$. The contours in the $(q_0,\, j_0)$ plane are depicted in Figs. \ref{fig:q0j0Taylor}-\ref{fig:q0j0Pade}. We assess the convergence of the chains employing the Gelman--Rubin criterion \cite{1992StaSc...7..457G} and requiring that $R-1<0.03$.

\begin{itemize}

    \item[-] $\mathcal{I}_1$ ($z_{\rm CMB}\leq 0.15$). Starting with the first interval, we infer that when using the Pantheon+\&SH0ES catalog our $H_0$ agrees at $1$-$\sigma$ with $H_0=73.6\pm 1.1\ \mathrm{km\,s^{-1}\,Mpc^{-1}}$  \cite{Brout:2022vxf} in both cases of Taylor or Pad\'e. 

    Focusing on the deceleration $q_0$ and jerk $j_0$ parameters in this interval we infer that the expected values from the $\Lambda$CDM scenario, i.e. $q_0=-0.55$ and $j_0=1$ fall inside the 95\% confidence region only when the Pantheon+\&SH0ES sample is adopted in both cases of using a third-order Taylor series or a $(1,2)$ Pad\'e approximant to parameterize the luminosity distance.

    \item[-] $\mathcal{I}_2$ ($z_{\rm CMB}\leq 0.30$). Expanding the number of data points we find that also in this case the inferred Hubble constant $H_0$ is compatible at $1$-$\sigma$ with the one probed by Pantheon+\&SH0ES. 

    Stopping at $z_{\text{CMB}}=0.30$ we find that the $(q_0,\, j_0)=(-0.55,\ 1)$ from the concordance paradigm do not lie in either confidence region for all three SNe Ia samples and in both scenarios of using Taylor or Pad\'e.
    
    \item[-] $\mathcal{I}_3$ ($z_{\rm CMB}\leq 0.45$). Expanding more the interval up to $z_{\rm CMB}= 0.45$ we find that also in this case regarding $H_0$ it still agrees at $1$-$\sigma$ with the Hubble constant inferred from Pantheon+\&SH0ES.

    Analogously to the previous case, even here the inferred values of the cosmographic parameters for the $\Lambda$CDM case lie outside the confidence regions.
     
    \item[-] $\mathcal{I}_4$ ($z_{\rm CMB}\leq 0.60$). Also for the fourth interval the compatibility between our $H_0$ and the estimation from Pantheon+\&SH0ES is at $1$-$\sigma$.

    The outcomes on $(q_0,\, j_0)$ reflect the previous with the exception of $z_{\text{CMB}}=0.15$ case. Specifically, also here whether we use Taylor or Pad\'e the values of the cosmographic parameters in the case of a perfect $\Lambda$CDM scenario fall outside the confidence regions.
    
    \item[-] $\mathcal{I}_5$ ($z_{\rm CMB}\leq 0.75$). Finally, enlarging even more our data sample, as for the previous cases, our Hubble constant is in agreement at $1$-$\sigma$ with $H_0=73.6\pm 1.1\ \mathrm{km\,s^{-1}\,Mpc^{-1}}$ \cite{Brout:2022vxf}. 

    As for the cosmographic parameters, we end up with the same considerations as before, i.e. $q_0=-0.55$ and $j_0=1$ do not fall into any confidence regions.
\end{itemize}

Summarizing the preliminary outcomes of this first analysis we find that for all intervals our $H_0$ is in agreement at $1$-$\sigma$ with $H_0=73.6\pm 1.1 \ \mathrm{km\,s^{-1}\,Mpc^{-1}}$ \cite{Brout:2022vxf} when both Taylor and Pad\'e are employed. 

Then, focusing on the $(q_0,\, j_0)$ parameters we find that with the exception of the interval up to $z_{\text{CMB}}=0.15$ in both cases of using the Taylor series up to the third order or the $(1,2)$ Pad\'e approximant to parameterize the luminosity distance when the correction due to the peculiar velocities is not taken into consideration, $(q_0,\, j_0)=(-0.55,\ 1)$ do not fall in any confidence regions even if the redshift interval is expanded.

\subsection{Second analysis results}\label{secondan}

\begin{figure}[htbp]
\centering

\begin{subfigure}{0.49\textwidth}
\centering
\includegraphics[width=\textwidth]{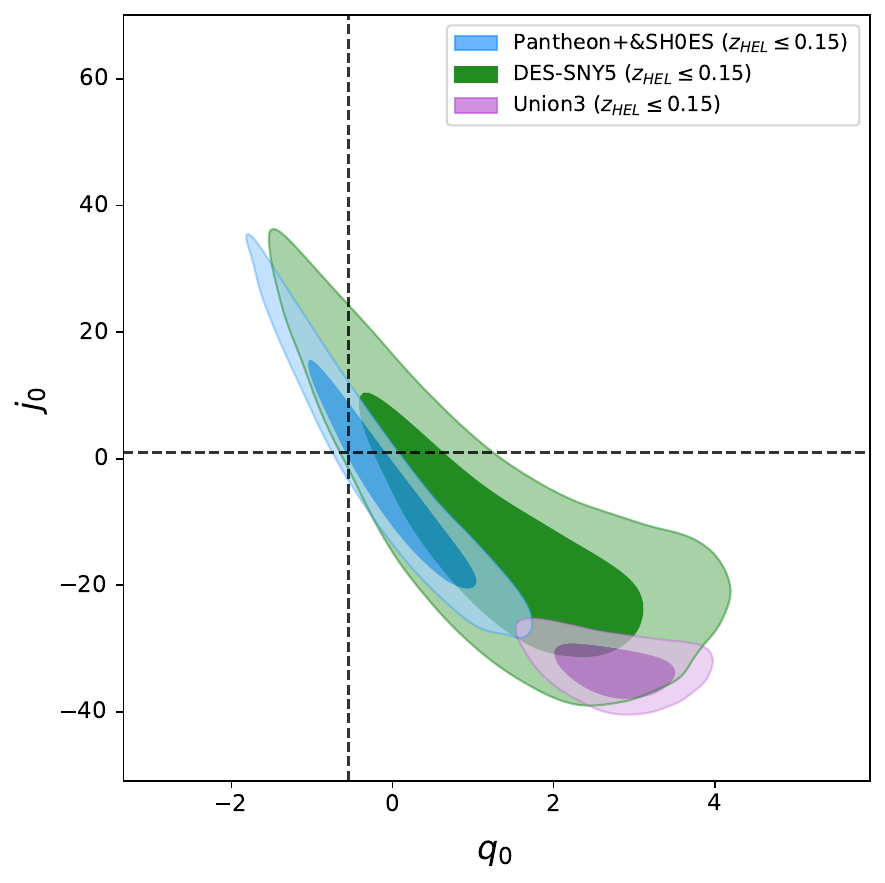}
\end{subfigure}
\begin{subfigure}{0.49\textwidth}
\centering
\includegraphics[width=\textwidth]{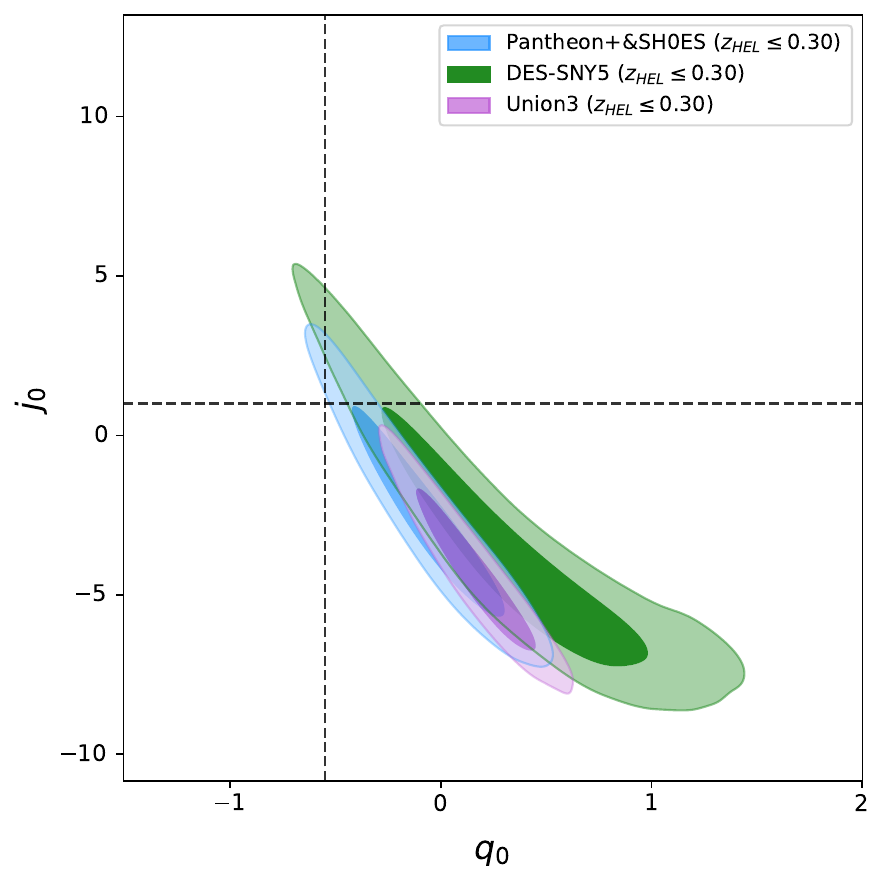}
\end{subfigure}
\begin{subfigure}{0.49\textwidth}
\centering
\includegraphics[width=\textwidth]{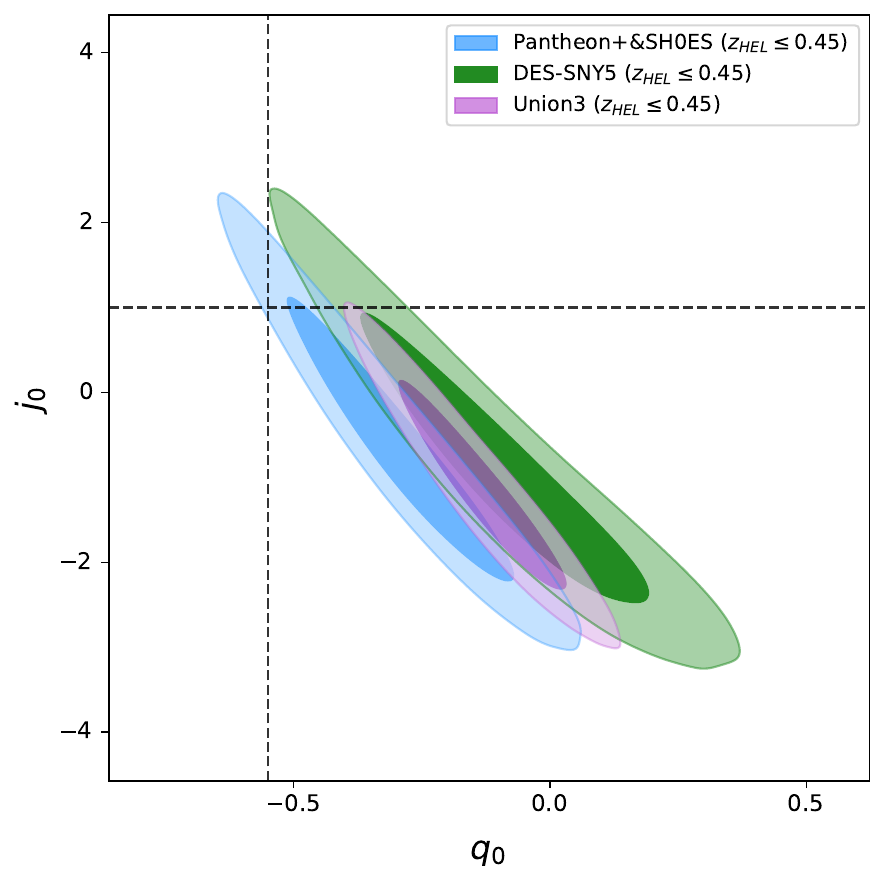}
\end{subfigure}
\begin{subfigure}{0.49\textwidth}
\centering
\includegraphics[width=\textwidth]{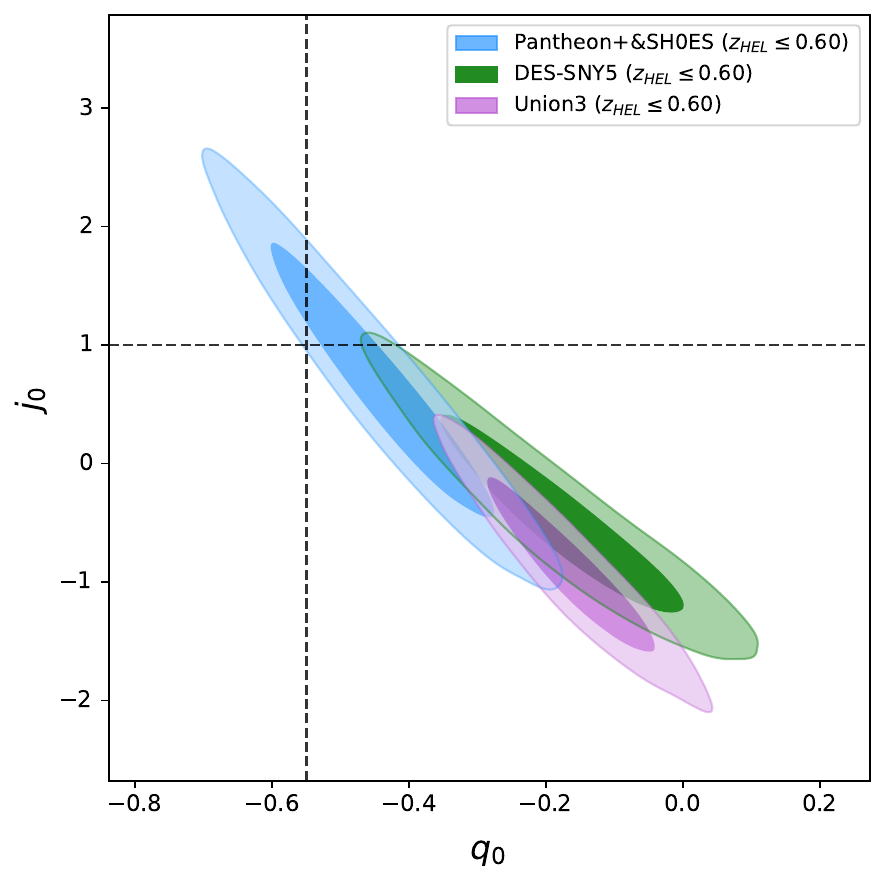}
\end{subfigure}
\begin{subfigure}{0.49\textwidth}
\centering
\includegraphics[width=\textwidth]{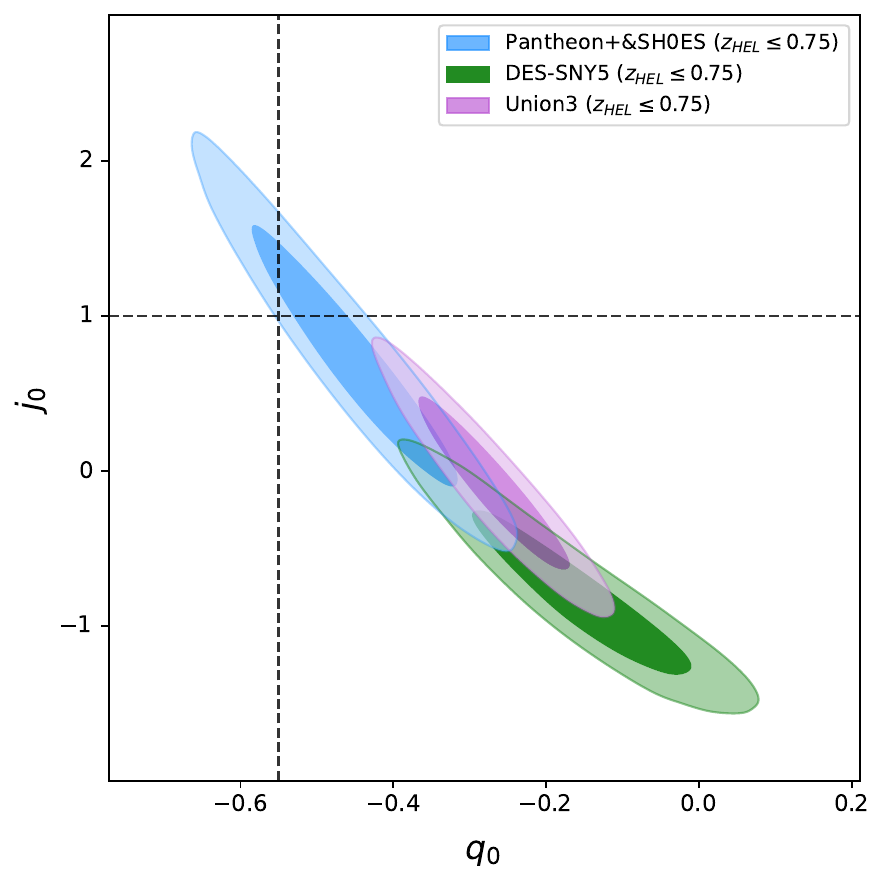}
\end{subfigure}

\caption{Confidence contours in the $(q_0,\, j_0)$ plane when the dipolar correction is considered with the luminosity distance written through the third order Taylor expansion. Also in this case the colors blue, green and purple refer to the Pantheon+\&SH0ES, DES-SNY5 and Union3 samples while the dashed lines represent $q_0=-0.55$ and $j_0=1$ for the fiducial $\Lambda$CDM model.}
\label{fig:q0j0dipoleTaylor}
\end{figure}

\begin{figure}[htbp]
\centering

\begin{subfigure}{0.49\textwidth}
\centering
\includegraphics[width=\textwidth]{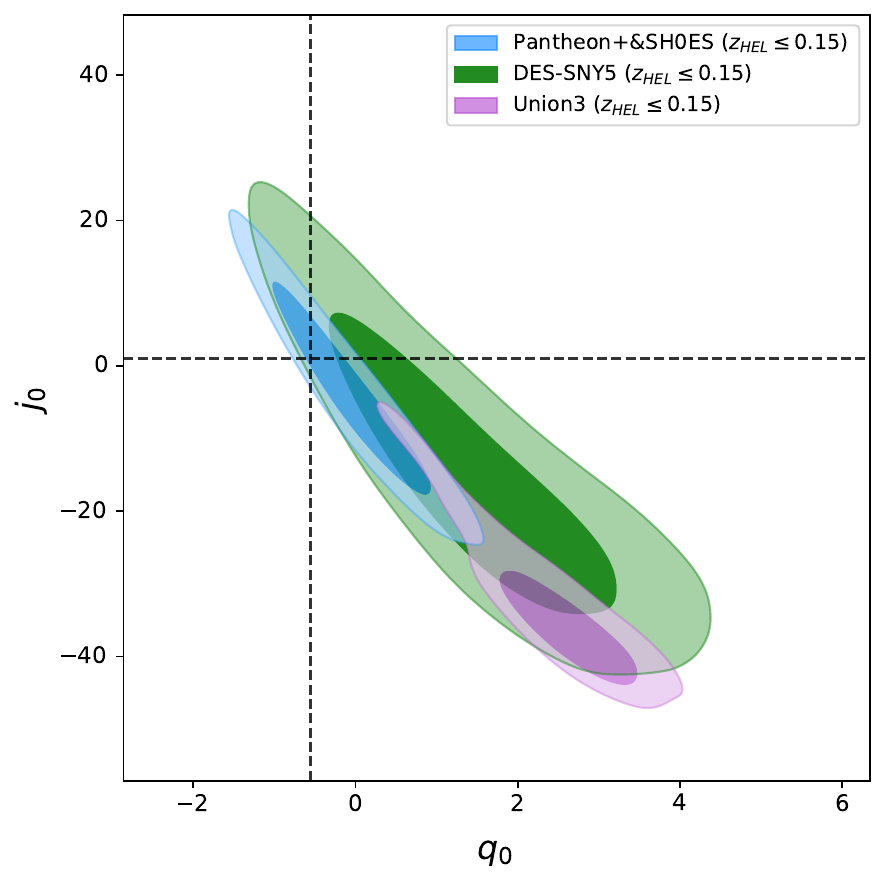}
\end{subfigure}
\begin{subfigure}{0.49\textwidth}
\centering
\includegraphics[width=\textwidth]{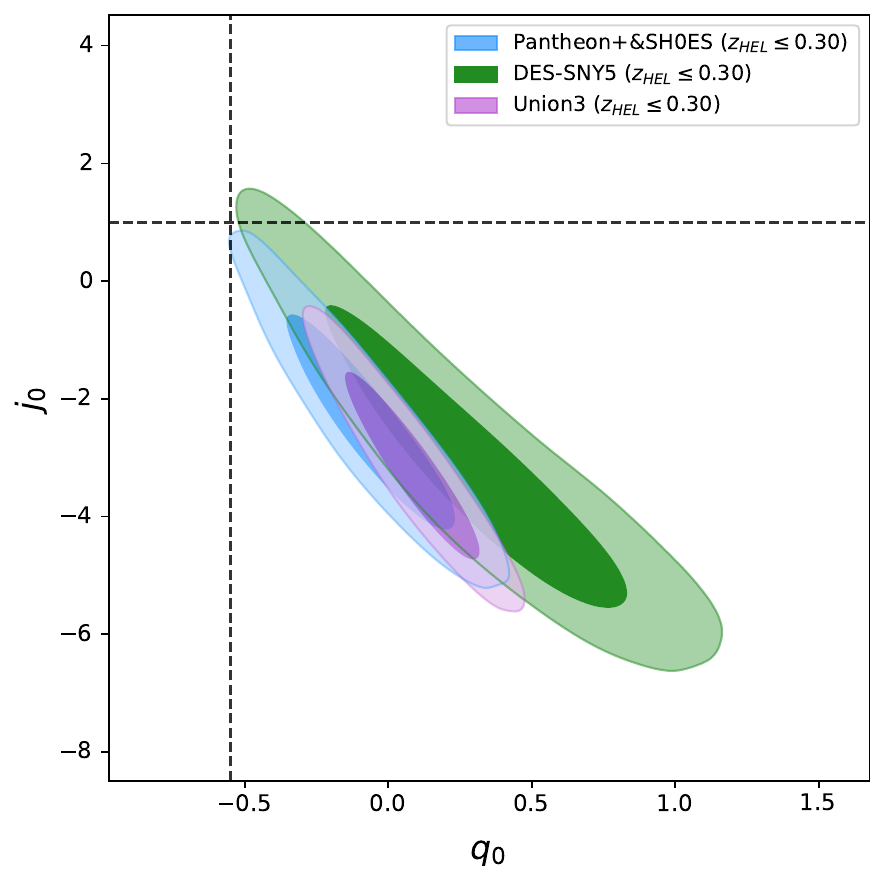}
\end{subfigure}
\begin{subfigure}{0.49\textwidth}
\centering
\includegraphics[width=\textwidth]{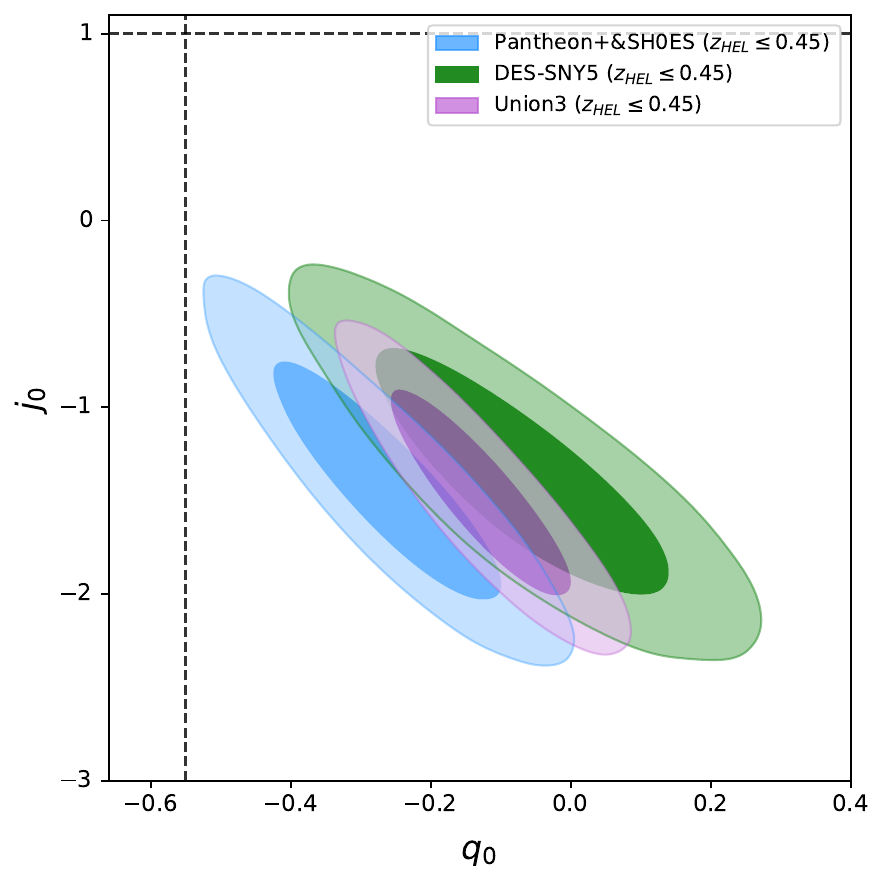}
\end{subfigure}
\begin{subfigure}{0.49\textwidth}
\centering
\includegraphics[width=\textwidth]{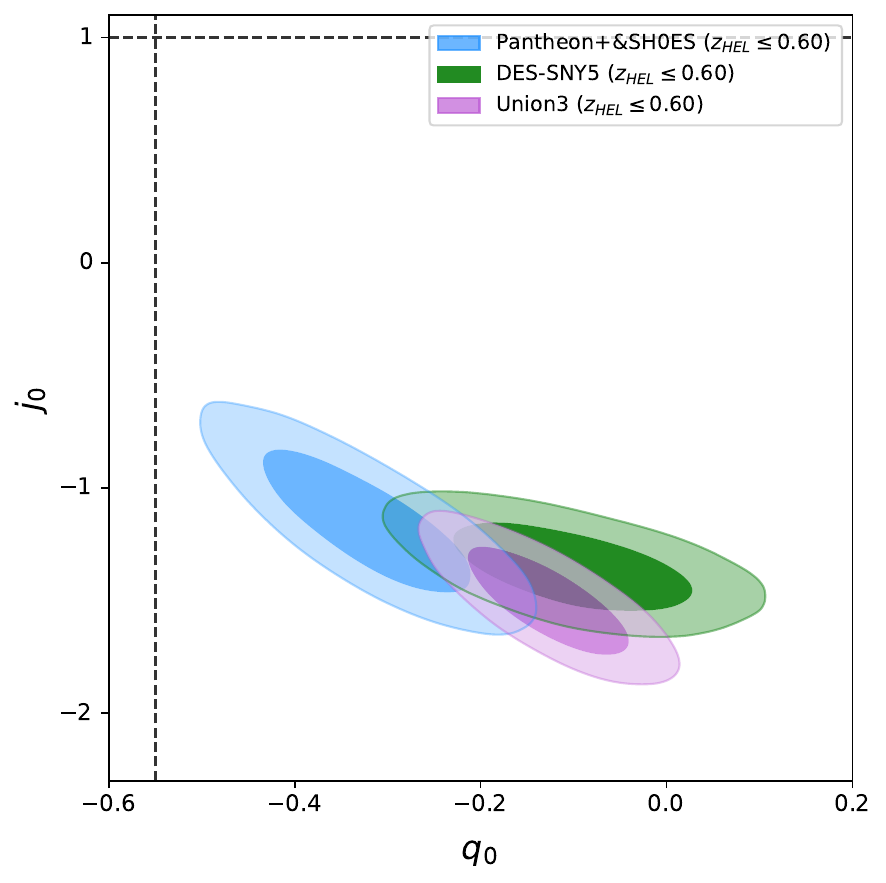}
\end{subfigure}
\begin{subfigure}{0.49\textwidth}
\centering
\includegraphics[width=\textwidth]{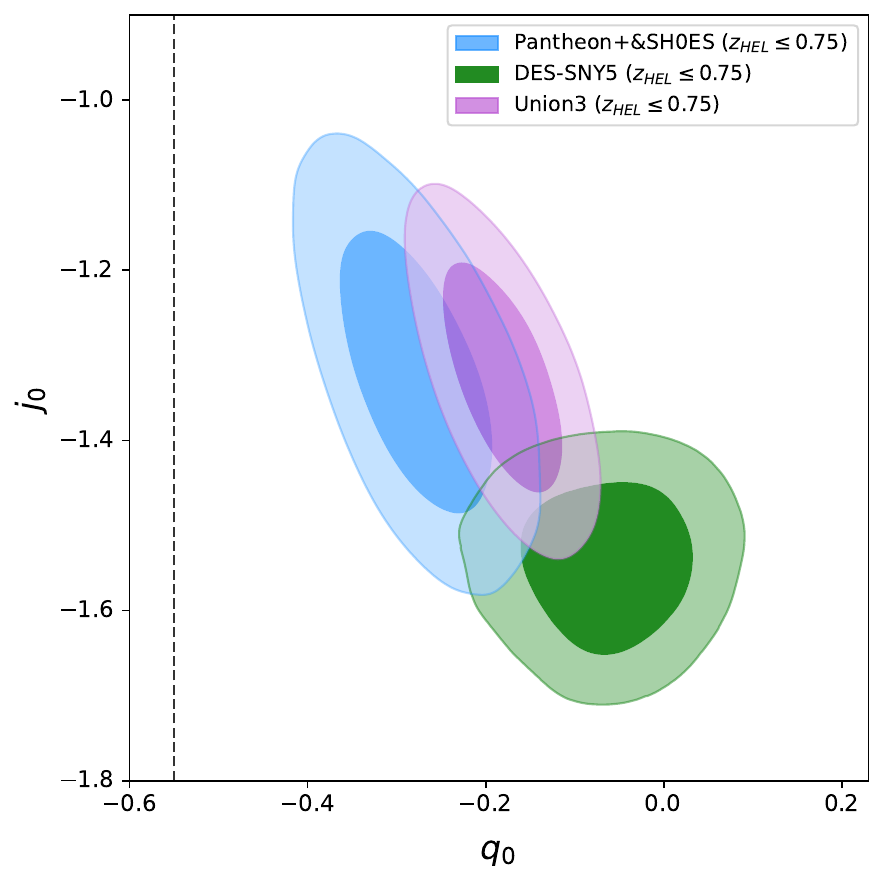}
\end{subfigure}

\caption{Same as Fig. \ref{fig:q0j0dipoleTaylor} but using the $(1,2)$ Pad\'e approximant to parameterize the luminosity distance.}
\label{fig:q0j0dipolePade}
\end{figure}

The second analysis focuses on constraining the cosmographic parameters using a dipole correction introducing three additional parameters, i.e. $(v_0,\, \text{ra},\, \text{dec})$ that are also constrained. In this last scenario, the interval subdivision is done in the same way as in the first analysis but considering the heliocentric redshift $z_{\rm HEL}$ instead of $z_{\rm CMB}$. The contours in the $(q_0,\, j_0)$ plane when the dipole correction is used are shown in Figs. \ref{fig:q0j0dipoleTaylor}-\ref{fig:q0j0dipolePade} and compared with the expected values of $q_0=-0.55$ and $j_0=1$ from the $\Lambda$CDM scenario. Also in this case, we require the convergence of the chains using the Gelman--Rubin criterion \cite{1992StaSc...7..457G} so that $R-1\lesssim 0.01$.

\begin{itemize}

\item[-] $\mathcal{I}_1$ ($z_{\rm HEL}\leq0.15$). In the first interval, we find that, when the Pantheon+\&SH0ES catalogue is employed, the inferred Hubble constant agrees at $1$-$\sigma$ with $H_0=73.6\pm1.1\,\mathrm{km\,s^{-1}\,Mpc^{-1}}$ \cite{Brout:2022vxf} for both the Taylor and Pad\'e parameterizations.

Concerning the remaining cosmographic quantities, the $\Lambda$CDM values $(q_0,j_0)=(-0.55,1)$ lie within the 95\% confidence region obtained with Pantheon+\&SH0ES and DES-SNY5 for both parameterizations of the luminosity distance. This is not the case for Union3, whose confidence regions are clearly separated from the fiducial $\Lambda$CDM cosmography already in this first interval.

\item[-] $\mathcal{I}_2$ ($z_{\rm HEL}\leq0.30$). Extending the sample to $z_{\rm HEL}=0.30$, the Pantheon+\&SH0ES determination of the Hubble constant remains consistent at $1$-$\sigma$ with the reference value.

For the Taylor expansion, Pantheon+\&SH0ES is still compatible with the fiducial $\Lambda$CDM values $(q_0,j_0)=(-0.55,1)$ within the 95\% confidence region. This is no longer the case for DES-SNY5 and Union3, whose confidence regions do not include the reference point. The departure becomes more pronounced for the Pad\'e parameterization, for which none of the three catalogues is compatible with the fiducial $\Lambda$CDM cosmography.

\item[-] $\mathcal{I}_3$ ($z_{\rm HEL}\leq0.45$). Increasing the upper redshift limit to $z_{\rm HEL}=0.45$ further reduces the uncertainties on the cosmographic parameters. At the same time, $H_0$ inferred from Pantheon+\&SH0ES continues to agree at the $1$-$\sigma$ level with the determination of Ref.~\cite{Brout:2022vxf}.

When the Taylor expansion is employed, Pantheon+\&SH0ES remains compatible with the fiducial $\Lambda$CDM cosmography at the $2$-$\sigma$ level, whereas the DES-SNY5 and Union3 confidence regions do not include the point $(q_0,j_0)=(-0.55,1)$. For the Pad\'e approximant, the discrepancy is more evident, as none of the three SNe Ia compilations is compatible with the reference $\Lambda$CDM values within the 95\% confidence regions.

\item[-] $\mathcal{I}_4$ ($z_{\rm HEL}\leq0.60$). In the fourth interval, the Hubble constant inferred from Pantheon+\&SH0ES remains in $1$-$\sigma$ agreement with the reference determination.

For the Taylor expansion, the Pantheon+\&SH0ES confidence region still includes the fiducial $\Lambda$CDM point at approximately the 95\% confidence level, while this is no longer the case for DES-SNY5 or Union3. When the Pad\'e parameterization is adopted, none of the three catalogues recovers the fiducial $\Lambda$CDM cosmography.

\item[-] $\mathcal{I}_5$ ($z_{\rm HEL}\leq0.75$). Finally, in the largest redshift interval considered in our analysis, the value of $H_0$ inferred from Pantheon+\&SH0ES remains compatible at $1$-$\sigma$ with $H_0=73.6\pm1.1\,\mathrm{km\,s^{-1}\,Mpc^{-1}}$ \cite{Brout:2022vxf}, independently of the adopted parameterization of the luminosity distance.

At the same time, this interval provides the tightest constraints on $(q_0,j_0)$. When the Taylor expansion is used, the Pantheon+\&SH0ES confidence region still reaches the fiducial $\Lambda$CDM point in approximately the 95\% confidence level, whereas DES-SNY5 and Union3 do not. For the Pad\'e approximant, also in this last case, the reference $\Lambda$CDM point lies outside the 95\% confidence regions obtained from all three SNe Ia catalogues.

\end{itemize}

The cosmographic results of this second analysis show that introducing the dipole correction modifies the comparison with the fiducial $\Lambda$CDM cosmography with respect to the previous analysis, while leaving the conclusions on the Hubble constant essentially unchanged. The inferred $H_0$ remains compatible at $1$-$\sigma$ with the Pantheon+\&SH0ES determination in all five redshift intervals for both the Taylor expansion and the Pad\'e approximant. Concerning the remaining cosmographic parameters, the comparison with the $\Lambda$CDM expectations depends strongly on both the SNe Ia sample and the adopted parameterization of the luminosity distance. When the Taylor expansion is employed, the Pantheon+\&SH0ES confidence contours remain compatible in approximately the 95\% confidence region with $(q_0,j_0)=(-0.55,1)$ throughout all five intervals. For DES-SNY5, this compatibility is found only in the first interval and is lost when the redshift range is enlarged. A different behaviour is obtained with Union3, whose confidence regions remain separated from the reference $\Lambda$CDM point in all five intervals. When the Pad\'e approximant is adopted, agreement with the fiducial cosmography is found only in the first interval for Pantheon+\&SH0ES and DES-SNY5, while from $\mathcal{I}_2$ onward none of the three catalogues is compatible with $(q_0,j_0)=(-0.55,1)$ within the 95\% confidence regions.

We now focus on the reconstructed dipole parameters. A remarkable feature emerging from Tab.~\ref{tab:cosmoDIPOLE} is their remarkable stability across the five redshift intervals. In all three SNe Ia compilations, the preferred velocity amplitude remains close to $v_0\simeq300~\mathrm{km\,s^{-1}}$, while the inferred sky direction shows only minor variations within each individual catalogue as the maximum redshift is increased. Such amplitudes are of the order expected for local bulk flows generated by the surrounding large-scale structure and are consistent with previous determinations based on SNe Ia \cite{Turnbull:2011ty,Davis:2010jq}. In our analysis, enlarging the redshift interval mainly improves the determination of the cosmographic parameters by increasing the available statistics, whereas it has only a limited impact on the reconstructed dipole parameters, as expected since peculiar velocities predominantly affect the nearby Universe \cite{Davis:2010jq}.

Pantheon+\&SH0ES and DES-SNY5 recover broadly consistent dipole directions, whereas Union3 favours a different preferred direction, especially in declination. Nevertheless, all three catalogues recover velocity amplitudes compatible with previous determinations based on SNe Ia and galaxy peculiar velocity surveys \cite{Turnbull:2011ty,Watkins:2008hf}. Therefore, the absence of any statistically significant evolution of $(v_0,\mathrm{ra},\mathrm{dec})$ from $\mathcal{I}_1$ to $\mathcal{I}_5$ indicates that the reconstructed dipole is robust with respect to the adopted redshift intervals.

\section{Comparison with the CMB dipole}\label{sec:6}

The dipole of the CMB results from our motion with respect to the rest frame of the CMB and represents its largest temperature anisotropy \cite{Sullivan:2021yms}. The Planck Collaboration has determined the amplitude of the dipole with extreme accuracy that exhibits an estimation of the systematic uncertainties of approximately $0.025\%$, which yields $v_0=369.82\pm 0.11~\mathrm{km\,s^{-1}}$ \cite{Planck:2018nkj}. Furthermore, also the direction of the dipole is computed giving the following values for the right ascension and declination in degree units, i.e. $\text{ra}=167.942^\circ\pm 0.007^\circ$ and $\text{dec}=-6.944^\circ\pm0.007^\circ$ \cite{Planck:2013kqc, Planck:2018nkj}.

Bearing these estimations in mind, we perform an additional analysis complementary to the one carried out in Sect. \ref{secondan}. We here do not consider $(v_0,\, \text{ra},\, \text{dec})$ as free parameters, while fixing them instead to the inferred Planck values described above. 

\begin{figure}[htbp]
\centering

\begin{subfigure}{0.49\textwidth}
\centering
\includegraphics[width=\textwidth]{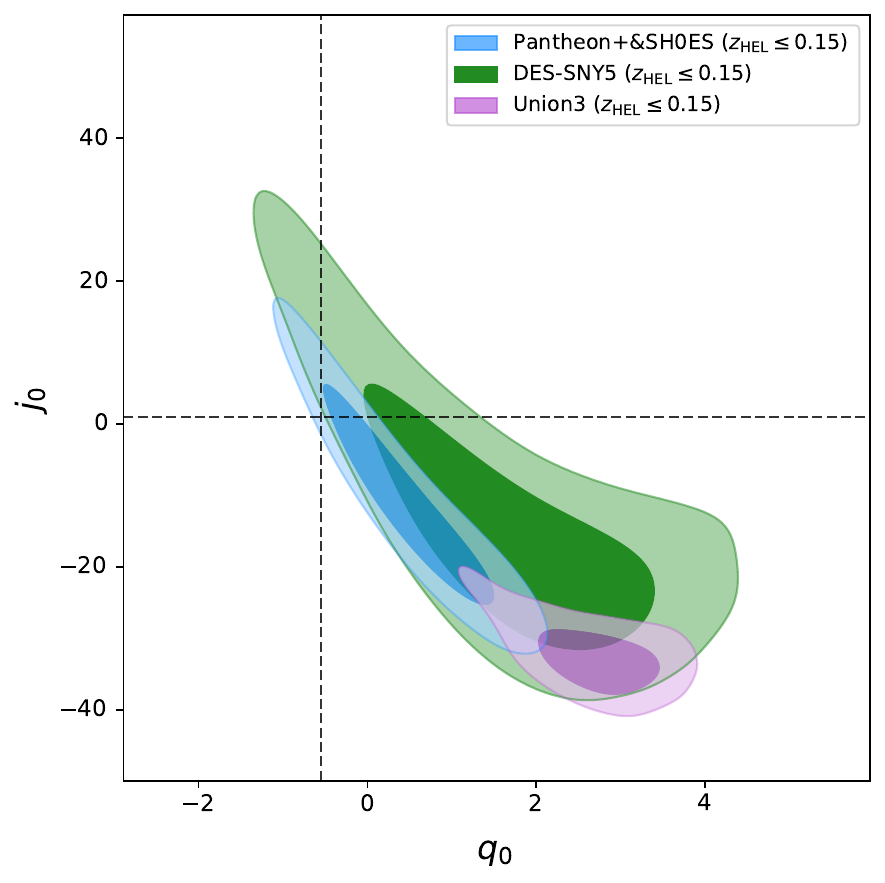}
\end{subfigure}
\begin{subfigure}{0.49\textwidth}
\centering
\includegraphics[width=\textwidth]{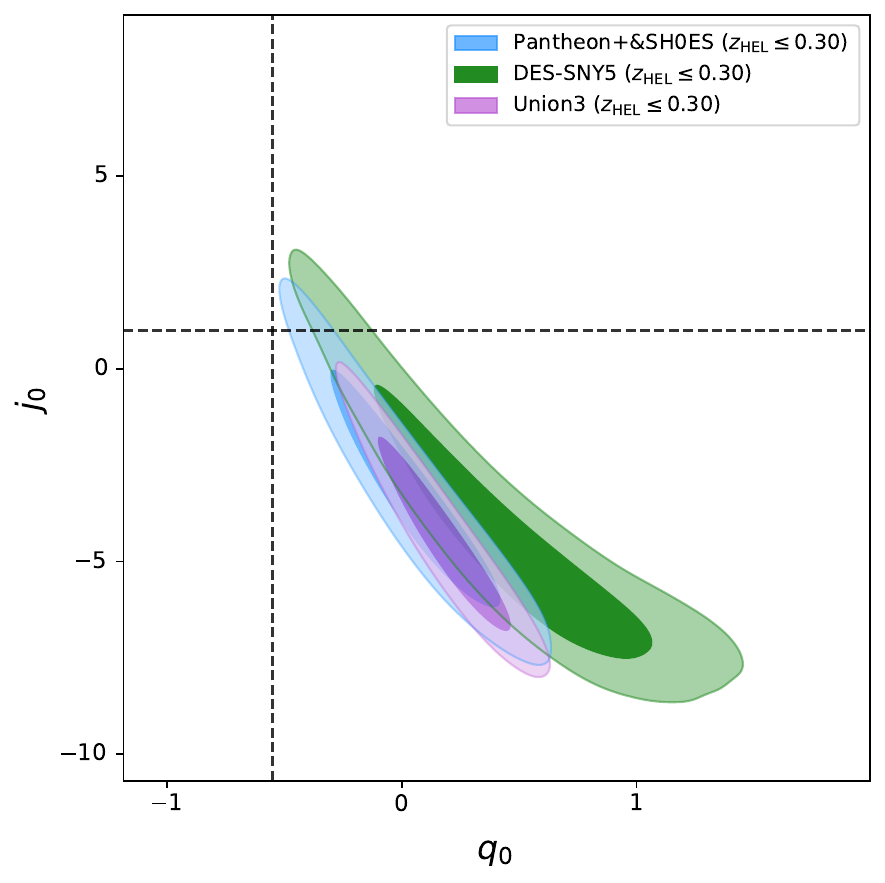}
\end{subfigure}
\begin{subfigure}{0.49\textwidth}
\centering
\includegraphics[width=\textwidth]{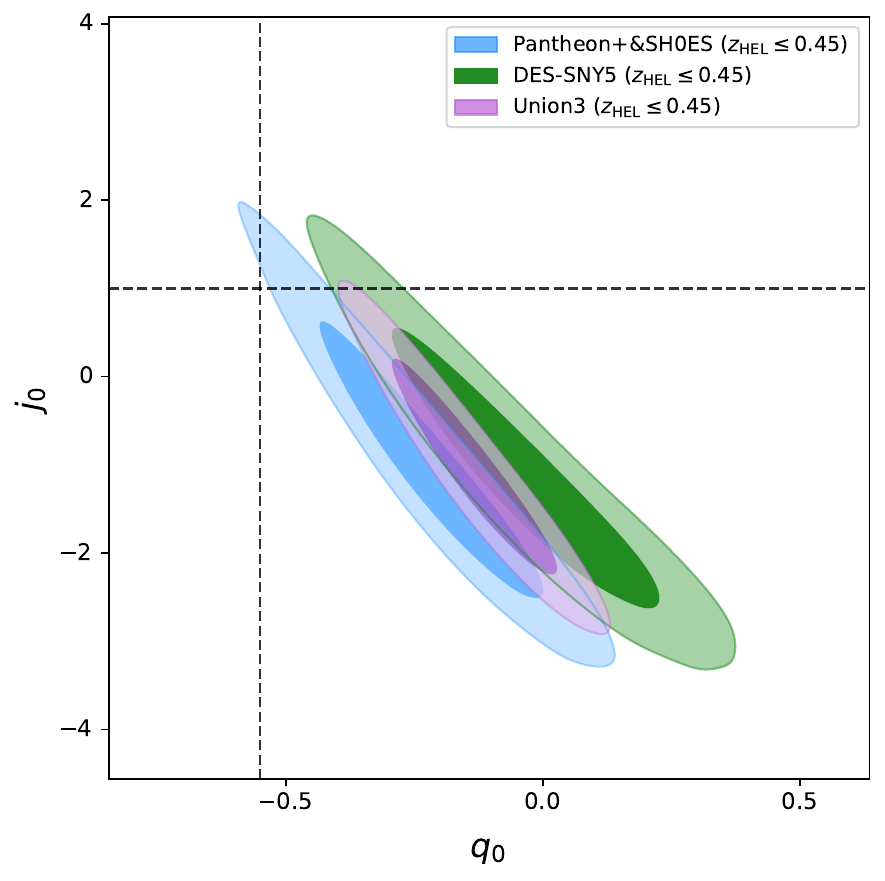}
\end{subfigure}
\begin{subfigure}{0.49\textwidth}
\centering
\includegraphics[width=\textwidth]{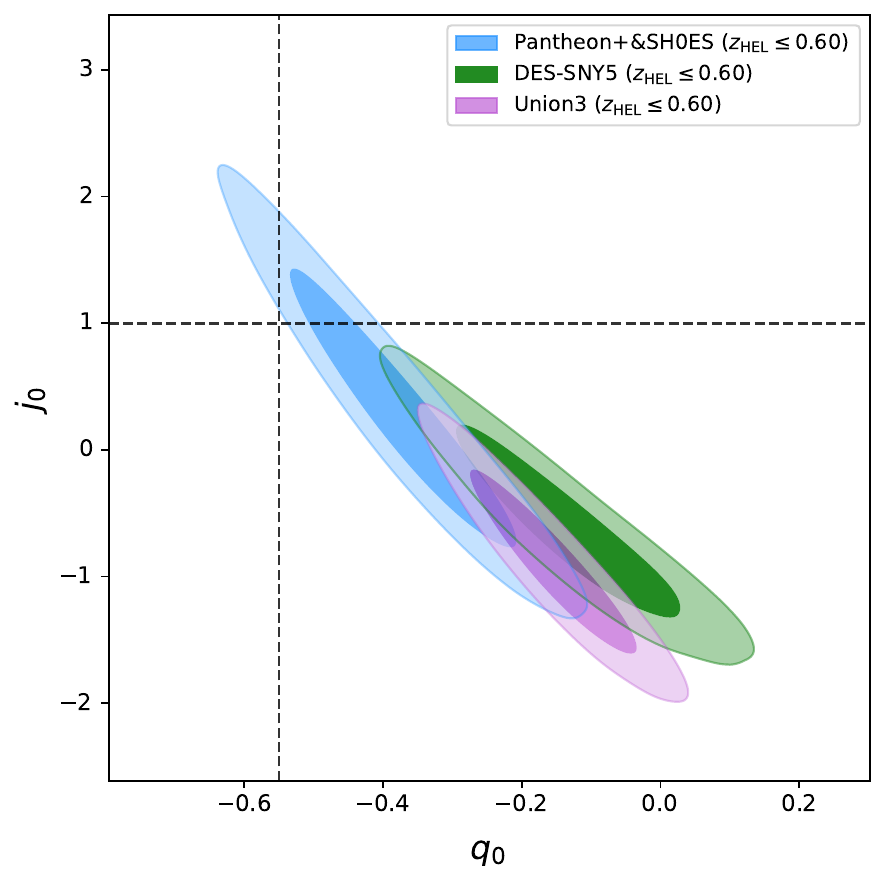}
\end{subfigure}
\begin{subfigure}{0.49\textwidth}
\centering
\includegraphics[width=\textwidth]{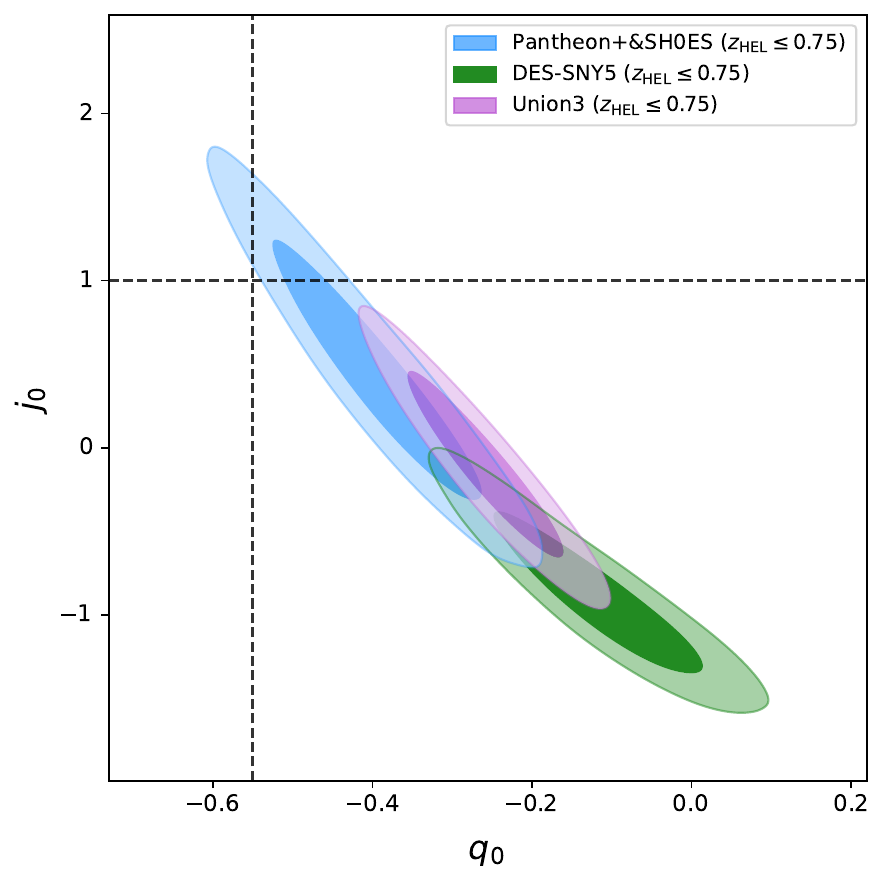}
\end{subfigure}

\caption{Confidence contours in the $(q_0,\, j_0)$ plane when $v_0 = 369.82~\mathrm{km\,s^{-1}}$, $\mathrm{ra} = 167.942^{\circ}$ and $\mathrm{dec} = -6.944^{\circ}$ \cite{Planck:2013kqc, Planck:2018nkj} and the luminosity distance written through the third order Taylor expansion. As always, the blue refers to the Pantheon+\&SH0ES, green to the DES-SNY5 and purple refers to the Union3 catalogs. The dashed lines represent $q_0=-0.55$ and $j_0=1$ for the fiducial $\Lambda$CDM model.}
\label{fig:q0j0CMBdipoleTaylor}
\end{figure}

\begin{figure}[htbp]
\centering

\begin{subfigure}{0.49\textwidth}
\centering
\includegraphics[width=\textwidth]{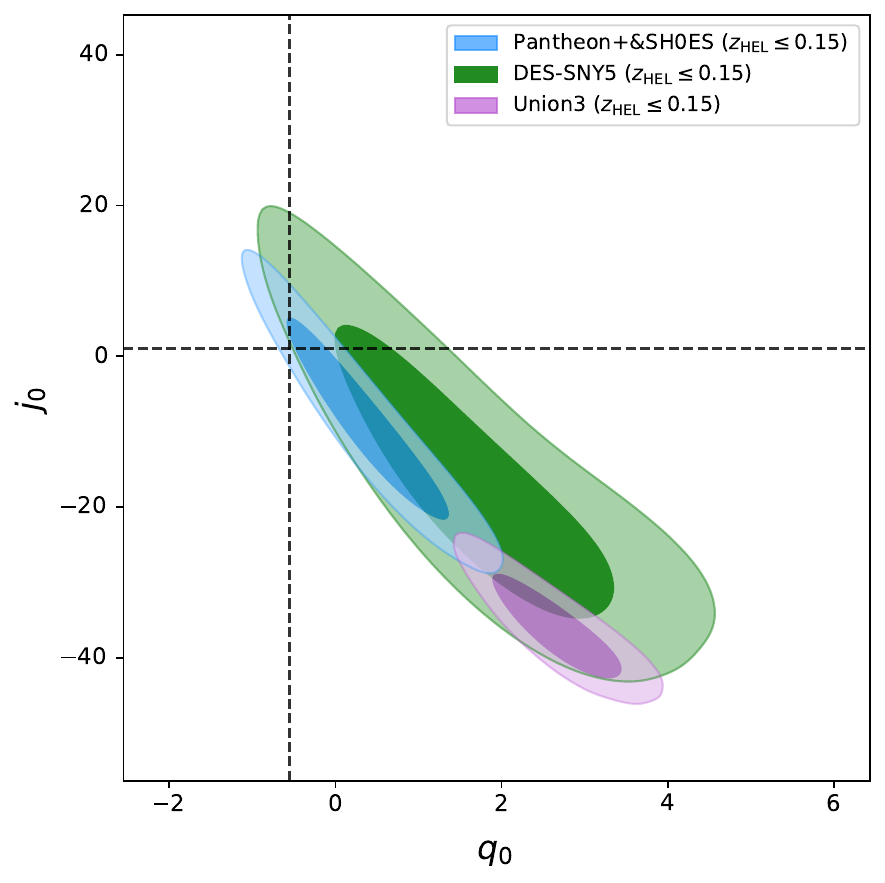}
\end{subfigure}
\begin{subfigure}{0.49\textwidth}
\centering
\includegraphics[width=\textwidth]{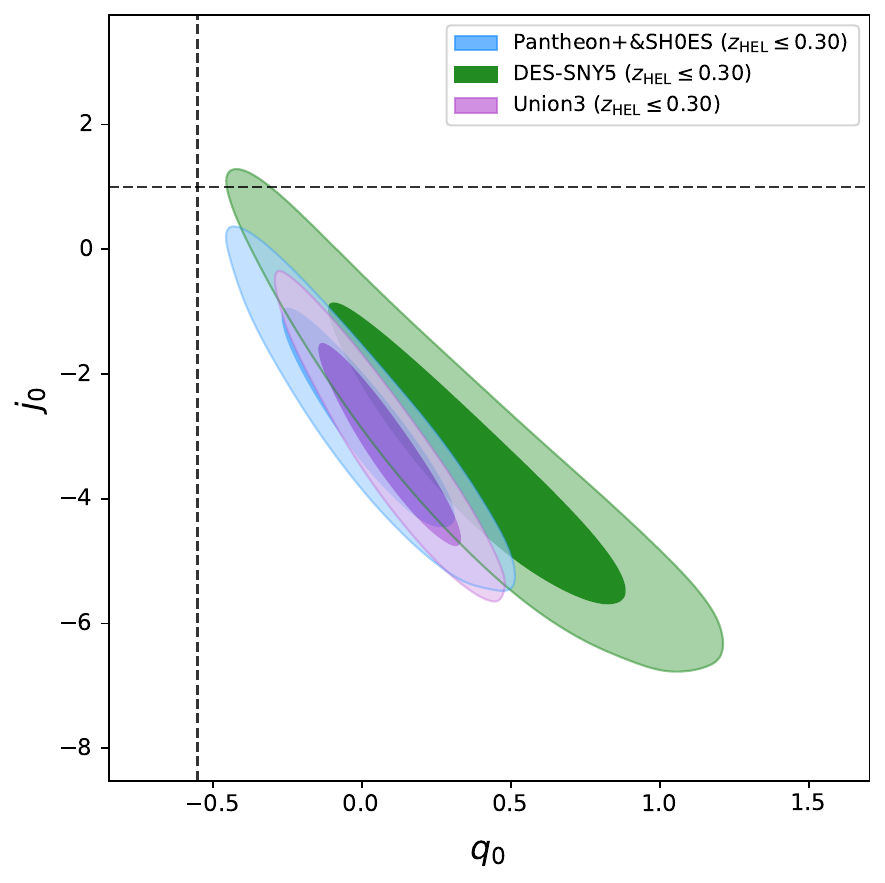}
\end{subfigure}
\begin{subfigure}{0.49\textwidth}
\centering
\includegraphics[width=\textwidth]{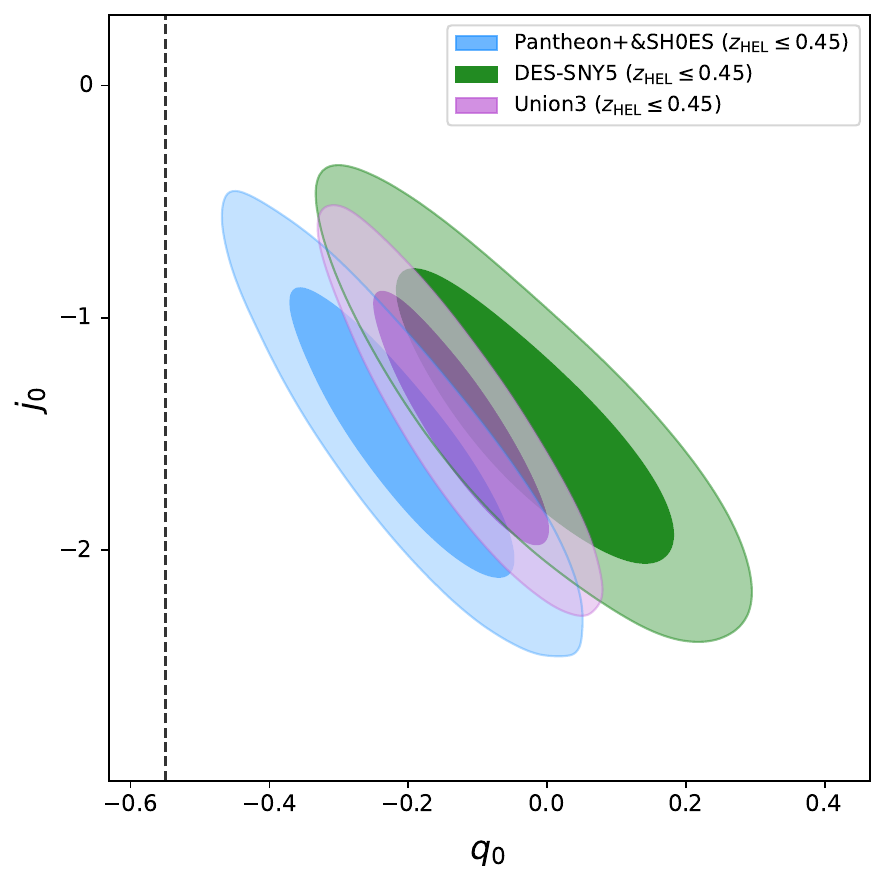}
\end{subfigure}
\begin{subfigure}{0.49\textwidth}
\centering
\includegraphics[width=\textwidth]{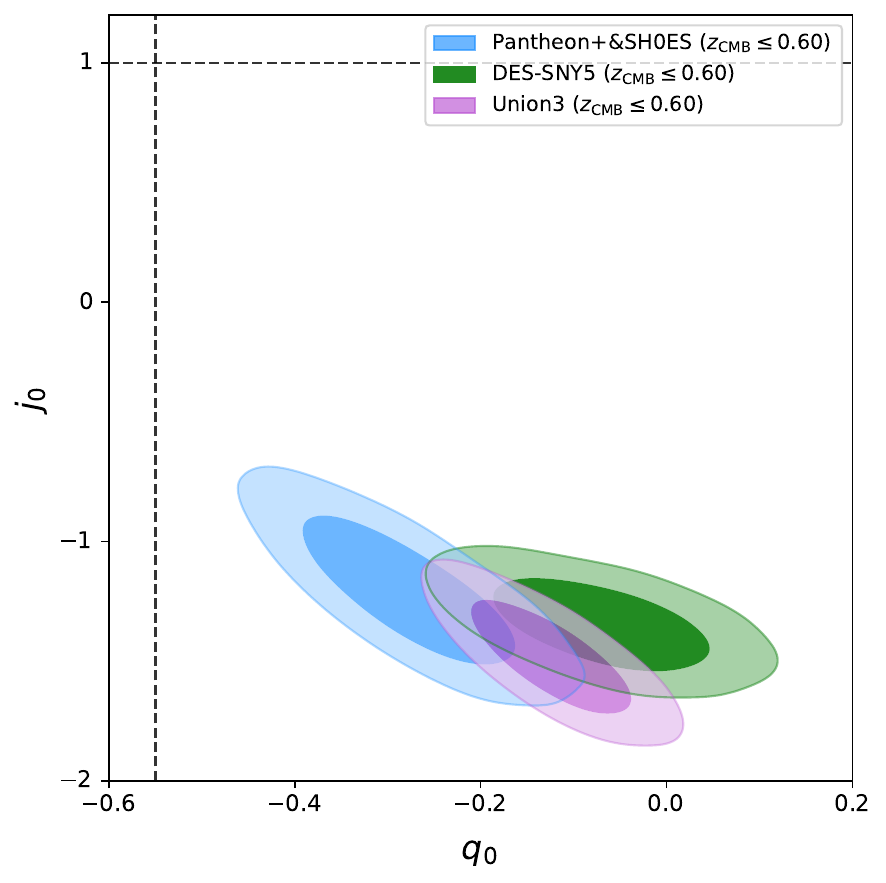}
\end{subfigure}
\begin{subfigure}{0.49\textwidth}
\centering
\includegraphics[width=\textwidth]{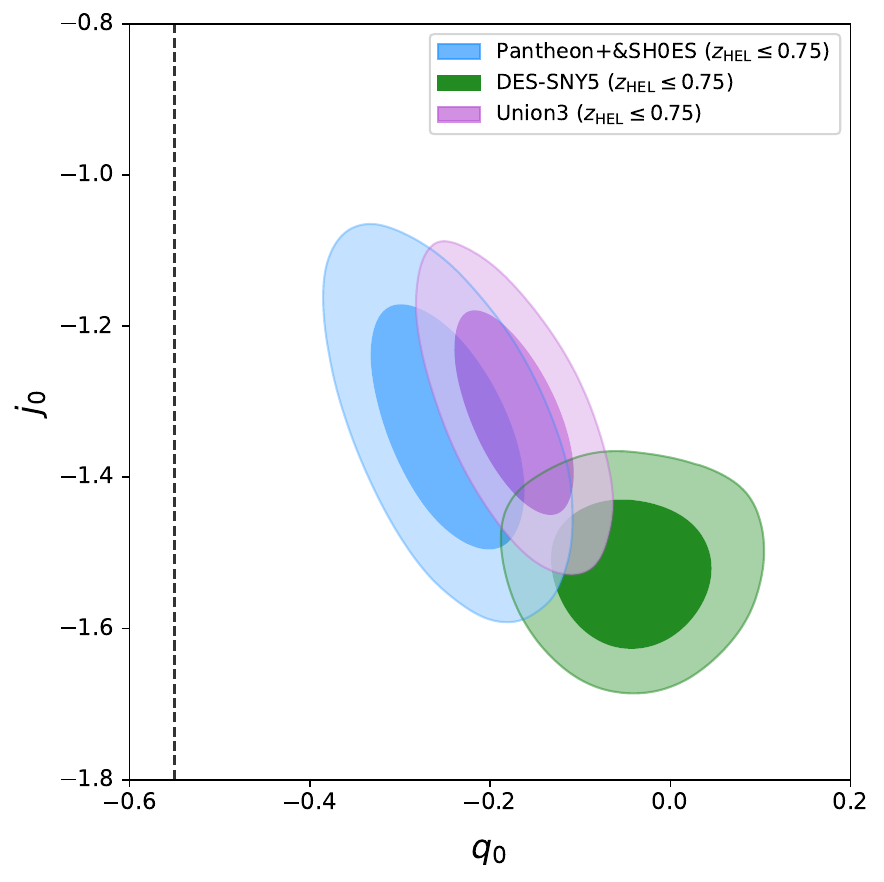}
\end{subfigure}

\caption{Same as Fig. \ref{fig:q0j0CMBdipoleTaylor} but with the luminosity distance written through the $(1,2)$ Pad\'e approximant.}
\label{fig:q0j0CMBdipolePade}
\end{figure}

The confidence regions in the $(q_0,\, j_0)$ plane are drawn in Figs. \ref{fig:q0j0CMBdipoleTaylor}-\ref{fig:q0j0CMBdipolePade} and compared with $q_0=-0.55$ and $j_0=1$, i.e., the fiducial values obtained from the $\Lambda$CDM model. 

To this end, we furthermore report in Appendix \ref{appendix} the table with the constrained parameters and the complete contour plots, in Tab. \ref{tab:cosmoCMBdipole} and Figs. \ref{fig:TaylorCMBDipole}-\ref{fig:PadeCMBDipole}, respectively.

Additionally, for these computations the convergence of the chains is reached through the Gelman--Rubin criterion \cite{1992StaSc...7..457G} once $R-1<0.01$. In particular, when 
\begin{equation}\label{CMBdipparam}
(v_0,\, \text{ra},\, \text{dec})=(369.82~\mathrm{km\,s^{-1}}, 167.942^\circ, -6.944^\circ),   
\end{equation}
the Hubble constant $H_0$ is in agreement at $1$-$\sigma$ with $H_0=73.6\pm 1.1~\mathrm{km\,s^{-1}\,Mpc^{-1}}$ in both cases of the luminosity distance written using Taylor or Pad\'e and in all redshift intervals. This same trend is also found when the dipole parameters are allowed to vary freely. 

Focusing on the deceleration $q_0$ and jerk $j_0$ parameters,  we find the following differences between this analysis and the one carried out in Sect. \ref{secondan}
\begin{itemize}
    \item [-] When the luminosity distance written using Taylor is adopted we observe a slight change in the confidence regions of the $(q_0,\, j_0)$ planes. Fixing the CMB dipole parameters through Eq. \eqref{CMBdipparam} gives a shift from the fiducial values of the deceleration and jerk parameters in the $\Lambda$CDM scenario, i.e. $q_0=-0.55$ and $j_0=1$ in all the redshift intervals considered as can be seen when Fig. \ref{fig:q0j0dipoleTaylor} and Fig. \ref{fig:q0j0CMBdipoleTaylor} are compared.

    Specifically, from Fig. \ref{fig:q0j0dipoleTaylor} in the intervals $\mathcal{I}_3,\ \mathcal{I}_4$ and $\mathcal{I}_5$ the expected values $(q_0,\, j_0)=(-0.55,\ 1)$ fall within the 95\% confidence region, while for $\mathcal{I}_1$ this also occurs in the 68\% confidence region. On the other hand, looking at Fig. \ref{fig:q0j0CMBdipoleTaylor} we observe a shift in compatibility with the $\mathcal{I}_2,\ \mathcal{I}_3,\ \mathcal{I}_4$ and $\mathcal{I}_5$ intervals where the fiducial $\Lambda$CDM values of $q_0$ and $j_0$ fall outside the 95\% confidence region and the first interval $\mathcal{I}_1$ shifts from the 68\% to the 95\% confidence region.
    
    \item [-] The behavior described above is also observed when the luminosity distance is parameterized adopting the $(1,2)$ Pad\'e approximant. Also in this case, comparing Fig. \ref{fig:q0j0dipolePade} presenting the confidence contours when $(v_0,\, \text{ra},\, \text{dec})$ are left free to vary with Fig. \ref{fig:q0j0CMBdipolePade} where they are fixed to inferred CMB values shows a shift from $q_0=-0.55$ and $j_0=1$. This is marked especially when the interval $\mathcal{I}_1$ is considered where we observe a departure from 68\% to the 95\% confidence region of the fiducial values of the deceleration and jerk parameter in the concordance paradigm.
\end{itemize}

\section{Final outlooks and perspectives}\label{conc}

In this work, we performed a cosmographic analysis of SNe Ia using the Pantheon+\&SH0ES, DES-SNY5 and Union3 compilations. We investigated the dependence of the inferred cosmographic parameters on the maximum redshift included in the analysis by considering five progressively enlarged redshift intervals. We checked the robustness of the reconstruction against the adopted representation of the luminosity distance and employed both a third-order Taylor expansion and a Pad\'e $(1,2)$ approximant, i.e., corresponding to the same order of Taylor expansion in a different representation.

In the first part of the analysis, we used the CMB-frame redshift and reconstructed the cosmographic parameters without introducing an additional dipolar correction. For Pantheon+\&SH0ES, the inferred value of the Hubble constant remained compatible with the corresponding Cepheid-calibrated determination over all the redshift intervals considered and for both cosmographic parameterizations. 

We argued that this stability indicated that the inferred value of $H_0$ was only weakly affected by the progressive enlargement of the supernova sample within the redshift range explored here, remaining overall independent from the interval used.

Conversely, the reconstructed values of the deceleration and jerk parameters showed a stronger dependence on both the adopted supernova compilation and the maximum redshift of the sample. At the lowest redshifts, the constraints remained relatively broad, particularly for the jerk parameter, as expected from the limited lever arm available to cosmography. 

As higher-redshift SNe were progressively included, the uncertainties decreased and the reconstructed cosmographic parameters became more tightly constrained. The comparison with the fiducial flat $\Lambda$CDM values showed the best overall agreement for Pantheon+\&SH0ES, particularly at the largest redshift intervals, while DES-SNY5 displayed only partial compatibility in selected cases, in agreement with previous literature that raised tension in the jerk parameter adopting the BAO data points \cite{Pourojaghi:2024bxa, Rodrigues:2025tfg}. Union3 instead showed more persistent offsets from the fiducial values, together with comparatively tighter constraints, indicating that more study is required in order to quantify the impact of dark energy at low redshifts.

We subsequently repeated the cosmographic analysis by introducing a free dipolar Doppler correction to the heliocentric redshift. The amplitude and direction of the corresponding velocity vector were constrained simultaneously with the cosmographic parameters. The inclusion of this additional dipolar degree of freedom did not qualitatively alter the main cosmographic results and, in particular, the Pantheon+\&SH0ES determination of $H_0$ remained stable, while the reconstruction of $q_0$ and $j_0$ continued to improve as the redshift interval was enlarged.

The reconstructed dipolar velocity amplitudes remained of order a few hundred kilometres per second for all three supernova compilations and showed only limited variations when the maximum redshift was increased. Pantheon+\&SH0ES and DES-SNY5 recovered broadly similar preferred directions, whereas Union3 favoured a different orientation, particularly in declination. The relative stability of the fitted dipole across the cumulative redshift intervals indicated that the addition of more distant supernovae did not substantially modify the preferred low-redshift dipolar component. 

Since the adopted redshift samples were nested, however, this stability was not interpreted as evidence that a coherent peculiar motion extended throughout the entire redshift range considered.

Accordingly, the fitted velocity vector has been interpreted as an effective relative-velocity dipole. Under our prescriptions, the physical bulk velocity could be reconstructed only after specifying the observer velocity with respect to the cosmological rest frame. This distinction allowed the dipolar correction to be separated conceptually from both the observer motion and the individual peculiar velocities of the supernova host galaxies.

Last but not least, we performed a complementary analysis in which the dipole amplitude and direction were fixed to the values inferred from the CMB dipole by the Planck Collaboration. We worked it out to check the sensitivity of the cosmographic reconstruction to the adopted kinematic frame. Hence, with the aim of determining whether the freedom associated with the dipolar correction played a significant role in the inferred cosmographic parameters, when the CMB dipole parameters were imposed, the Hubble constant inferred from Pantheon+\&SH0ES remained compatible at $1$-$\sigma$ with the corresponding reference determination, for all the redshift intervals considered. 

The same has been found for both the Taylor and Pad\'e parameterizations, reproducing the same behavior obtained when the dipole parameters were allowed to vary freely. The constraints on the deceleration and jerk parameters remained more sensitive to the adopted supernova compilation and to the maximum redshift of the sample. Moreover, fixing the dipole to the CMB values resulted in a minimal shift in the confidence contours with respect to the second analysis.

As future developments, we intend to extend this work by considering disjoint redshift shells in order to better isolate the radial dependence of the dipolar signal and to consider the possible scale dependence of local motions. A more complete treatment of the peculiar velocity covariance could also improve the separation between coherent flows and individual host-galaxy motions. In addition, the reconstructed dipole could be directly converted into a physical bulk-flow estimate and compared with independent measurements from galaxy peculiar velocity surveys. Additionally, we intend to repeat the analysis adopting wider SNe Ia data catalogs through the construction of mock samples. Finally, we will also complement our analyses with other low-redshift probes, e.g. cosmic chronometers, BAO measurements, etc., to test the stability of the cosmographic reconstruction. A deep study on the high-redshift cosmography will also be subject of our future investigations. In particular, we will wonder whether multi-field cosmology \cite{DAgostino:2021vvv,Luongo:2026ndg,Luongo:2025ovo,Luongo:2024opv} may reconcile late- and early-times through a single picture.

\section*{Acknowledgements}

ACA is grateful for the support of the Istituto Nazionale di Fisica Nucleare (INFN) Sezione di Napoli, Iniziativa Specifica QGSKY. YC and OL acknowledge financial support from the Brera Astronomical Observatory of the National Institute for Astrophysics (INAF). OL is grateful to Marek Biesiada and Anjan Ananda Sen for interesting discussions on the topic of evolving dark energy at low redshifts and acknowledges the hospitality of the National Centre for Nuclear Research (NCBJ), Warsaw, where this work was carried out. FP acknowledges support from the INFN InDark project. This paper is based upon work from COST Action CA21136 - Addressing observational tensions in cosmology with systematics and fundamental physics (CosmoVerse), supported by COST (European Cooperation in Science and Technology).

\bibliography{bibliography}

\appendix

\onecolumngrid

\section{MCMC analysis results and contour plots}\label{appendix}

\begin{figure*}
    \centering

    \begin{subfigure}{0.45\textwidth}
\centering
\includegraphics[width=\textwidth]{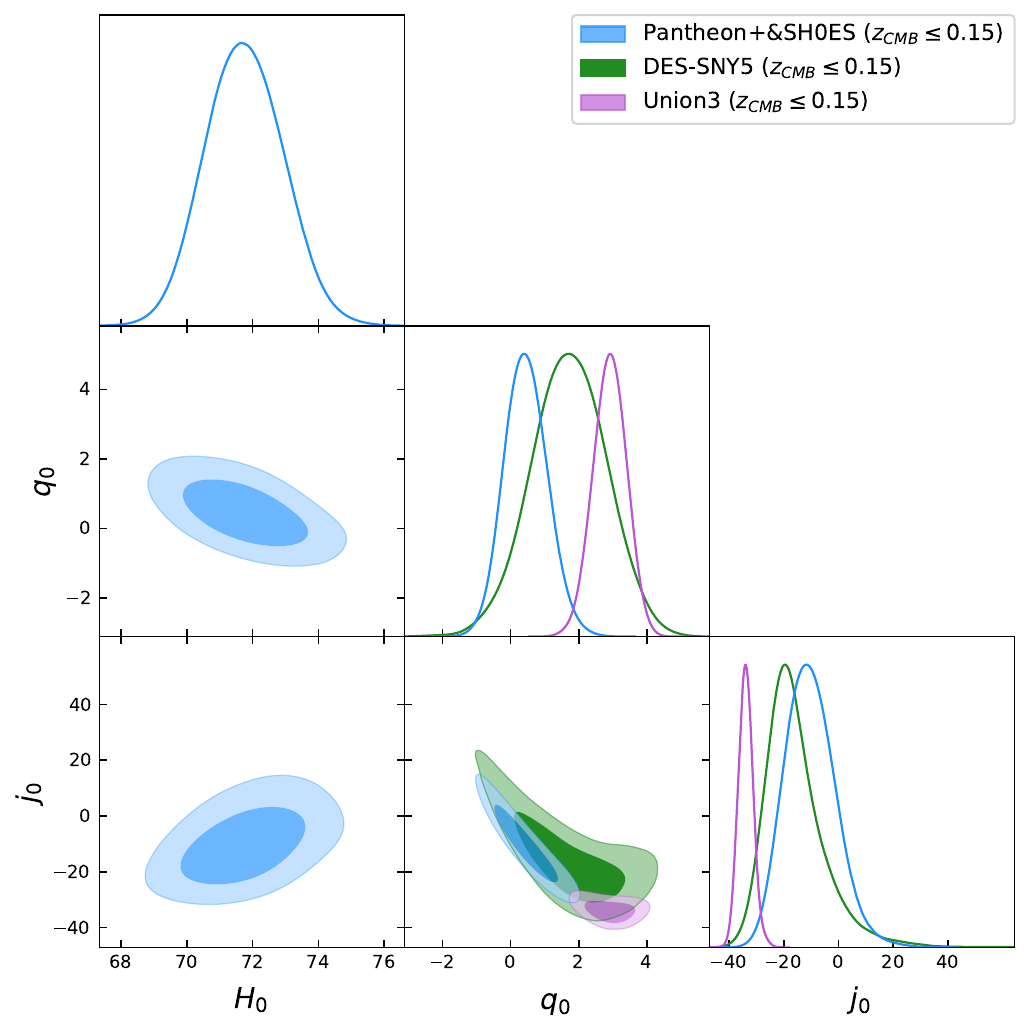}
\end{subfigure}
\begin{subfigure}{0.45\textwidth}
\centering
\includegraphics[width=\textwidth]{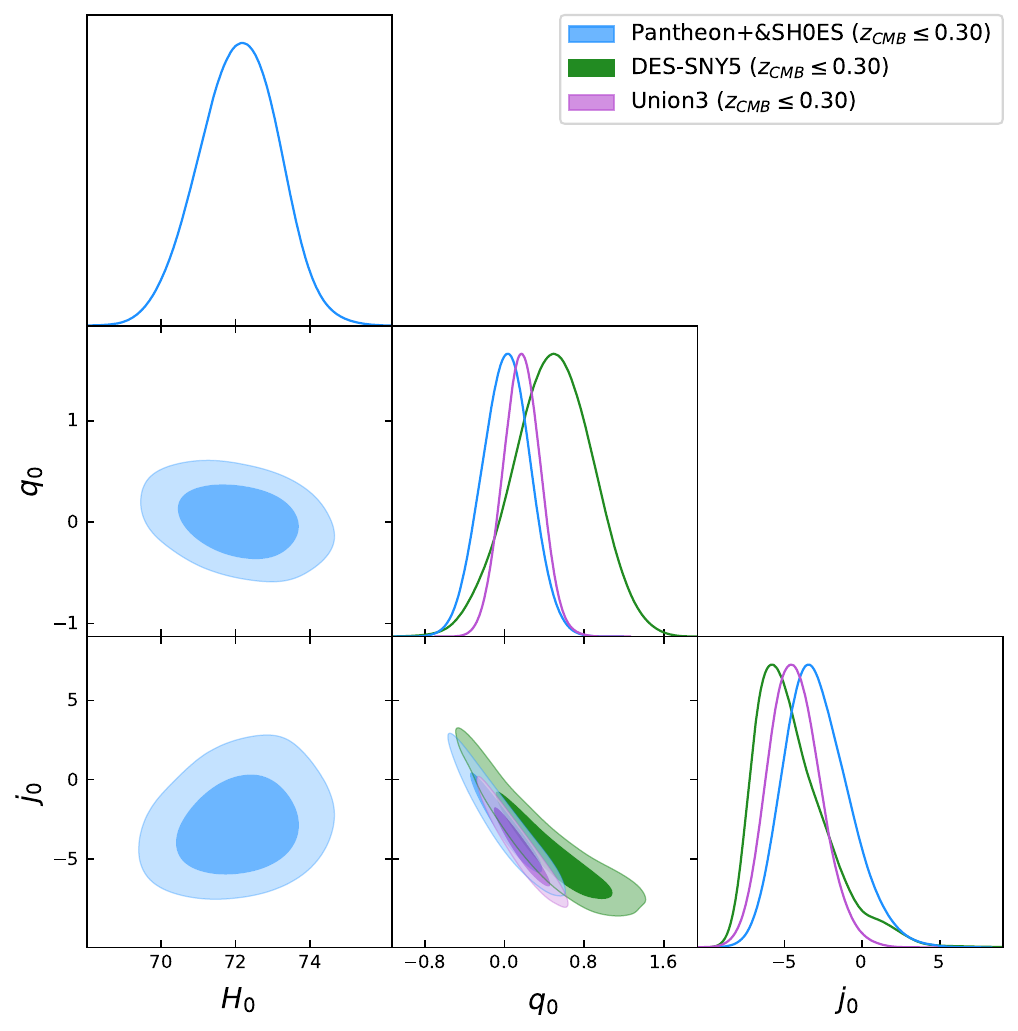}
\end{subfigure}
\begin{subfigure}{0.45\textwidth}
\centering
\includegraphics[width=\textwidth]{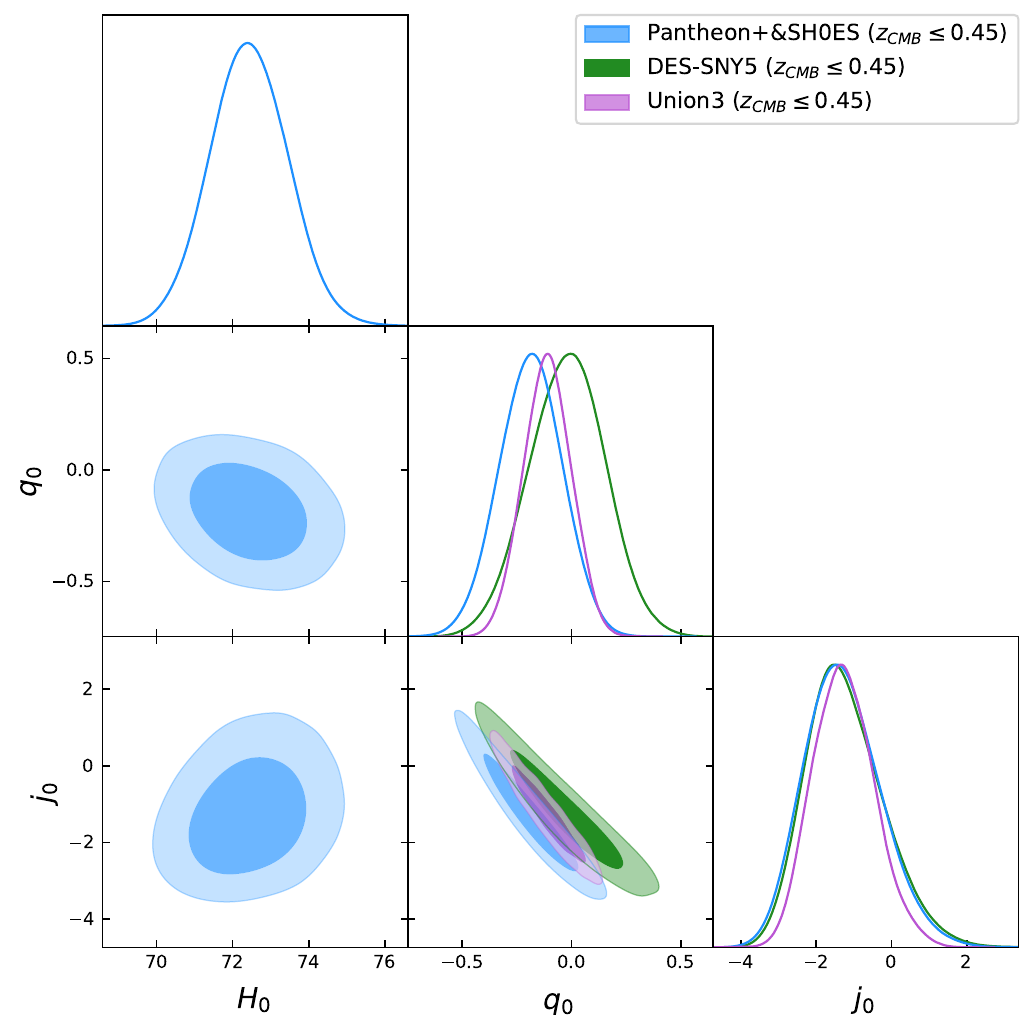}
\end{subfigure}
\begin{subfigure}{0.45\textwidth}
\centering
\includegraphics[width=\textwidth]{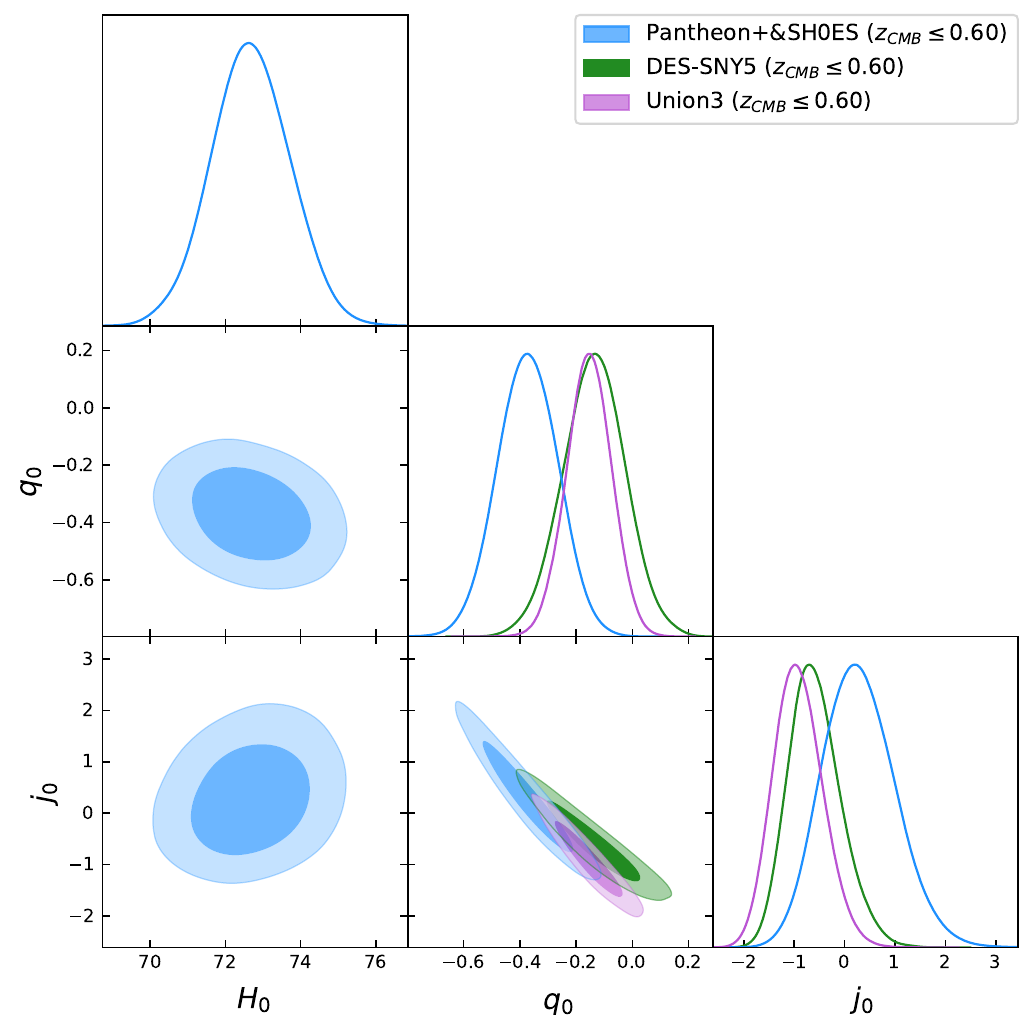}
\end{subfigure}
\begin{subfigure}{0.45\textwidth}
\centering
\includegraphics[width=\textwidth]{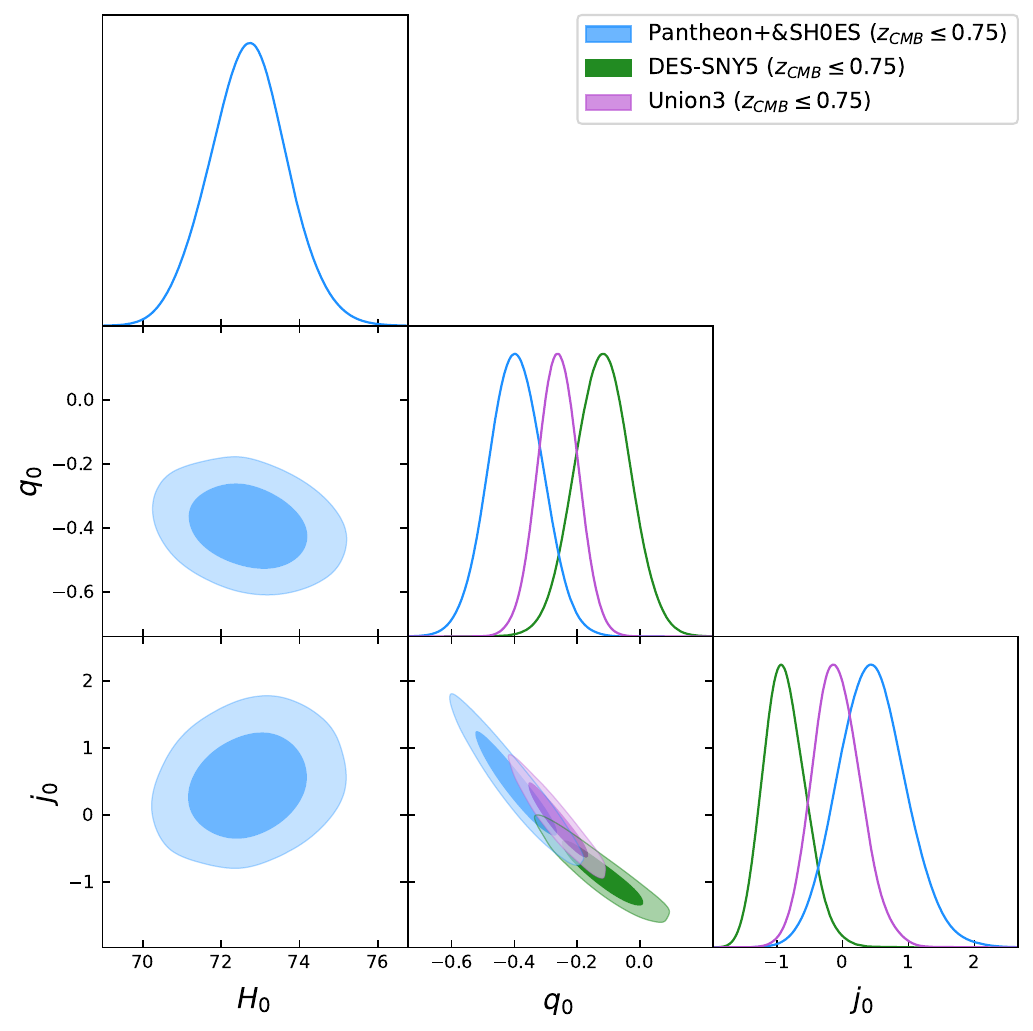}
\end{subfigure}

    \caption{Contour plots for the $H_0$, $q_0$, $j_0$ cosmographic parameters depicted considering the luminosity distance $d_L$ written using a third order Taylor expansion. The different colors represent the SNe Ia sample adopted in our analyses, the Pantheon+\&SH0ES catalog in blue, the DES-SNY5 catalog in green and the Union3 catalog in purple.}
    \label{fig:TaylorzCMB}
\end{figure*}

\begin{figure*}
    \centering

    \begin{subfigure}{0.45\textwidth}
\centering
\includegraphics[width=\textwidth]{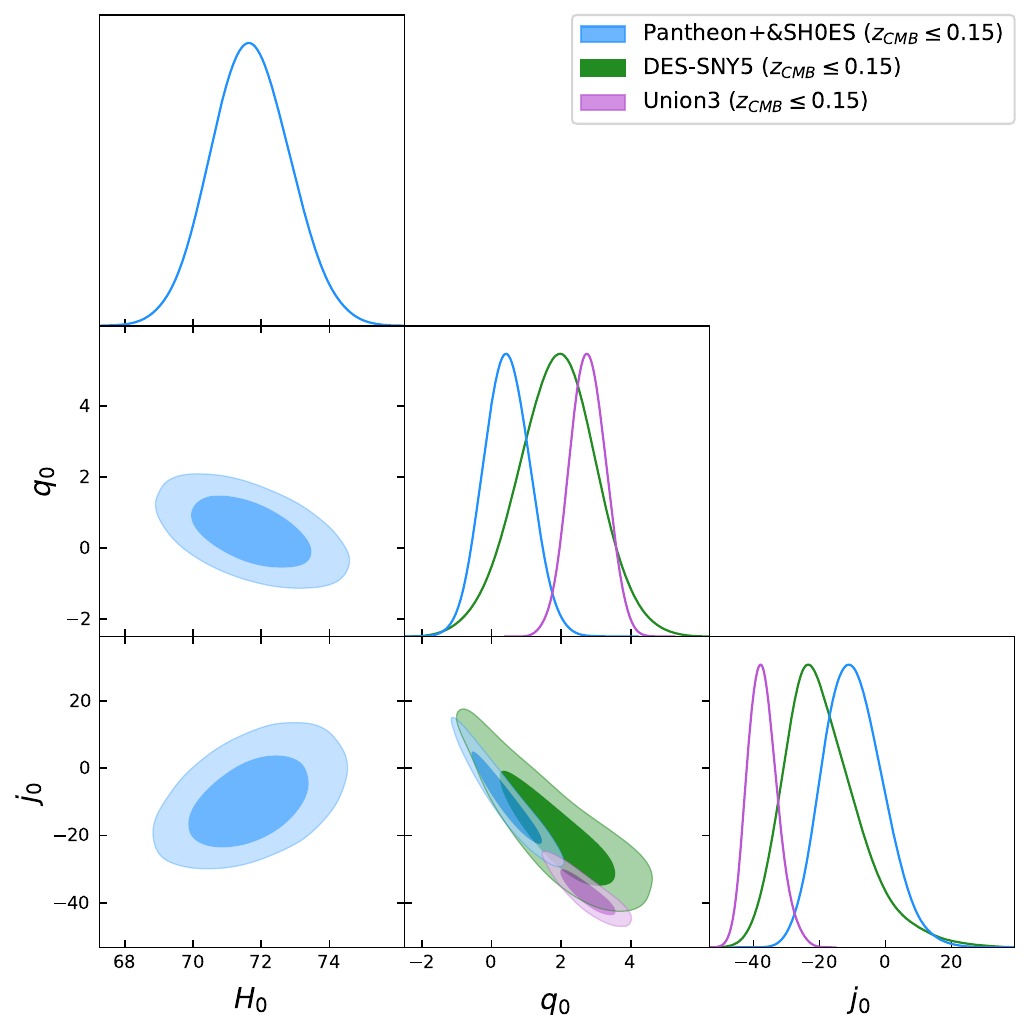}
\end{subfigure}
\begin{subfigure}{0.45\textwidth}
\centering
\includegraphics[width=\textwidth]{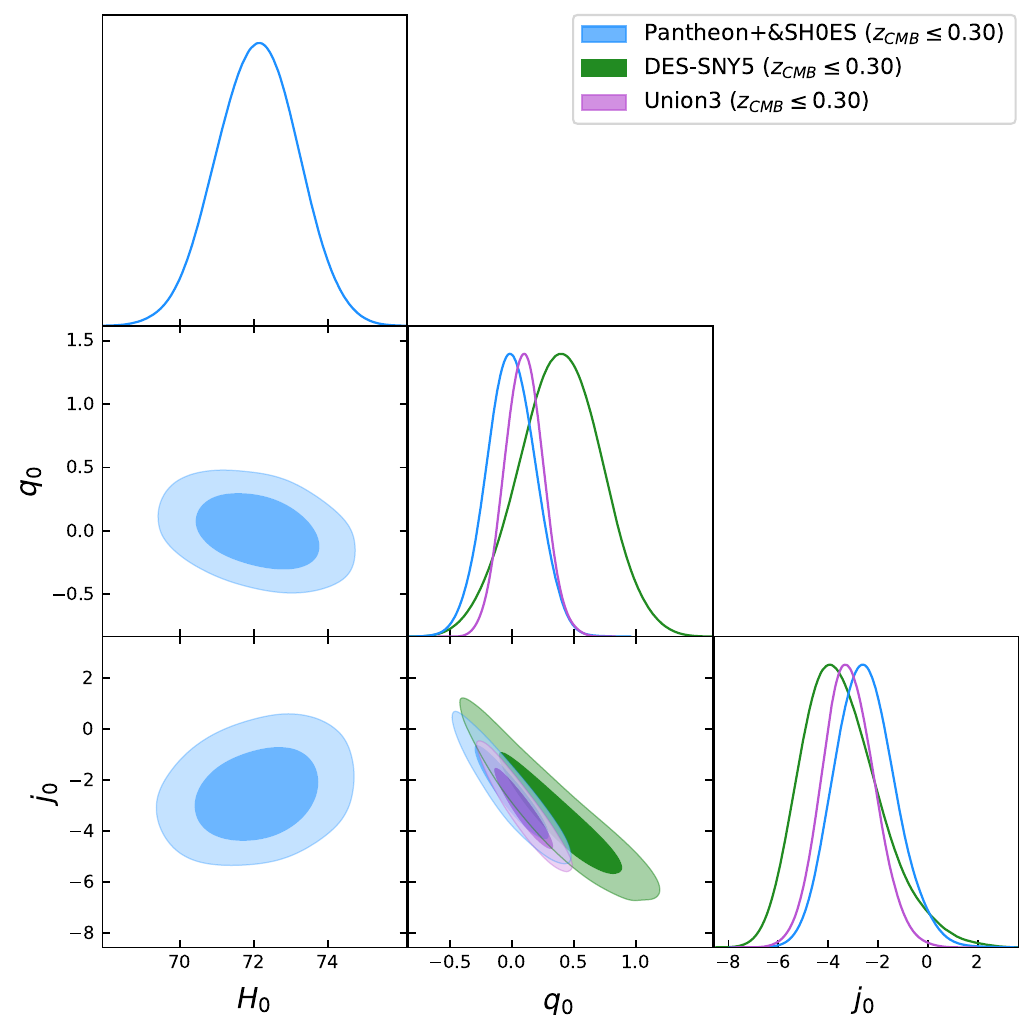}
\end{subfigure}
\begin{subfigure}{0.45\textwidth}
\centering
\includegraphics[width=\textwidth]{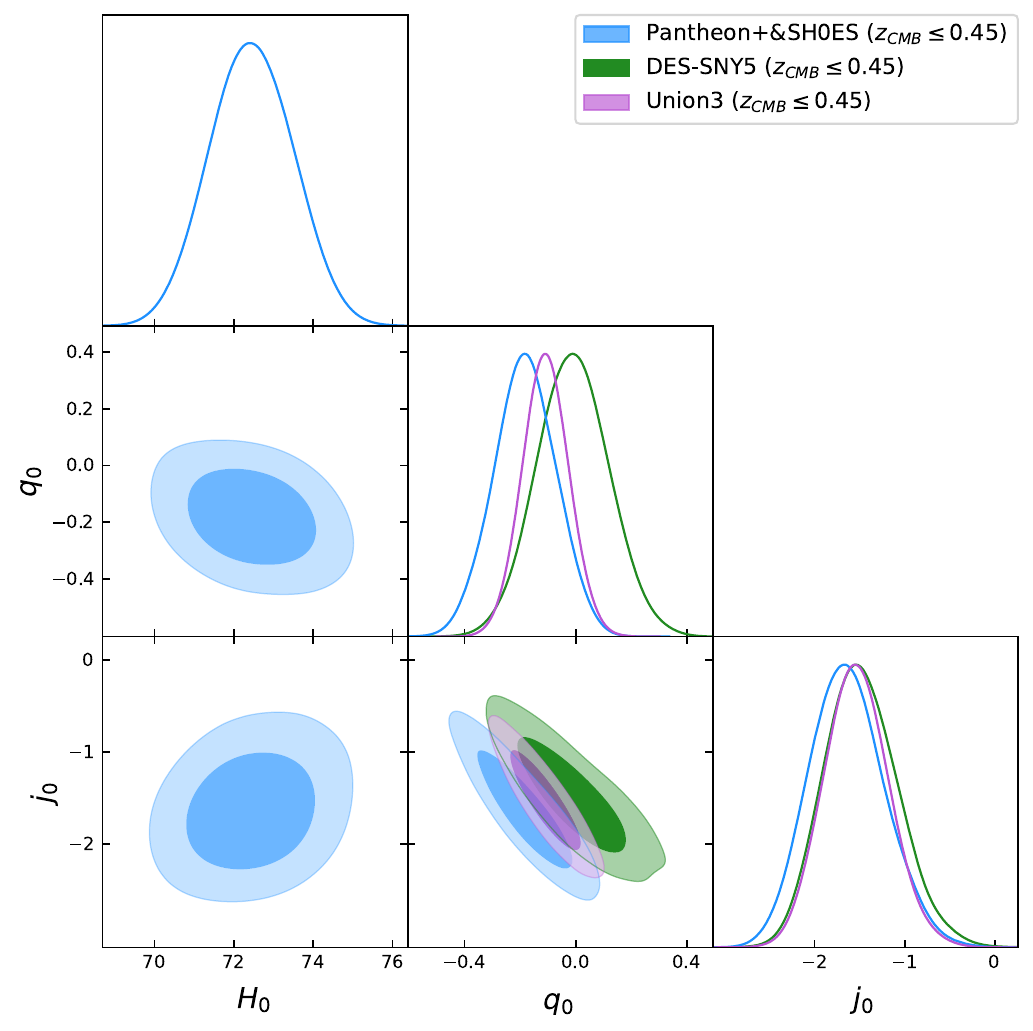}
\end{subfigure}
\begin{subfigure}{0.45\textwidth}
\centering
\includegraphics[width=\textwidth]{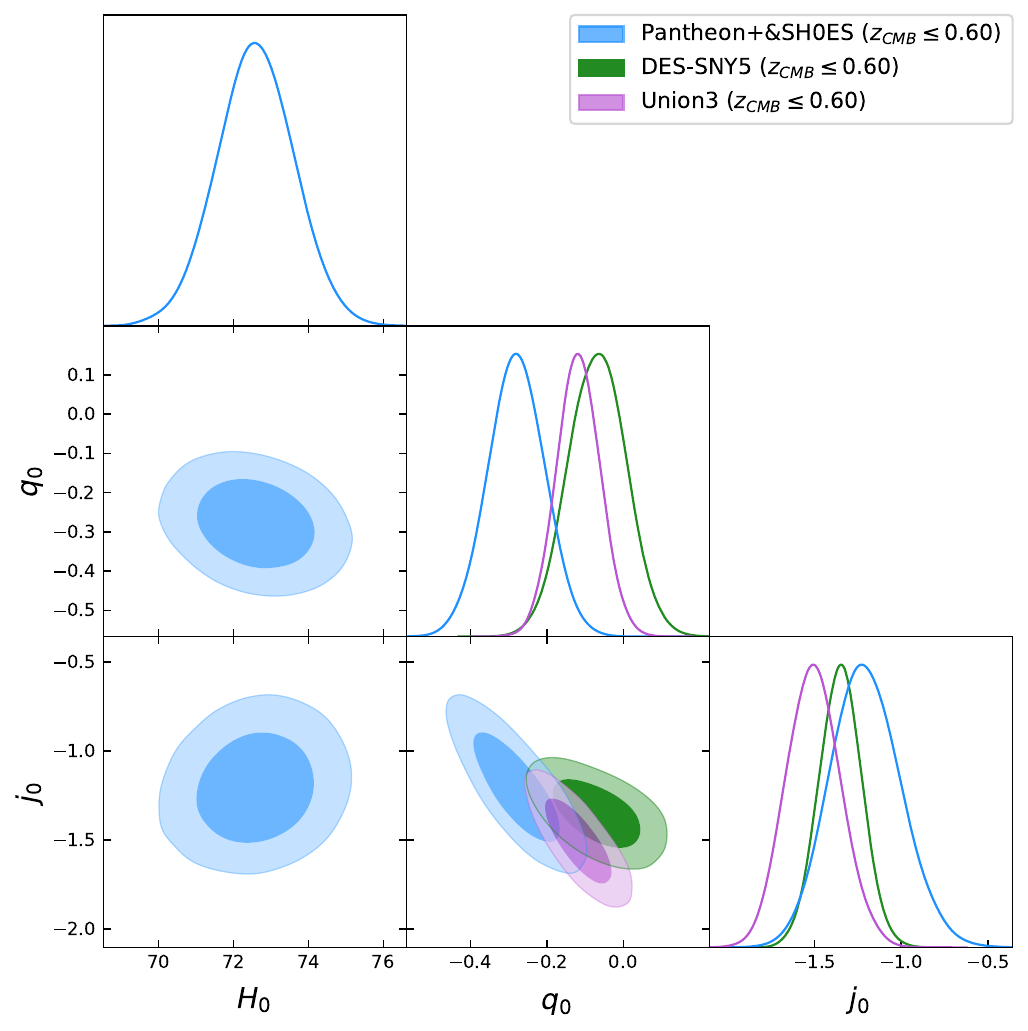}
\end{subfigure}
\begin{subfigure}{0.45\textwidth}
\centering
\includegraphics[width=\textwidth]{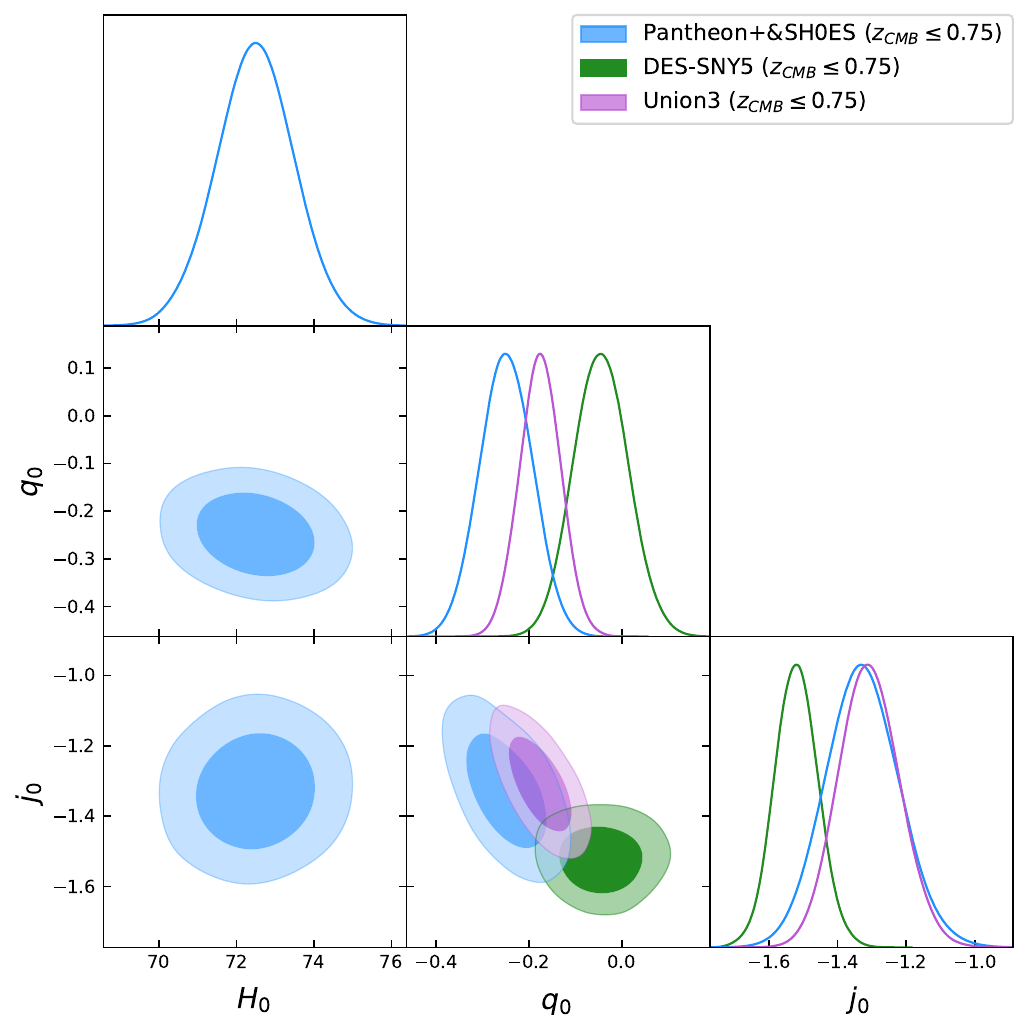}
\end{subfigure}

    \caption{Same as Fig. \ref{fig:TaylorzCMB} but considering the luminosity distance parameterized with a $(1,2)$ Pad\'e approximant.}
    \label{fig:PadezCMB}
\end{figure*}

\begin{figure*}
    \centering

    \begin{subfigure}{0.45\textwidth}
\centering
\includegraphics[width=\textwidth]{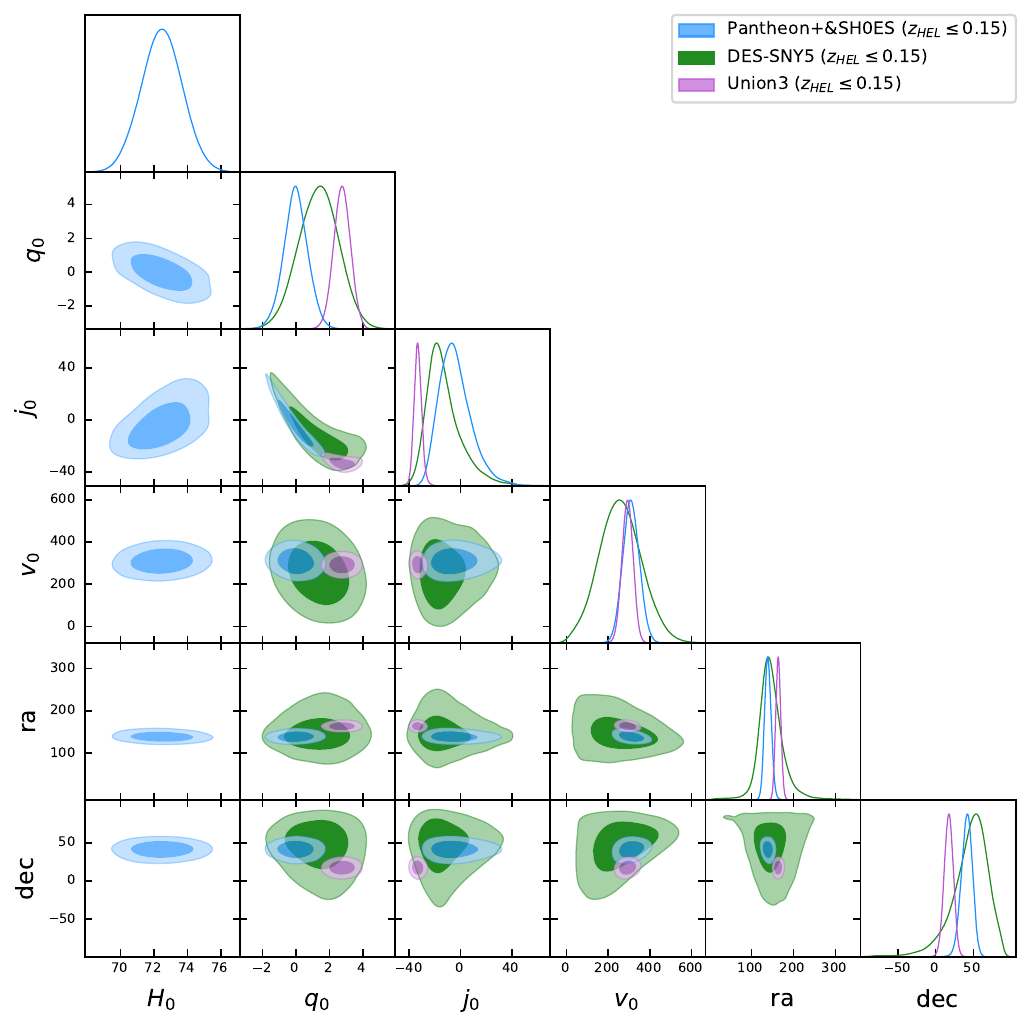}
\end{subfigure}
\begin{subfigure}{0.45\textwidth}
\centering
\includegraphics[width=\textwidth]{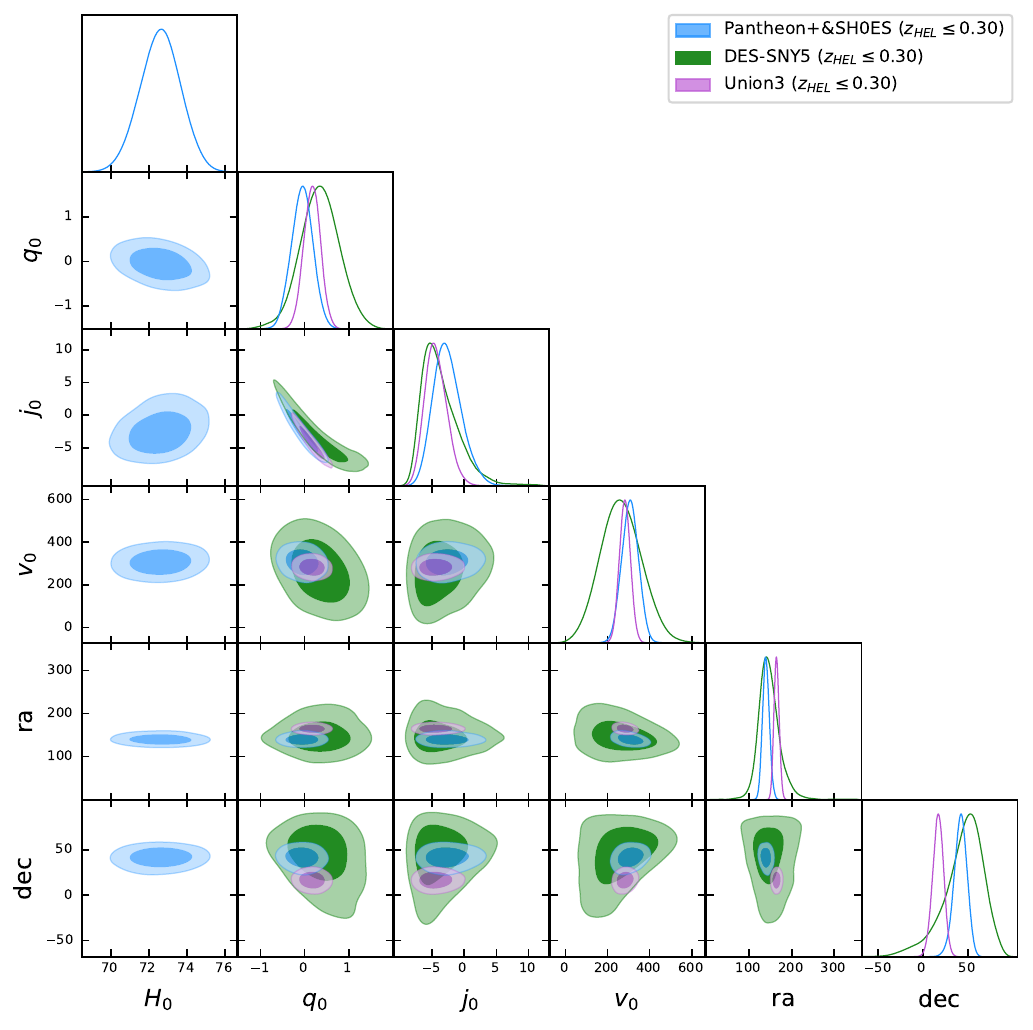}
\end{subfigure}
\begin{subfigure}{0.45\textwidth}
\centering
\includegraphics[width=\textwidth]{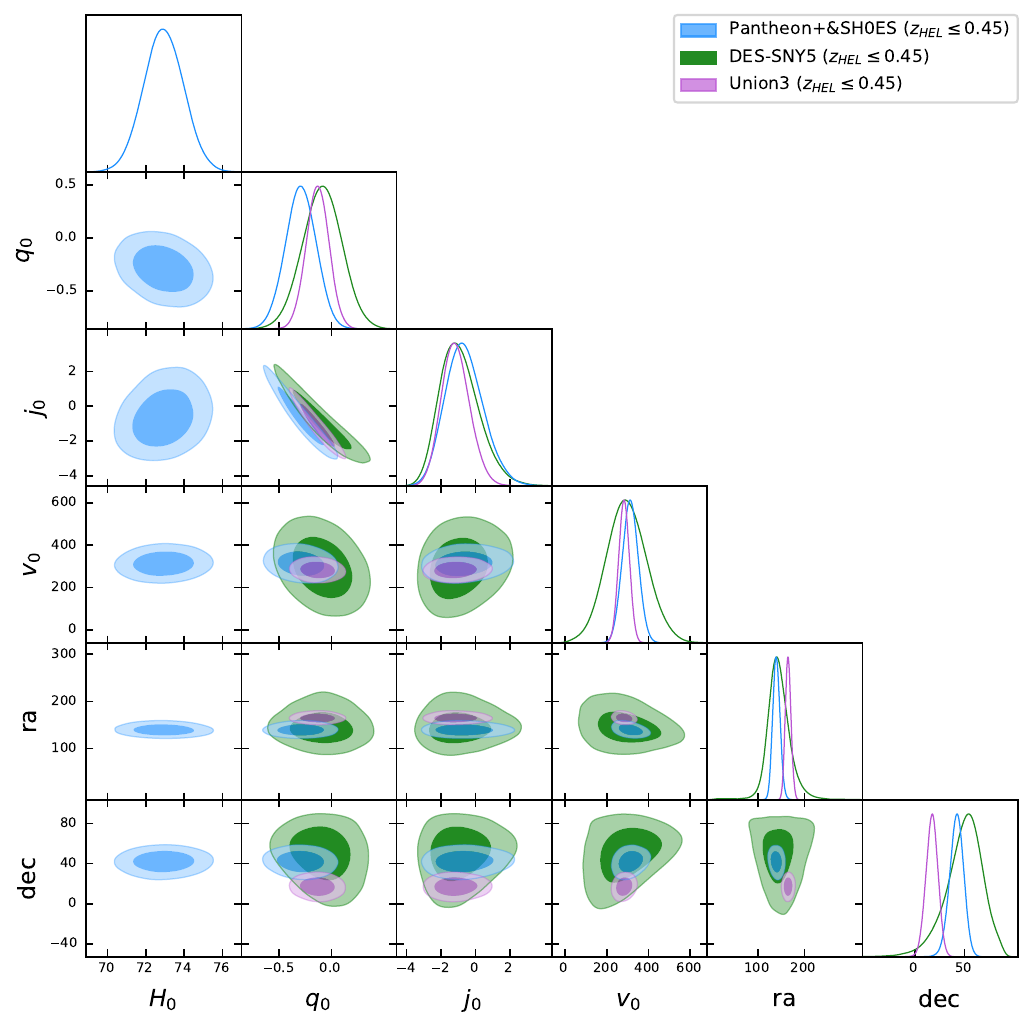}
\end{subfigure}
\begin{subfigure}{0.45\textwidth}
\centering
\includegraphics[width=\textwidth]{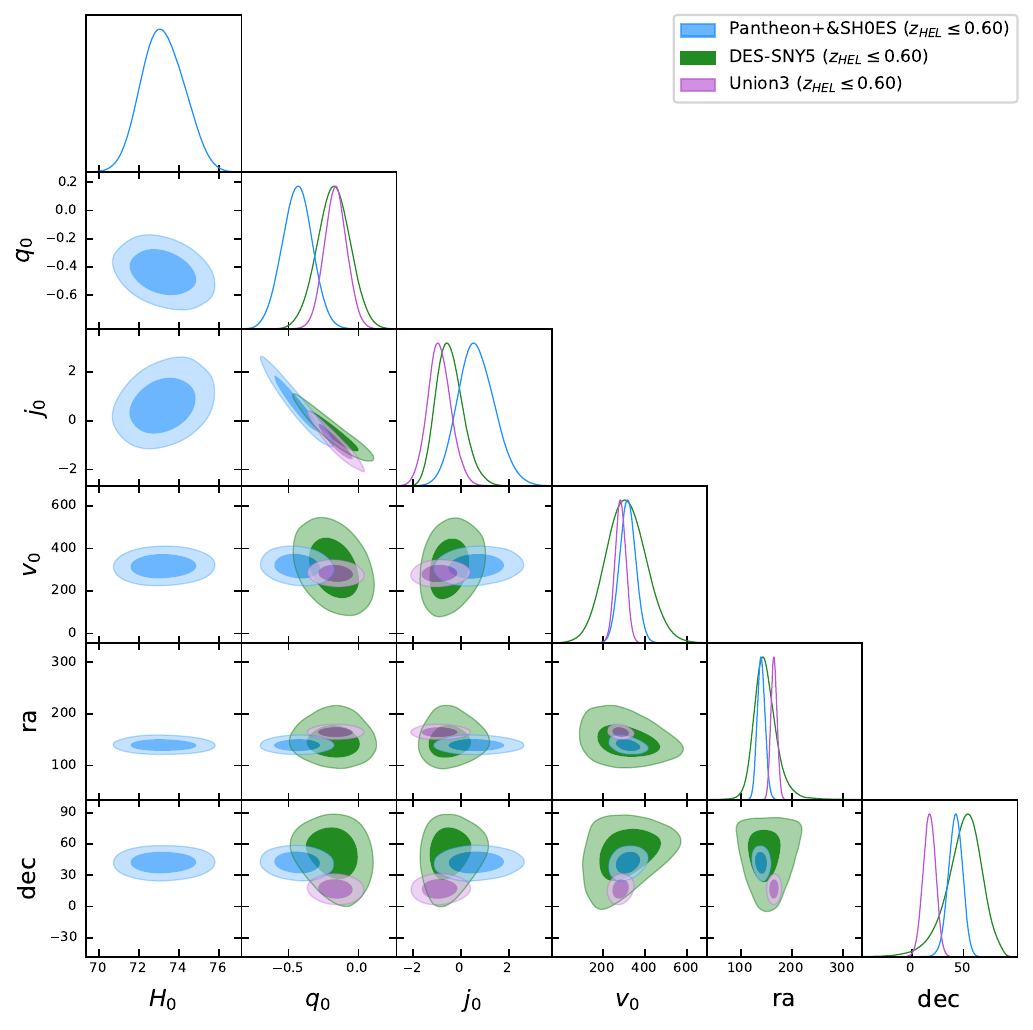}
\end{subfigure}
\begin{subfigure}{0.45\textwidth}
\centering
\includegraphics[width=\textwidth]{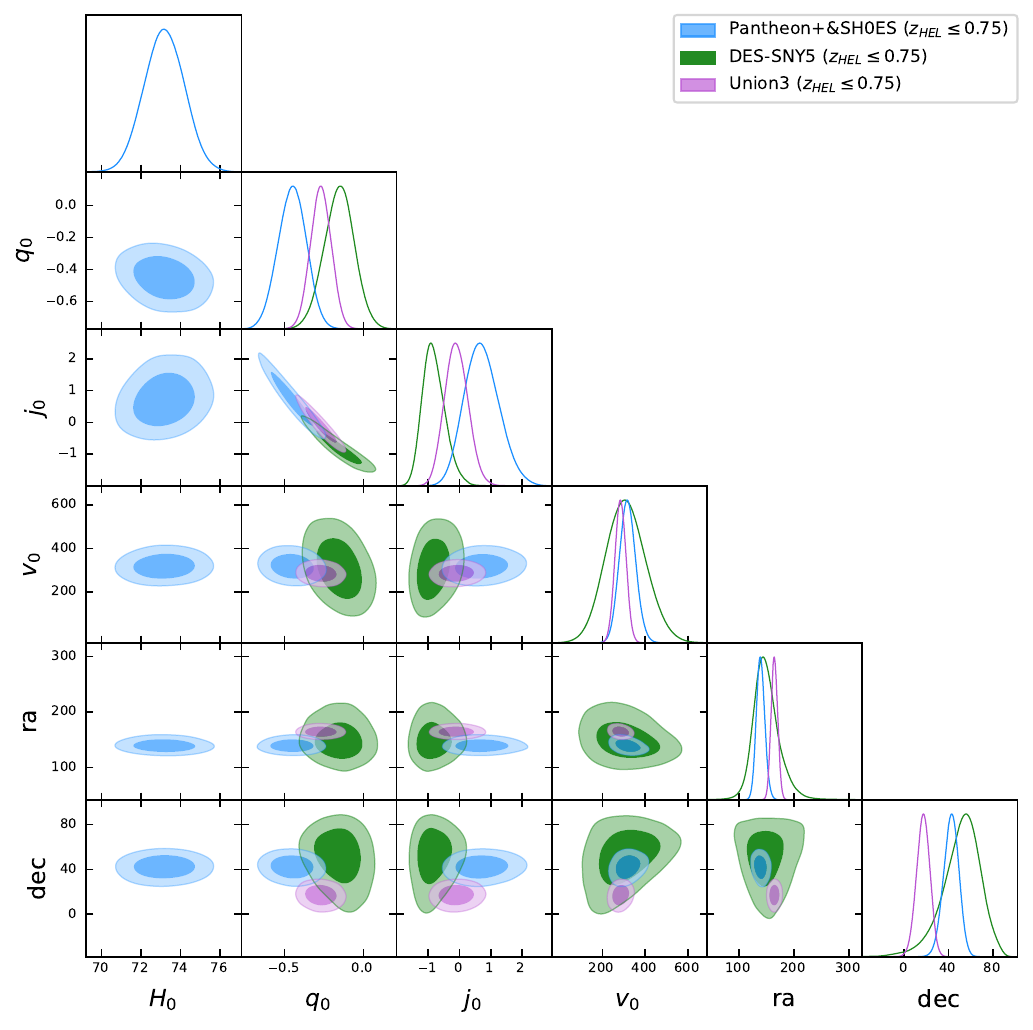}
\end{subfigure}

    \caption{Contour plots for the $H_0$, $q_0$, $j_0$ cosmographic parameters and for the velocity amplitude $v_0$ and direction across the sky $(\text{ra},\ \text{dec})$ depicted considering the luminosity distance $d_L$ written using a third order Taylor expansion. The different colors represent the SNe Ia sample adopted in our analyses, the Pantheon+\&SH0ES catalog in blue, the DES-SNY5 catalog in green and the Union3 catalog in purple.}
    \label{fig:TaylorDipole}
\end{figure*}

\begin{figure*}
    \centering

    \begin{subfigure}{0.45\textwidth}
\centering
\includegraphics[width=\textwidth]{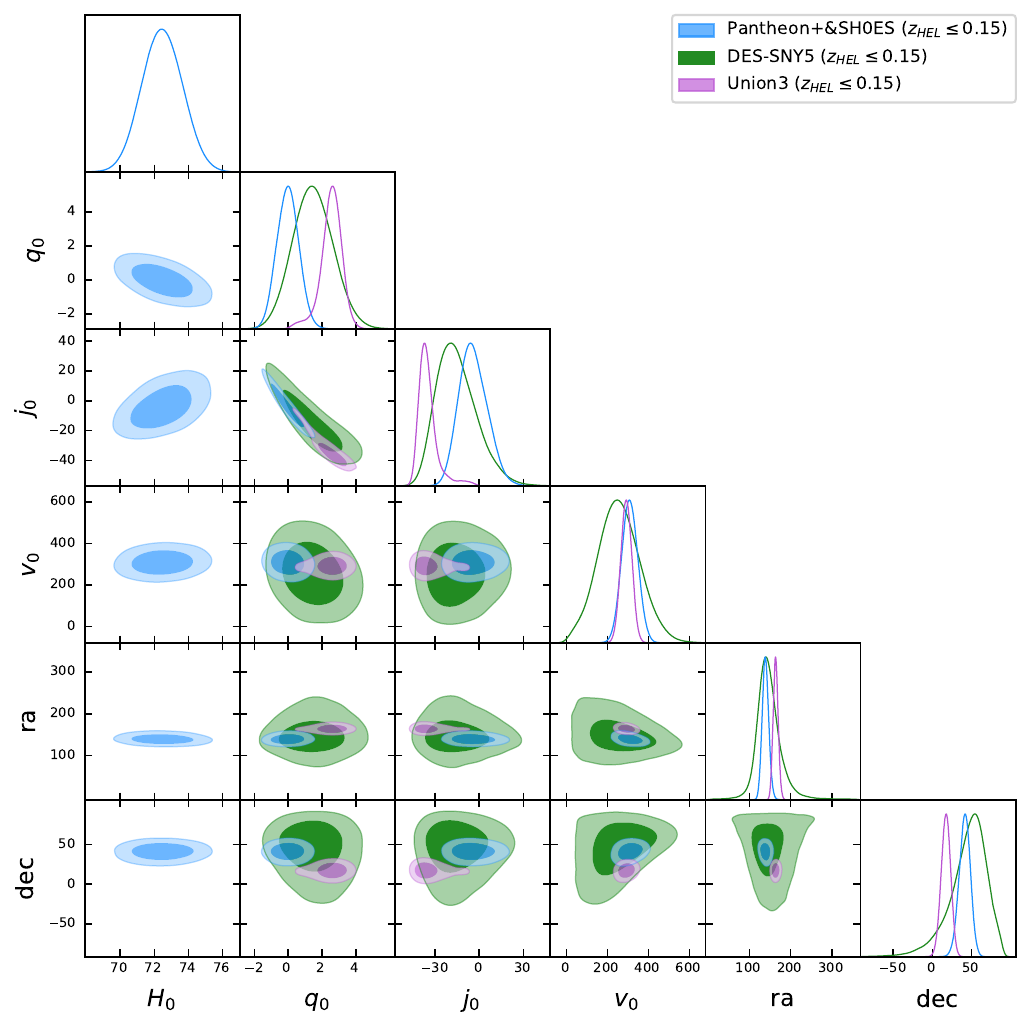}
\end{subfigure}
\begin{subfigure}{0.45\textwidth}
\centering
\includegraphics[width=\textwidth]{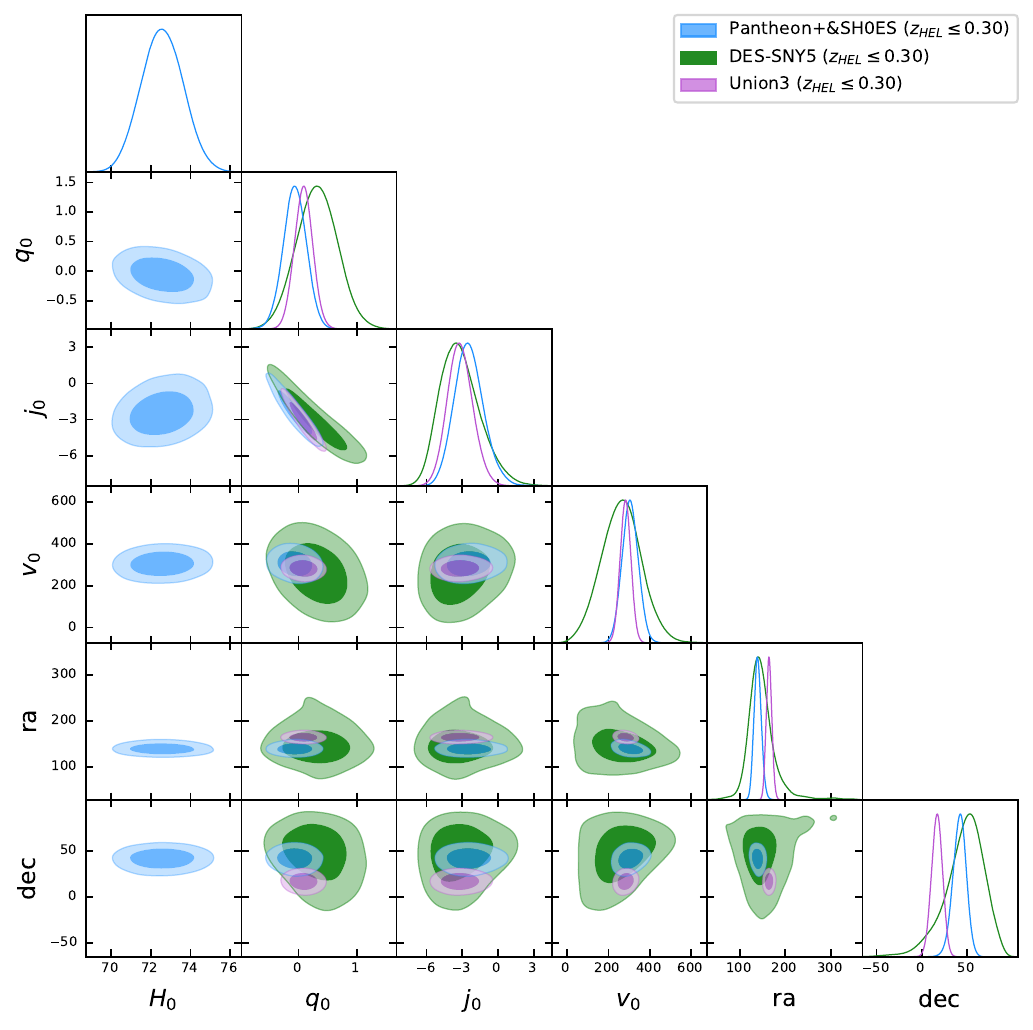}
\end{subfigure}
\begin{subfigure}{0.45\textwidth}
\centering
\includegraphics[width=\textwidth]{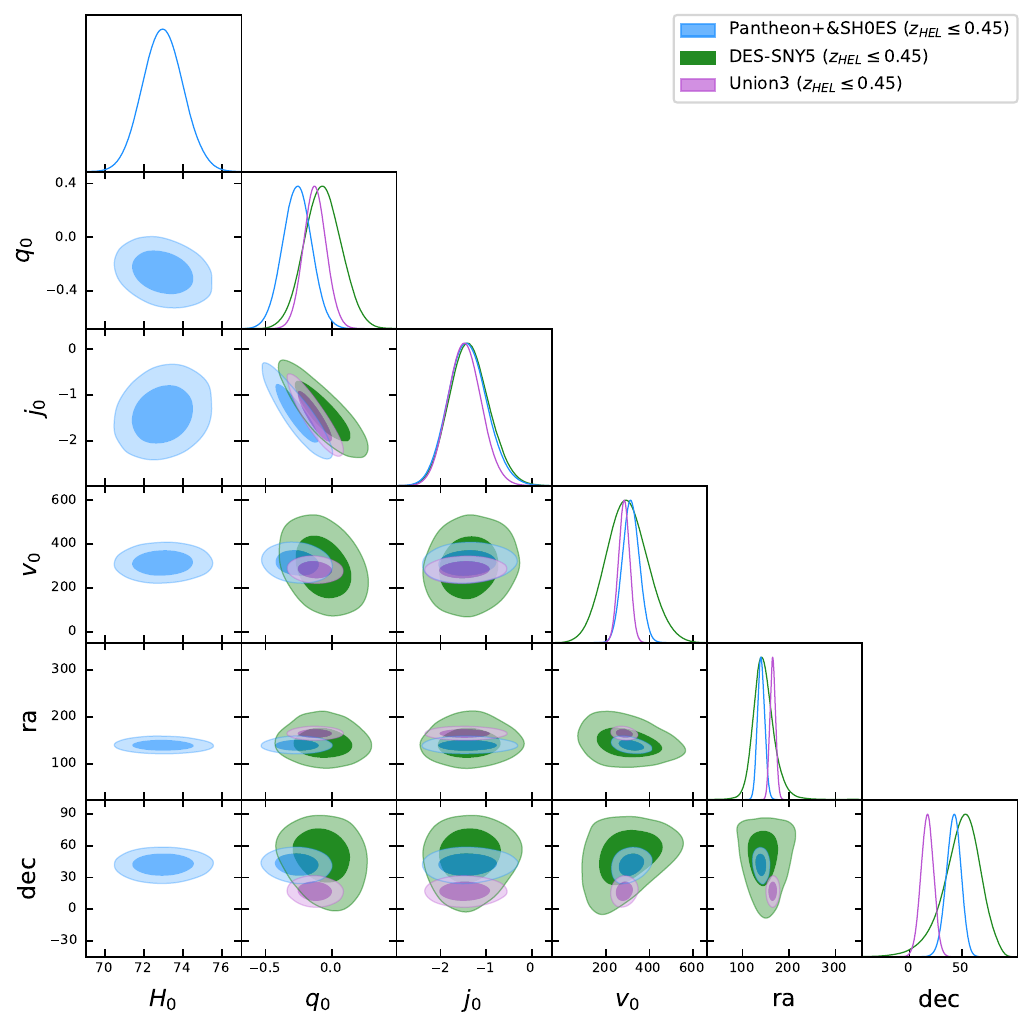}
\end{subfigure}
\begin{subfigure}{0.45\textwidth}
\centering
\includegraphics[width=\textwidth]{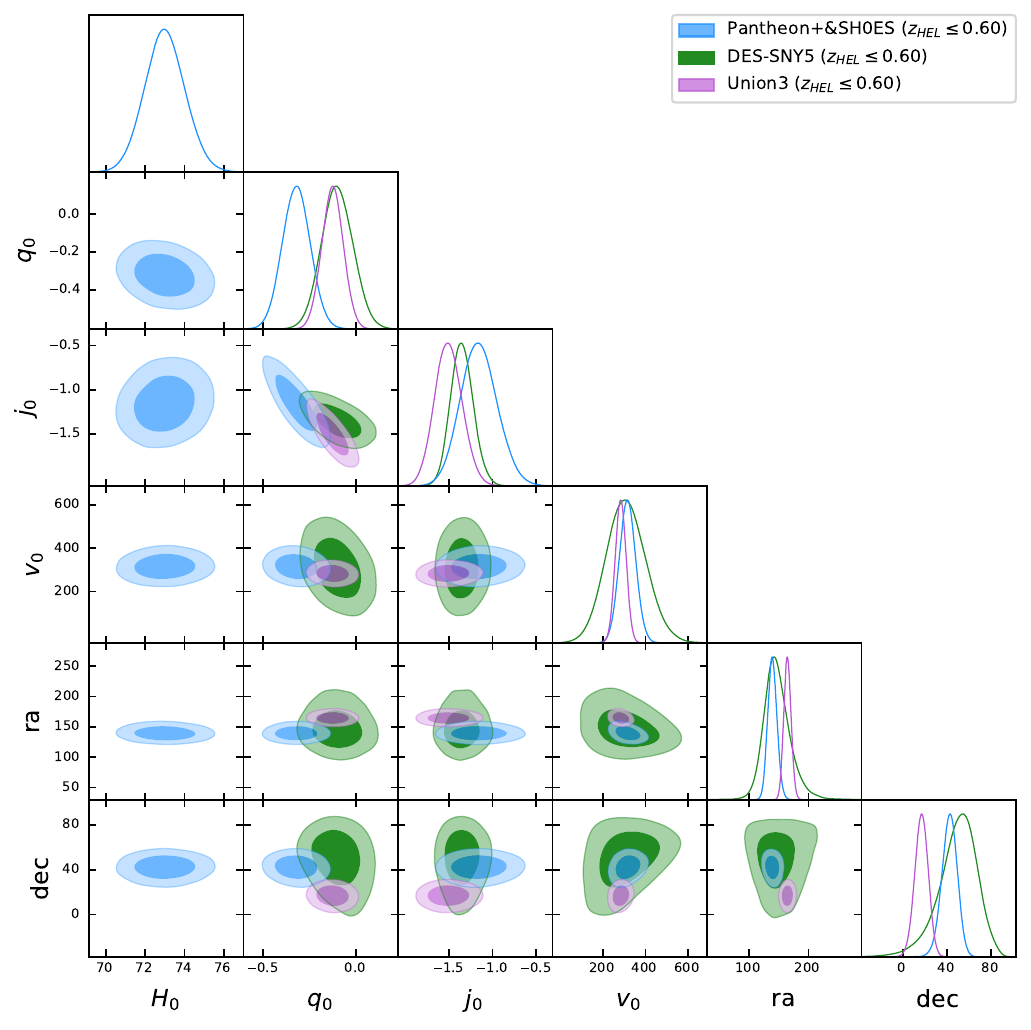}
\end{subfigure}
\begin{subfigure}{0.45\textwidth}
\centering
\includegraphics[width=\textwidth]{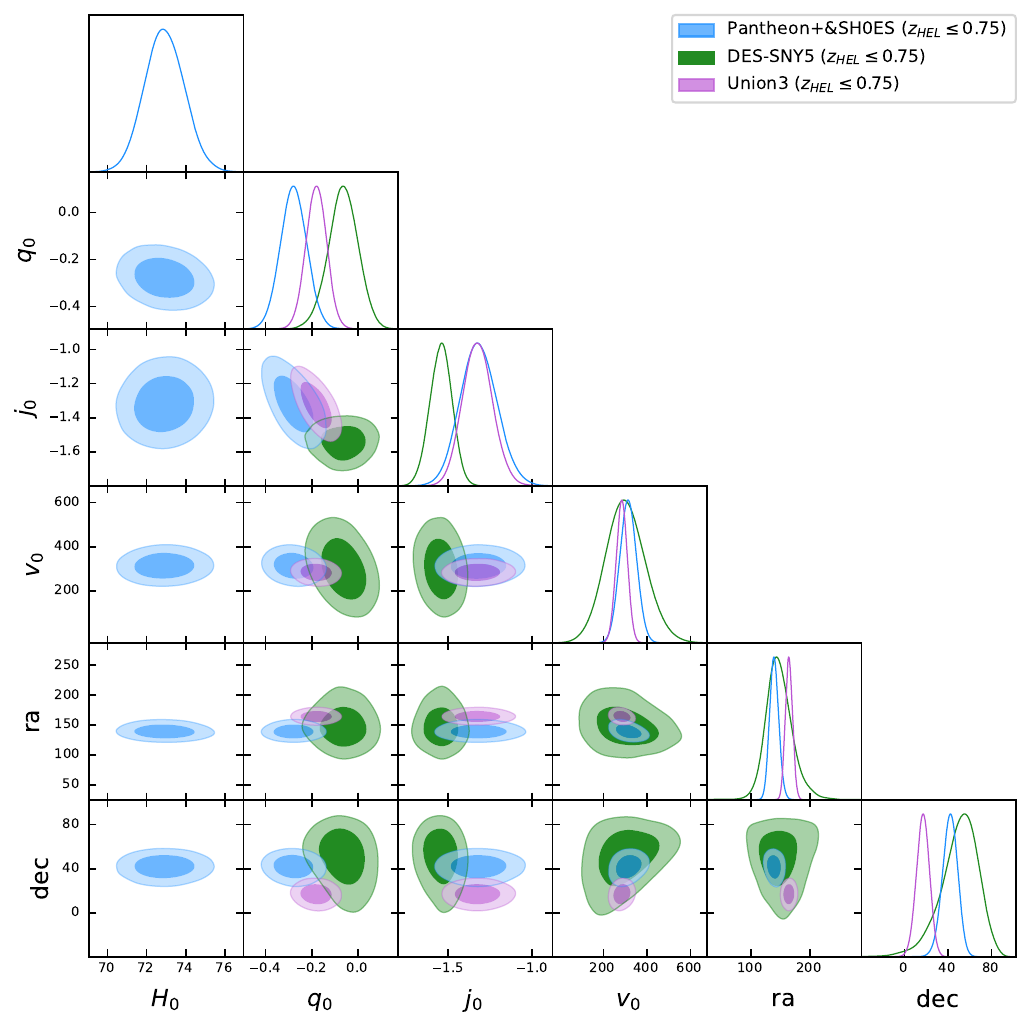}
\end{subfigure}

    \caption{Same as Fig. \ref{fig:TaylorDipole} but with the luminosity distance written through the Pad\'e parameterization of order $(1,2)$.}
    \label{fig:PadeDipole}
\end{figure*}

\begin{figure*}
    \centering

    \begin{subfigure}{0.45\textwidth}
\centering
\includegraphics[width=\textwidth]{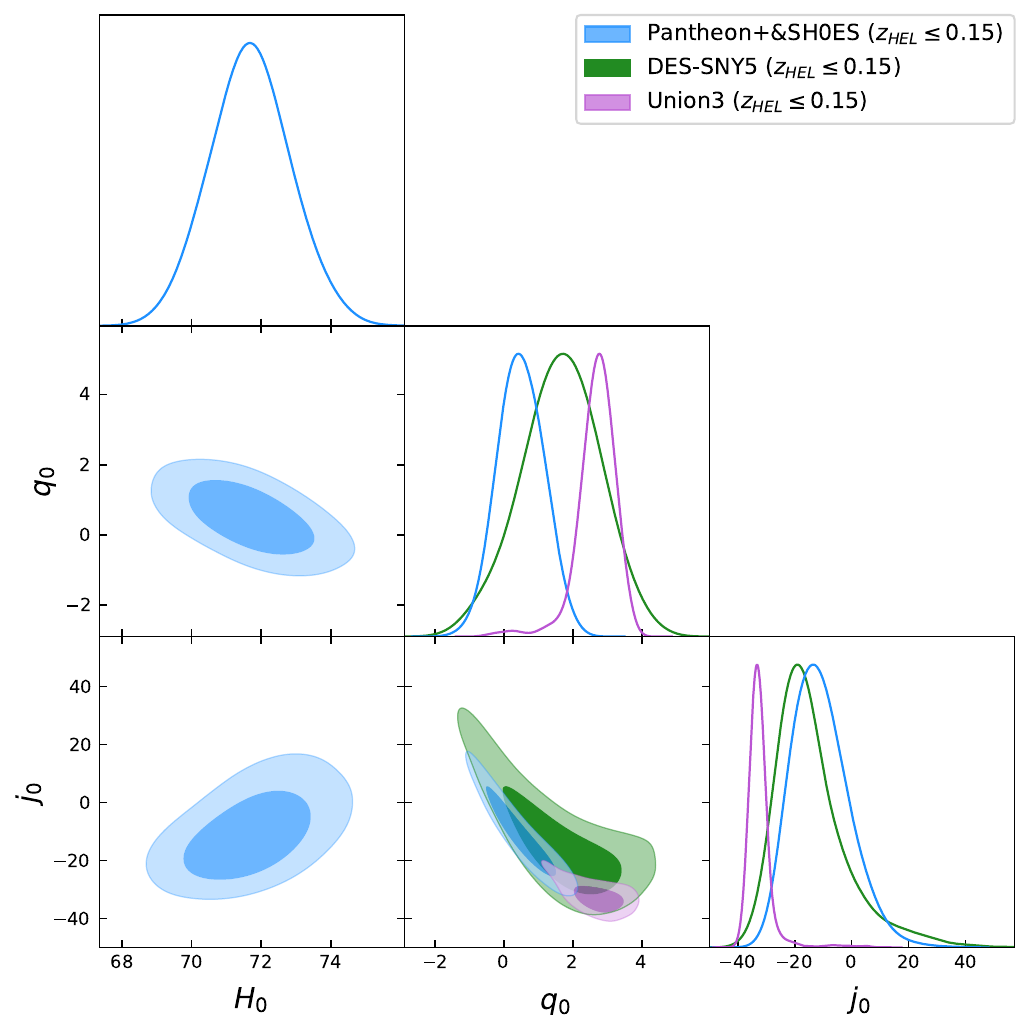}
\end{subfigure}
\begin{subfigure}{0.45\textwidth}
\centering
\includegraphics[width=\textwidth]{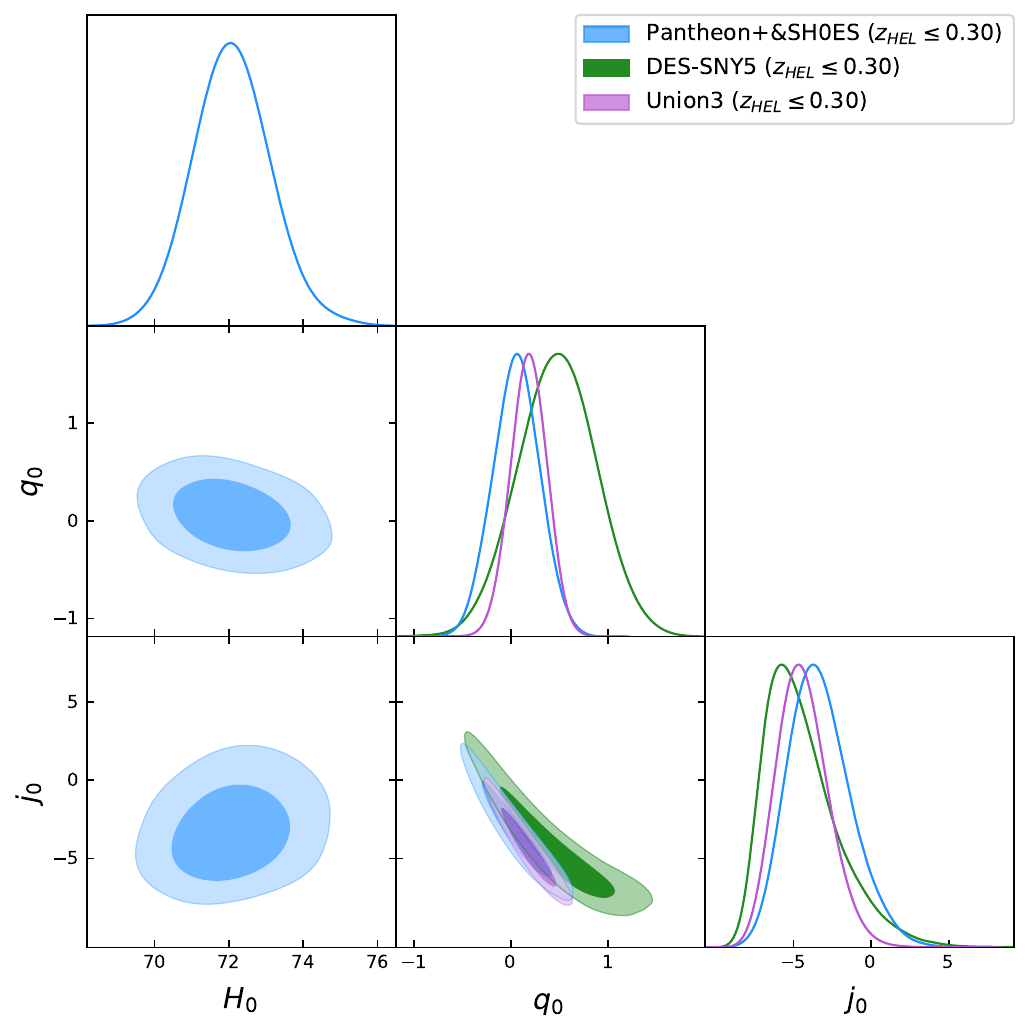}
\end{subfigure}
\begin{subfigure}{0.45\textwidth}
\centering
\includegraphics[width=\textwidth]{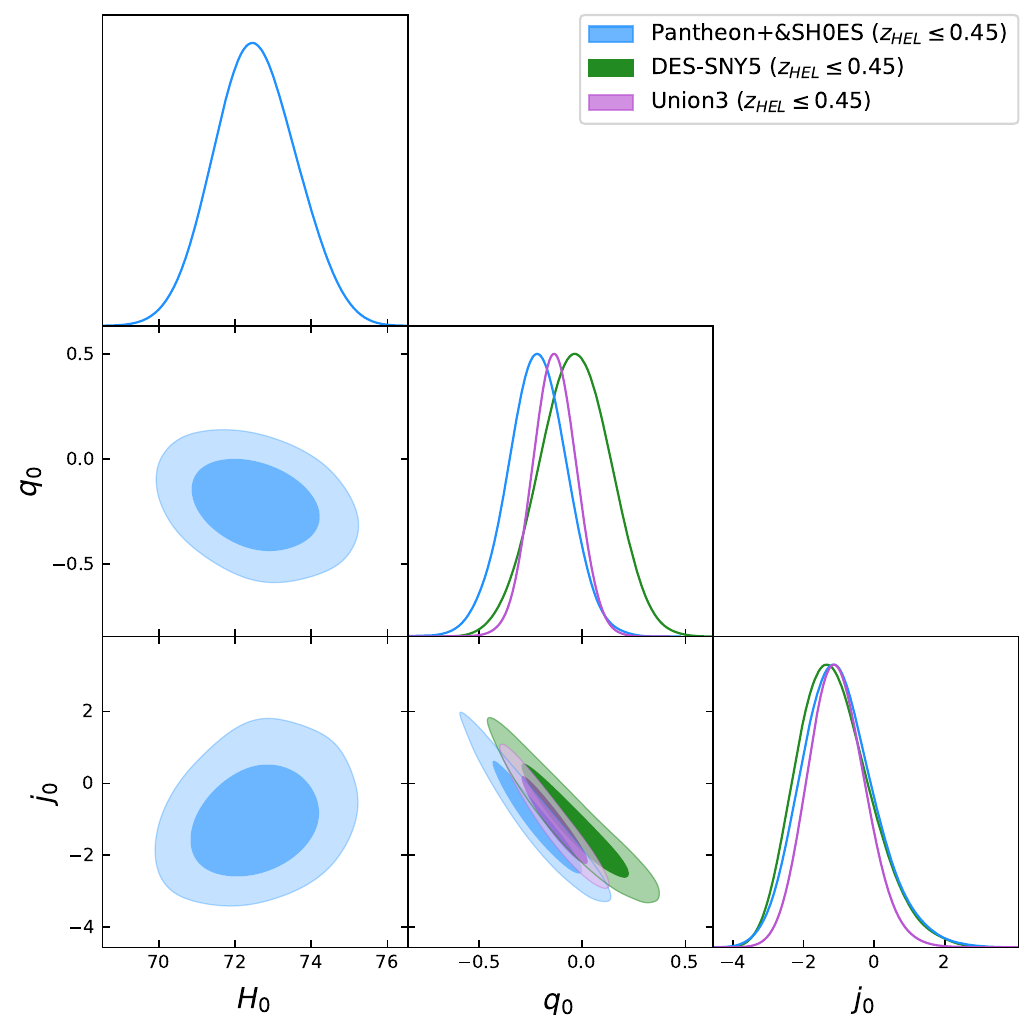}
\end{subfigure}
\begin{subfigure}{0.45\textwidth}
\centering
\includegraphics[width=\textwidth]{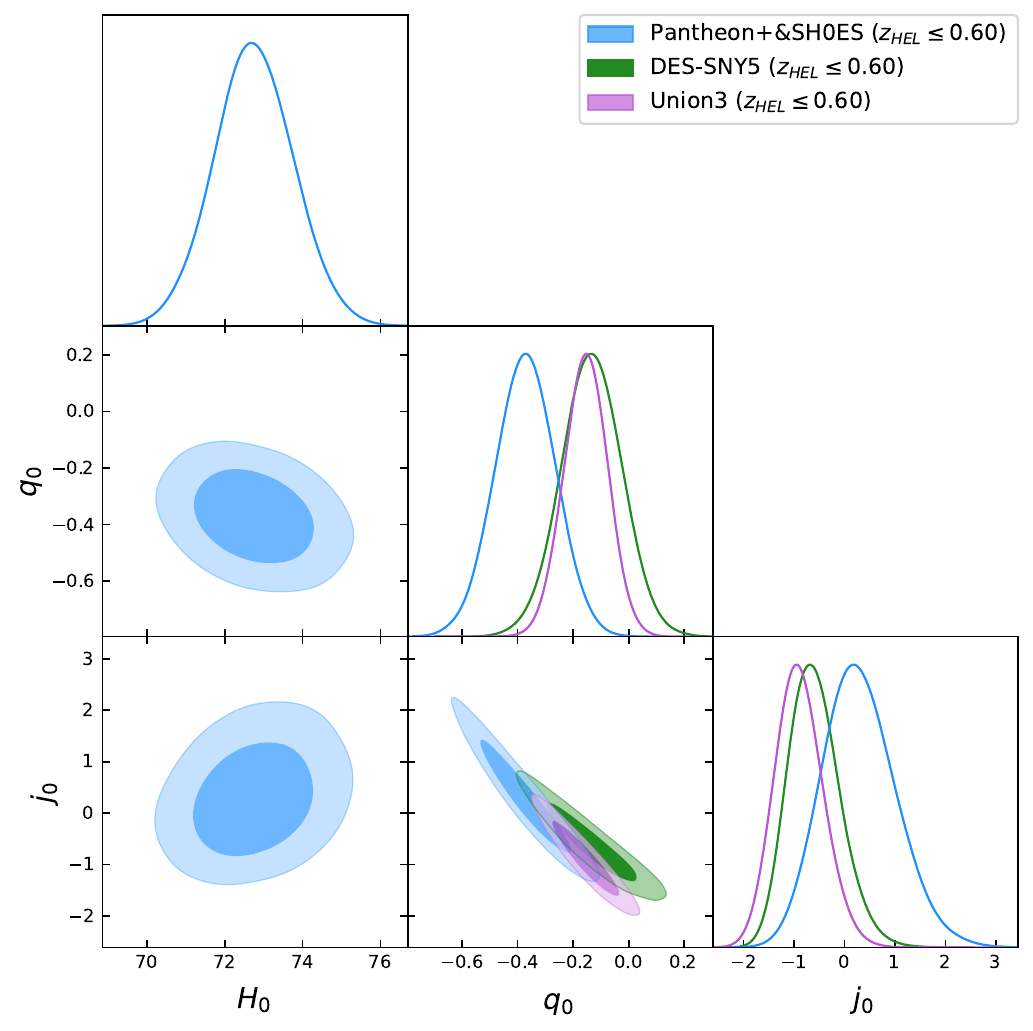}
\end{subfigure}
\begin{subfigure}{0.45\textwidth}
\centering
\includegraphics[width=\textwidth]{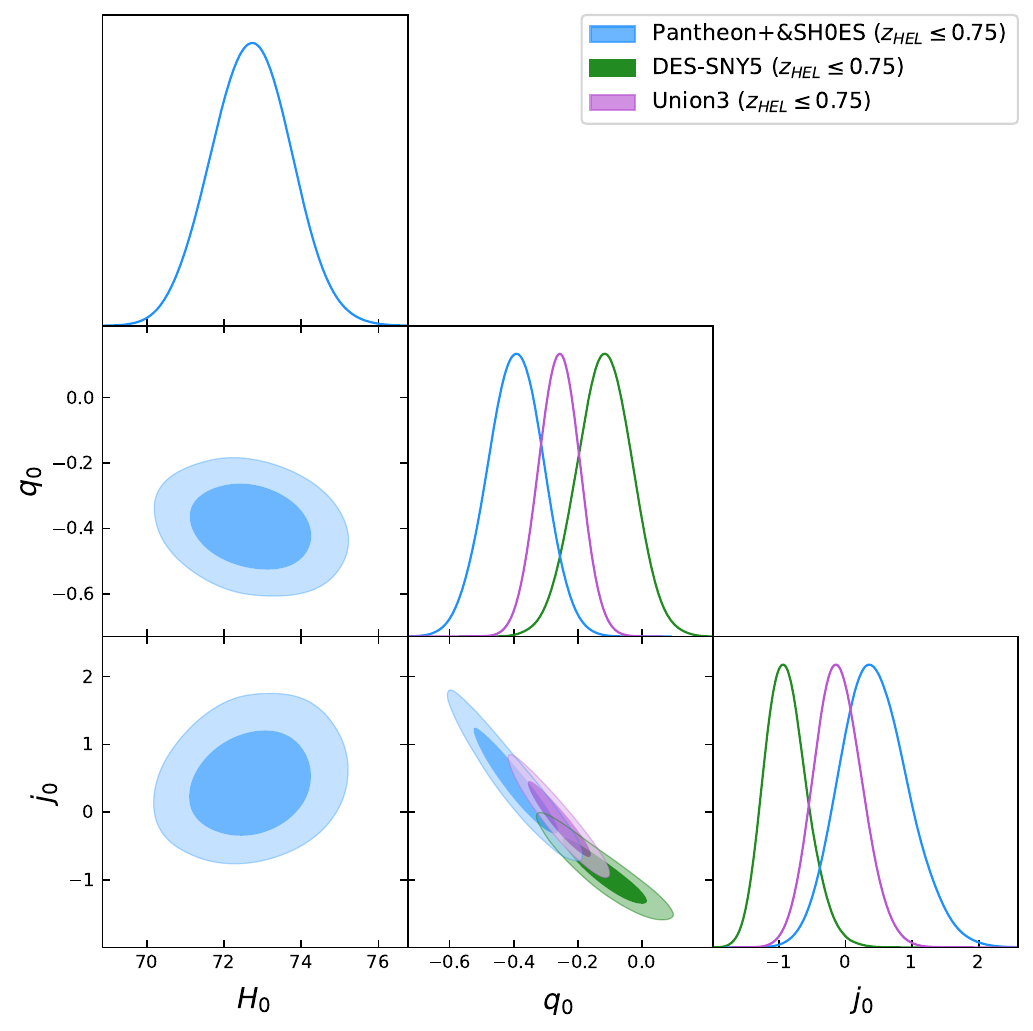}
\end{subfigure}

    \caption{Contour plots for the $H_0$, $q_0$, $j_0$ cosmographic parameters fixing $v_0 = 369.82~\mathrm{km\,s^{-1}}$, $\mathrm{ra} = 167.942^{\circ}$ and $\mathrm{dec} = -6.944^{\circ}$ to the inferred CMB values \cite{Planck:2013kqc, Planck:2018nkj}. The luminosity distance $d_L$ is written using the Taylor series. Also in this case, the Pantheon+\&SH0ES catalog is represented in blue, the DES-SNY5 catalog in green and the Union3 catalog in purple.}
    \label{fig:TaylorCMBDipole}
\end{figure*}

\begin{figure*}
    \centering

    \begin{subfigure}{0.45\textwidth}
\centering
\includegraphics[width=\textwidth]{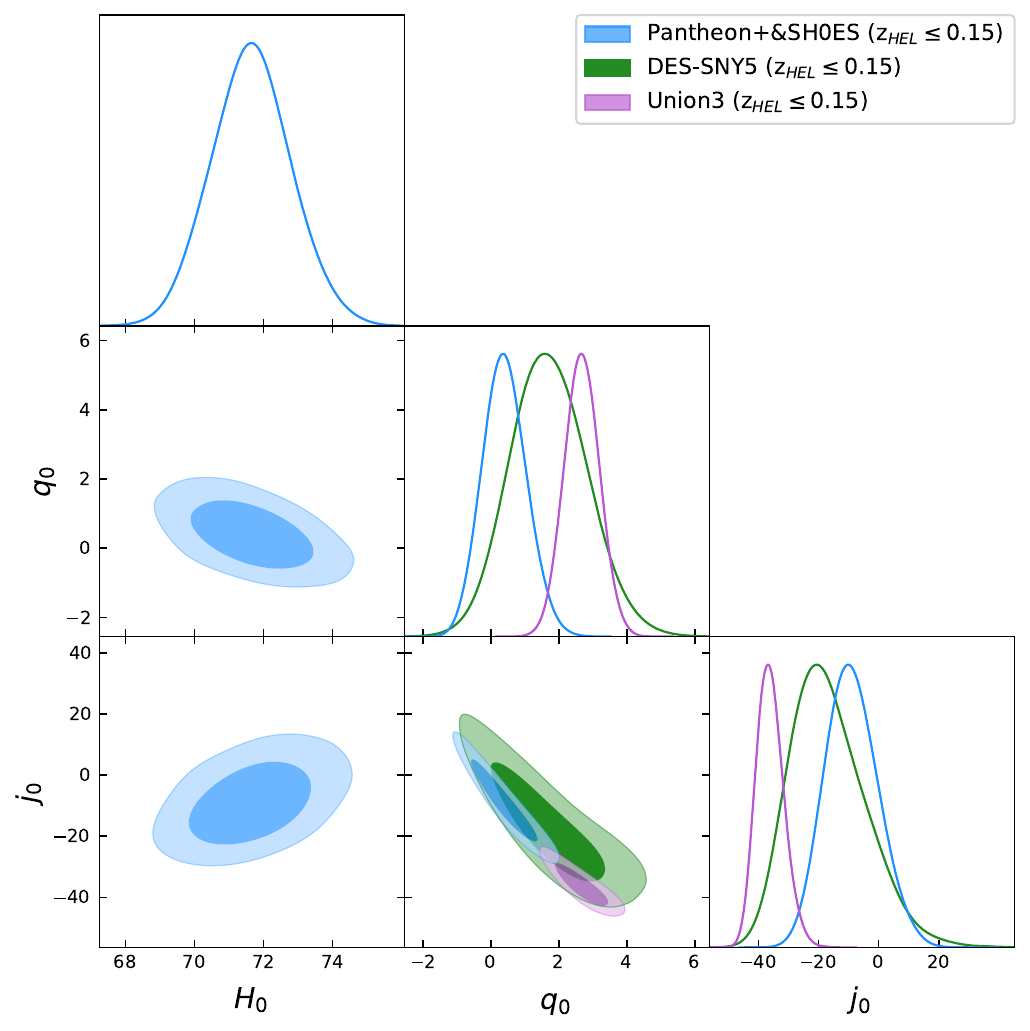}
\end{subfigure}
\begin{subfigure}{0.45\textwidth}
\centering
\includegraphics[width=\textwidth]{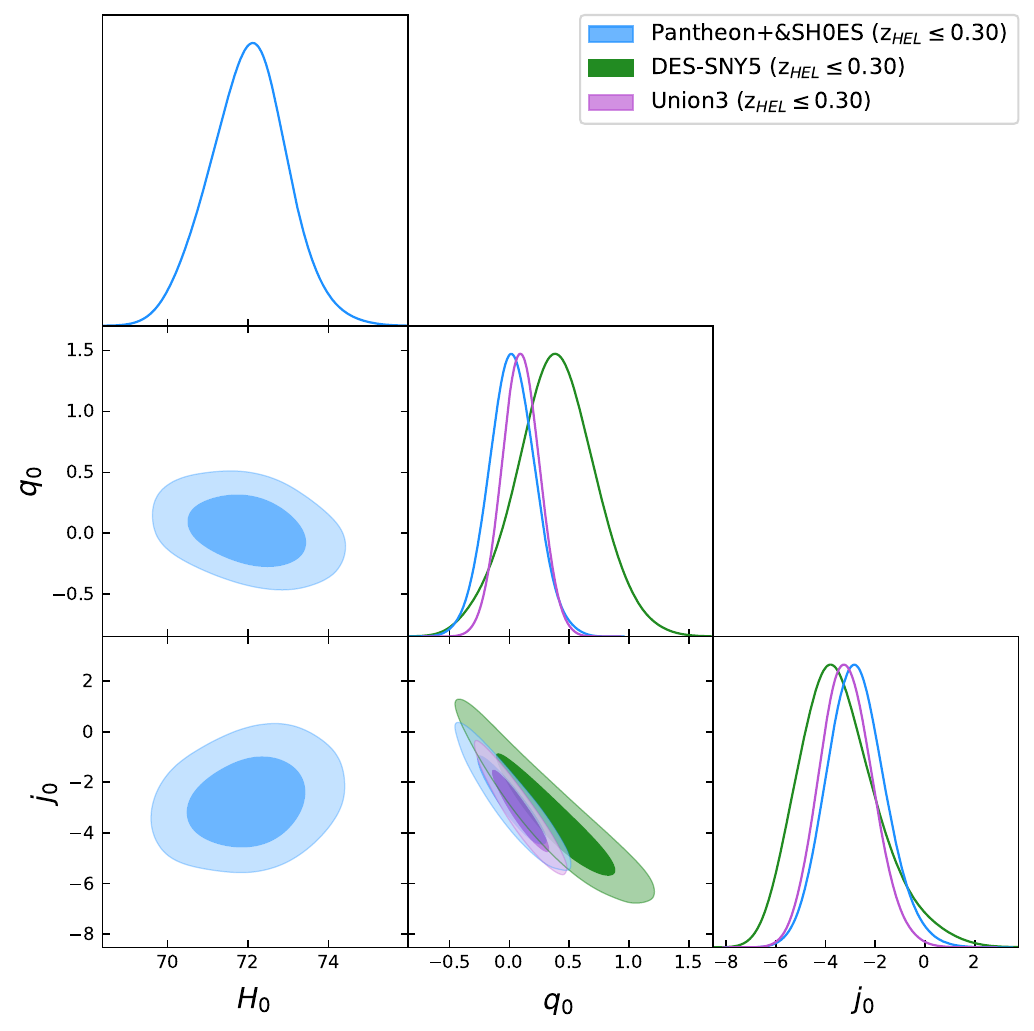}
\end{subfigure}
\begin{subfigure}{0.45\textwidth}
\centering
\includegraphics[width=\textwidth]{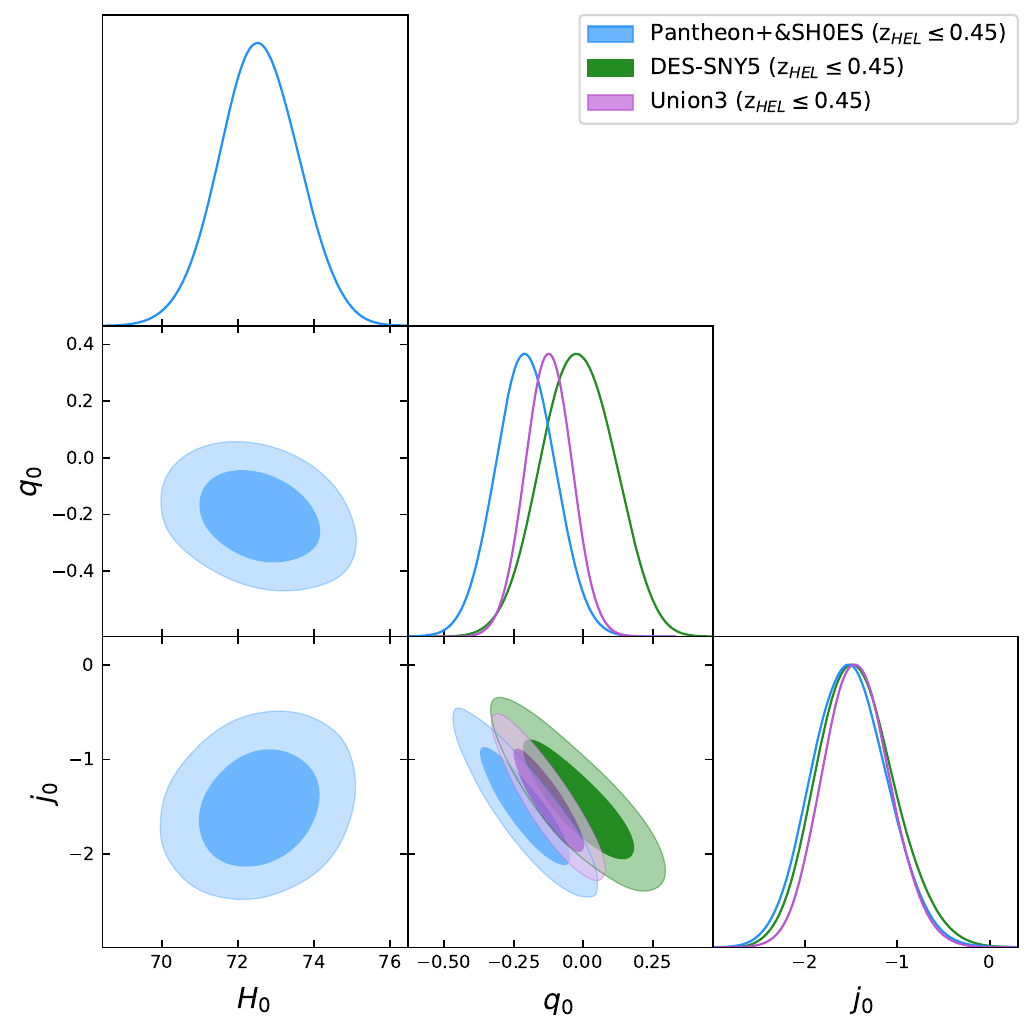}
\end{subfigure}
\begin{subfigure}{0.45\textwidth}
\centering
\includegraphics[width=\textwidth]{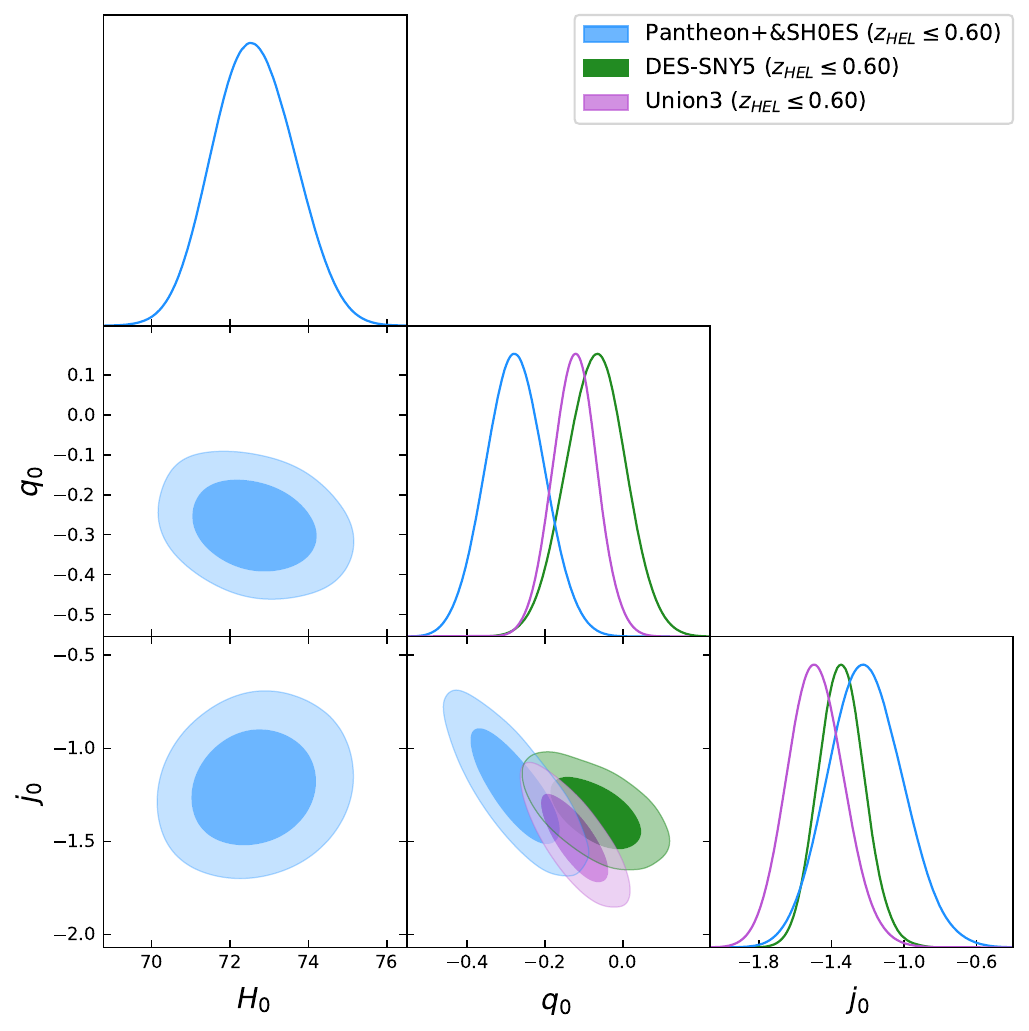}
\end{subfigure}
\begin{subfigure}{0.45\textwidth}
\centering
\includegraphics[width=\textwidth]{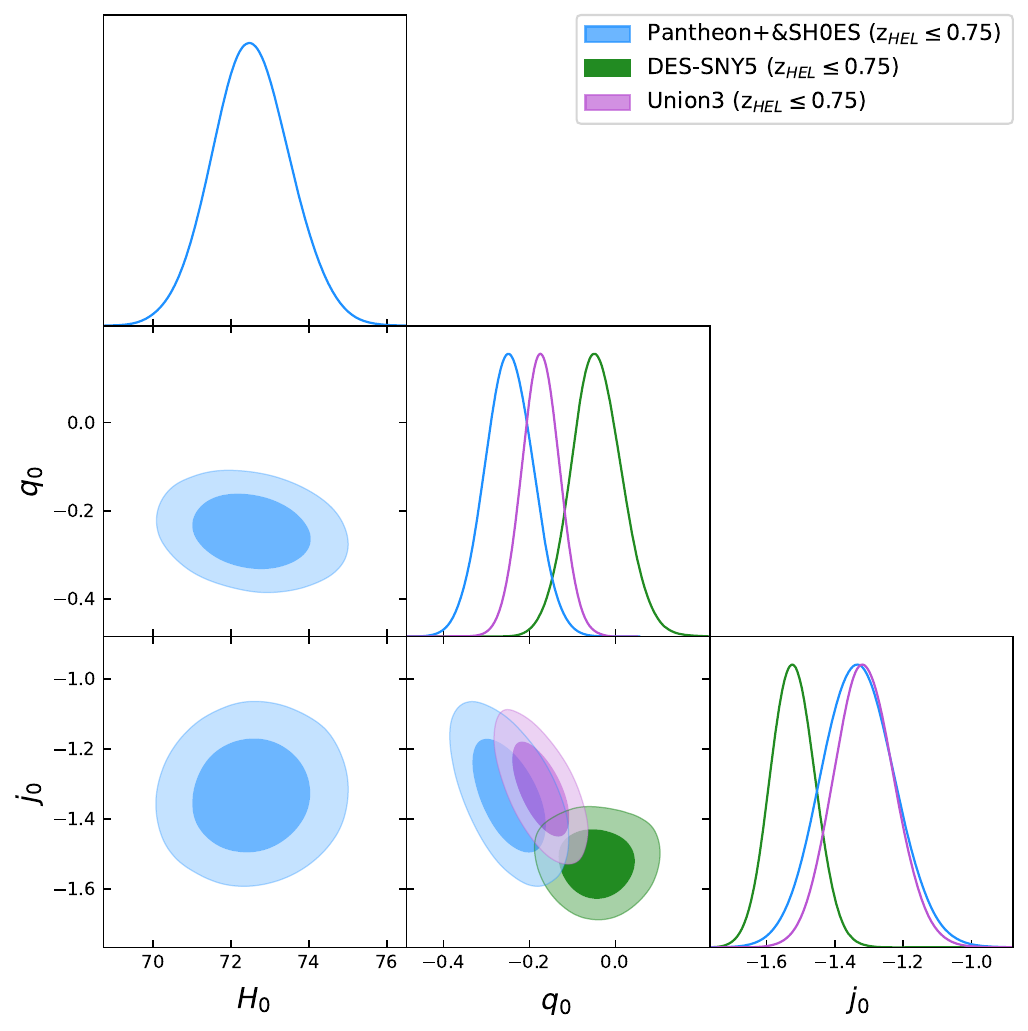}
\end{subfigure}

    \caption{Same as Fig. \ref{fig:TaylorCMBDipole} but employing the parameterized luminosity distance through a $(2,1)$ Pad\'e approximant.}
    \label{fig:PadeCMBDipole}
\end{figure*}

In this first appendix we show the tables with constraints on $H_0$ and on the cosmographic parameters $(q_0,\, j_0)$ for the three supernova samples considering the CMB-frame redshift $z_{\text{CMB}}$ in Tab. \ref{tab:cosmozCMB}, the dipole corrections and the dipole terms $(v_0,\, \text{ra},\, \text{dec})$ in Tab. \ref{tab:cosmoDIPOLE} fixed to the values of the CMB in Tab. \ref{tab:cosmoCMBdipole} presented in Sects. \ref{sec:5}-\ref{sec:6}, respectively. Furthermore, Figs. \ref{fig:TaylorzCMB}-\ref{fig:PadezCMB} depict the contours for the first analysis, Figs. \ref{fig:TaylorDipole}-\ref{fig:PadeDipole} the contours for the second analysis and Figs. \ref{fig:TaylorCMBDipole}-\ref{fig:PadeCMBDipole} show the contours when the CMB dipole is fixed. 

\begin{table*}
    \centering
    \setlength{\tabcolsep}{3.em}
    \renewcommand{\arraystretch}{1.2}
    \begin{tabular}{cccc}
    \hline\hline
    $z_{\rm CMB}$ & $H_0$ & $q_0$ & $j_0$  \\
    \hline\hline
    \multicolumn{4}{c}{{\bf Pantheon+\&SH0ES}}
    \\
    \hline
    \multicolumn{4}{c}{Taylor}\\
    $0.15$ & $71.75^{+1.186(2.397)}_{-1.271(2.348)}$ & $+0.446^{+0.566(1.333)}_{-0.698(1.222)}$ & $-10.37^{+8.459(18.87)}_{-10.35(18.26)}$\\
    $0.30$ & $72.09^{+1.163(2.103)}_{-1.040(2.271)}$ & $+0.017^{+0.253(0.491)}_{-0.246(0.498)}$ & $-2.887^{+1.830(4.499)}_{-2.431(4.056)}$\\
    $0.45$ & $72.43^{+1.036(2.143)}_{-1.079(2.093)}$ & $-0.187^{+0.151(0.301)}_{-0.149(0.296)}$ & $-1.284^{+0.940(2.160)}_{-1.156(2.020)}$\\
    $0.60$ & $72.67^{+1.072(2.193)}_{-1.103(2.174)}$ & $-0.372^{+0.112(0.221)}_{-0.112(0.221)}$ & $+0.290^{+0.691(1.510)}_{-0.806(1.458)}$\\
    $0.75$ & $72.70^{+1.013(2.046)}_{-1.040(2.102)}$ & $-0.396^{+0.089(0.181)}_{-0.091(0.179)}$ & $+0.451^{+0.504(1.075)}_{-0.560(1.069)}$\\
    \multicolumn{4}{c}{Pad\'e (1,2)}\\
    $0.15$ & $71.69^{+1.189(2.438)}_{-1.239(2.418)}$ & $
    +0.443^{+0.692(1.392)}_{-0.716(1.404)}$ & $-9.566^{+8.552(19.20)}_{-10.35(18.23)}$\\
    $0.30$ & $72.10^{+1.152(2.225)}_{-1.119(2.260)}$ & $-0.005^{+0.200(0.412)}_{-0.207(0.401)}$ & $-2.519^{+1.161(2.540)}_{-1.361(2.475)}$\\
    $0.45$ & $72.45^{+1.093(2.080)}_{-1.088(2.138)}$ & $-0.182^{+0.117(0.230)}_{-0.114(0.233)}$ & $-1.639^{+0.401(0.886)}_{-0.467(0.844)}$\\
    $0.60$ & $72.59^{+1.086(2.176)}_{-1.071(2.173)}$ & $-0.279^{+0.079(0.156)}_{-0.078(0.158)}$ & $-1.205^{+0.205(0.429)}_{-0.224(0.428)}$\\
    $0.75$ & $72.49^{+1.027(2.079)}_{-1.041(2.135)}$ & $-0.249^{+0.059(0.118)}_{-0.059(0.116)}$ & $-1.328^{+0.109(0.226)}_{-0.114(0.220)}$\\
    \hline
    \multicolumn{4}{c}{{\bf DES-SNY5}}\\
    \hline
    \multicolumn{4}{c}{Taylor}\\
    $0.15$ & $-$ & $+1.708^{+1.157(2.298)}_{-1.126(2.275)}$ & $-15.62^{+7.467(24.78)}_{-13.21(22.24)}$ \\
    $0.30$ & $-$ & $+0.483^{+0.428(0.806)}_{-0.399(0.831)}$ & $-4.449^{+1.563(5.264)}_{-2.940(5.122)}$\\
    $0.45$ & $-$ & $-0.021^{+0.185(0.353)}_{-0.174(0.358)}$ & $-1.225^{+0.892(2.193)}_{-1.195(2.004)}$\\
    $0.60$ & $-$ & $-0.136^{+0.116(0.228)}_{-0.111(0.231)}$ & $-0.584^{+0.463(1.104)}_{-0.598(1.041)}$\\
    $0.75$ & $-$ & $-0.119^{+0.093(0.183)}_{-0.091(0.183)}$ & $-0.876^{+0.304(0.668)}_{-0.368(0.671)}$\\
    \multicolumn{4}{c}{Pad\'e (1,2)}\\
    $0.15$ & $-$ & $+1.883^{+1.172(2.324)}_{-1.148(2.400)}$ & $-18.76^{+8.846(25.75)}_{-13.66(22.31)}$\\
    $0.30$ & $-$ & $+0.389^{+0.354(0.667)}_{-0.333(0.689)}$ & $-3.439^{+1.314(3.464)}_{-1.896(3.083)}$\\
    $0.45$ & $-$ & $-0.010^{+0.131(0.271)}_{-0.138(0.268)}$ & $-1.482^{+0.396(0.858)}_{-0.453(0.838)}$\\
    $0.60$ & $-$ & $-0.069^{+0.077(0.158)}_{-0.080(0.155)}$ & $-1.351^{+0.127(0.262)}_{-0.132(0.258)}$\\
    $0.75$ & $-$ & $-0.045^{+0.060(0.122)}_{-0.062(0.122)}$ & $-1.520^{+0.063(0.133)}_{-0.067(0.132)}$\\
    \hline
    \multicolumn{4}{c}{{\bf Union3}}\\
    \hline
    \multicolumn{4}{c}{Taylor}\\
    $0.15$ & $-$ & $+2.905^{+0.522(1.004)}_{-0.494(1.009)}$ & $-33.84^{+2.728(5.709)}_{-2.917(5.820)}$\\
    $0.30$ & $-$ & $+0.173^{+0.194(0.387)}_{-0.197(0.387)}$ & $-4.331^{+1.561(3.504)}_{-1.879(3.339)}$\\
    $0.45$ & $-$ & $-0.111^{+0.109(0.217)}_{-0.118(0.224)}$ & $-1.265^{+0.784(1.721)}_{-0.903(1.616)}$\\
    $0.60$ & $-$ & $-0.154^{+0.083(0.160)}_{-0.080(0.164)}$ & $-0.921^{+0.459(0.968)}_{-0.529(0.979)}$\\
    $0.75$ & $-$ & $-0.261^{+0.066(0.134)}_{-0.067(0.131)}$ & $-0.088^{+0.365(0.761)}_{-0.389(0.739)}$\\
    \multicolumn{4}{c}{Pad\'e (1,2)}\\
    $0.15$ & $-$ & $+2.756^{+0.537(1.084)}_{-0.551(1.056)}$ & $-37.18^{+3.993(9.689)}_{-5.180(8.775)}$\\
    $0.30$ & $-$ & $+0.097^{+0.160(0.316)}_{-0.161(0.319)}$ & $-3.173^{+1.040(2.190)}_{-1.119(2.121)}$\\
    $0.45$ & $-$ & $-0.108^{+0.085(0.176)}_{-0.088(0.172)}$ & $-1.534^{+0.359(0.759)}_{-0.386(0.735)}$\\
    $0.60$ & $-$ & $-0.118^{+0.059(0.121)}_{-0.060(0.118)}$ & $-1.505^{+0.160(0.331)}_{-0.170(0.324)}$\\
    $0.75$ & $-$ & $-0.176^{+0.047(0.092)}_{-0.046(0.094)}$ & $-1.308^{+0.090(0.189)}_{-0.097(0.183)}$\\
    \hline\hline
    \end{tabular}
\caption{Constraints on the Hubble constant $H_0$ when Pantheon+\&SH0ES sample is adopted and on the cosmographic parameters $(q_0,\, j_0)$ for all three SNe Ia catalogs with $1$-$\sigma$($2$-$\sigma$) error bars for $z\leq z_{\rm CMB}$ adopting a third order Taylor expansion and $(1, 2)$ Pad\'e approximant, respectively for Pantheon+\&SH0ES, DES-SNY5 and Union3 data catalogs.}
    \label{tab:cosmozCMB}
\end{table*}

\begin{table*}
    \centering
    \setlength{\tabcolsep}{0.7em}
    \renewcommand{\arraystretch}{1.2}
    \begin{tabular}{lcccccc}
    \hline\hline
    $z_{\rm HEL}$ & $H_0$ & $q_0$ & $j_0$ & $v_0$ & $\text{ra}$ & $\text{dec}$ \\
    \hline\hline
    \multicolumn{7}{c}{{\bf Pantheon+\&SH0ES}}
    \\
    \hline
    \multicolumn{7}{c}{Taylor}\\
    $0.15$ & $72.47^{+1.274(2.518)}_{-1.250(2.515)}$ & $-0.013^{+0.725(1.509)}_{-0.737(1.494)}$ & $-3.858^{+9.674(27.04)}_{-15.14(24.35)}$ & $309.4^{+40.10(80.94)}_{-41.46(81.14)}$ & $139.1^{+7.703(16.41)}_{-8.475(16.11)}$ & $41.20^{+7.990(15.01)}_{-7.060(15.18)}$\\
    $0.30$ & $72.61^{+1.123(2.207)}_{-1.086(2.187)}$ & $-0.057^{+0.255(0.495)}_{-0.242(0.494)}$ & $-2.565^{+1.900(4.705)}_{-2.565(4.327)}$ & $307.1^{+41.11(81.88)}_{-41.28(80.70)}$ & $139.4^{+7.974(16.78)}_{-8.610(16.56)}$ & $41.48^{+7.922(14.93)}_{-6.972(15.47)}$\\
    $0.45$ & $72.93^{+1.087(2.182)}_{-1.089(2.179)}$ & $-0.291^{+0.153(0.302)}_{-0.151(0.304)}$ & $-0.627^{+1.030(2.378)}_{-1.247(2.192)}$ & $313.3^{+39.20(81.35)}_{-40.79(79.10)}$ & $139.8^{+7.775(16.49)}_{-8.536(16.20)}$ & $41.90^{+7.682(14.41)}_{-6.894(15.02)}$\\
    $0.60$ & $73.20^{+1.039(2.172)}_{-1.180(2.069)}$ & $-0.438^{+0.112(0.223)}_{-0.111(0.222)}$ & $+0.654^{+0.739(1.635)}_{-0.831(1.531)}$ & $317.1^{+38.41(79.44)}_{-40.07(78.49)}$ & $139.6^{+7.542(15.95)}_{-8.244(15.81)}$ & $42.05^{+7.532(14.46)}_{-6.884(14.57)}$\\
    $0.75$ & $73.18^{+1.073(2.037)}_{-1.045(2.118)}$ & $-0.450^{+0.094(0.184)}_{-0.091(0.181)}$ & $+0.720^{+0.525(1.164)}_{-0.617(1.106)}$ & $318.2^{+37.71(78.81)}_{-40.03(78.61)}$ & $139.5^{+7.309(16.16)}_{-8.214(15.61)}$ & $42.22^{+7.406(14.06)}_{-6.826(14.38)}$\\
    \multicolumn{7}{c}{Pad\'e (1,2)}\\
    $0.15$ & $72.47^{+1.261(2.445)}_{-1.209(2.417)}$ & $-0.019^{+0.662(1.338)}_{-0.693(1.337)}$ & $-4.082^{+8.885(20.17)}_{-10.99(19.00)}$ & $308.8^{+39.94(81.17)}_{-41.32(81.35)}$ & $139.0^{+7.650(16.99)}_{-8.629(16.16)}$ & $41.12^{+7.806(14.77)}_{-7.047(15.13)}$\\
    $0.30$ & $72.58^{+1.097(2.134)}_{-1.101(2.189)}$ & $-0.063^{+0.205(0.408)}_{-0.198(0.406)}$ & $-2.415^{+1.147(2.597)}_{-1.362(2.514)}$ & $305.6^{+39.95(81.41)}_{-41.28(80.72)}$ & $139.4^{+7.739(16.71)}_{-8.582(16.23)}$ & $41.36^{+7.924(15.30)}_{-7.216(15.61)}$\\
    $0.45$ & $72.97^{+1.064(2.138)}_{-1.083(2.143)}$ & $-0.262^{+0.112(0.226)}_{-0.112(0.225)}$ & $-1.403^{+0.416(0.893)}_{-0.465(0.874)}$ & $314.5^{+39.22(80.37)}_{-41.07(79.50)}$ & $139.7^{+7.783(16.52)}_{-8.434(16.15)}$ & $42.04^{+7.593(14.23)}_{-6.918(14.86)}$\\
    $0.60$ & $73.01^{+1.035(2.132)}_{-1.082(2.113)}$ & $-0.321^{+0.076(0.155)}_{-0.079(0.152)}$ & $-1.154^{+0.207(0.438)}_{-0.227(0.429)}$ & $315.6^{+39.04(80.73)}_{-40.61(80.07)}$ & $139.7^{+7.692(16.39)}_{-8.360(15.87)}$ & $42.22^{+7.541(14.38)}_{-6.865(14.66)}$\\
    $0.75$ & $72.91^{+1.041(2.142)}_{-1.078(2.096)}$ & $-0.278^{+0.058(0.119)}_{-0.060(0.118)}$ & $-1.318^{+0.111(0.232)}_{-0.118(0.226)}$ & $313.7^{+39.91(81.81)}_{-41.00(80.22)}$ & $139.7^{+7.590(16.61)}_{-8.556(16.01)}$ & $41.79^{+7.589(14.76)}_{-7.065(14.71)}$\\
    \hline
    \multicolumn{7}{c}{{\bf DES-SNY5}}\\
    \hline
    \multicolumn{7}{c}{Taylor}\\
    $0.15$ & $-$ & $+1.353^{+1.256(2.485)}_{-1.234(2.421)}$ & $-12.52^{+8.994(32.00)}_{-17.31(27.88)}$ & $257.7^{+99.28(207.6)}_{-108.1(206.9)}$ & $147.2^{+25.65(78.42)}_{-31.95(64.12)}$ & $45.13^{+27.51(43.64)}_{-16.57(48.81)}$ \\
    $0.30$ & $-$ & $+0.345^{+0.439(0.936)}_{-0.455(0.883)}$ & $-3.624^{+1.789(6.134)}_{-3.548(5.002)}$ & $264.6^{+95.79(199.4)}_{-103.7(193.8)}$ & $147.6^{+22.15(62.42)}_{-29.70(58.43)}$ & $44.58^{+26.32(41.73)}_{-15.84(48.53)}$\\
    $0.45$ & $-$ & $-0.088^{+0.195(0.383)}_{-0.191(0.391)}$ & $-0.893^{+0.976(2.452)}_{-1.343(2.235)}$ & $296.4^{+95.24(199.7)}_{-102.0(196.6)}$ & $145.0^{+23.12(60.07)}_{-27.18(56.82)}$ & $49.18^{+21.23(36.28)}_{-15.64(38.80)}$\\
    $0.60$ & $-$ & $-0.177^{+0.123(0.244)}_{-0.121(0.244)}$ & $-0.460^{+0.497(1.187)}_{-0.641(1.111)}$ & $310.2^{+93.76(197.8)}_{-101.7(191.3)}$ & $148.9^{+20.67(54.88)}_{-27.92(55.28)}$ & $49.76^{+20.17(34.83)}_{-14.66(37.12)}$\\
    $0.75$ & $-$ & $-0.153^{+0.100(0.196)}_{-0.096(0.198)}$ & $-0.808^{+0.314(0.776)}_{-0.412(0.706)}$ & $310.2^{+90.63(194.7)}_{-101.5(187.7)}$ & $150.0^{+22.38(55.22)}_{-27.21(52.66)}$ & $51.07^{+19.74(34.06)}_{-13.80(36.04)}$\\
    \multicolumn{7}{c}{Pad\'e (1,2)}\\
    $0.15$ & $-$ & $+1.470^{+1.200(2.455)}_{-1.275(2.463)}$ & $-15.10^{+11.17(29.20)}_{-16.46(26.42)}$ & $256.3^{+99.15(205.6)}_{-108.4(209.7)}$ & $147.7^{+24.06(76.30)}_{-32.50(64.92)}$ & $44.84^{+27.82(43.55)}_{-16.51(49.77)}$\\
    $0.30$ & $-$ & $+0.315^{+0.369(0.705)}_{-0.351(0.725)}$ & $-3.149^{+1.348(3.569)}_{-2.032(3.229)}$ & $262.9^{+96.64(197.9)}_{-99.49(195.6)}$ & $148.5^{+22.51(73.98)}_{-32.15(64.42)}$ & $46.53^{+25.16(41.30)}_{-16.80(44.70)}$\\
    $0.45$ & $-$ & $-0.068^{+0.139(0.286)}_{-0.147(0.285)}$ & $-1.366^{+0.414(0.903)}_{-0.472(0.873)}$ & $296.6^{+93.82(198.0)}_{-101.5(190.8)}$ & $146.2^{+22.77(56.03)}_{-26.69(52.20)}$ & $49.06^{+20.76(35.69)}_{-15.10(38.29)}$\\
    $0.60$ & $-$ & $-0.102^{+0.085(0.175)}_{-0.089(0.171)}$ & $-1.349^{+0.140(0.277)}_{-0.130(0.267)}$ & $309.4^{+90.85(192.7)}_{-99.29(187.5)}$ & $147.7^{+20.17(54.06)}_{-27.40(51.01)}$ & $49.39^{+20.05(34.64)}_{-13.90(36.72)}$\\
    $0.75$ & $-$ & $-0.064^{+0.068(0.131)}_{-0.063(0.133)}$ & $-1.543^{+0.070(0.132)}_{-0.067(0.138)}$ & $301.5^{+88.18(190.8)}_{-98.87(186.6)}$ & $148.4^{+22.69(55.37)}_{-25.74(48.79)}$ & $50.25^{+20.19(33.65)}_{-14.09(36.72)}$\\
    \hline
    \multicolumn{7}{c}{{\bf Union3}}\\
    \hline
    \multicolumn{7}{c}{Taylor}\\
    $0.15$ & $-$ & $+2.752^{+0.519(1.049)}_{-0.520(1.034)}$ & $-33.15^{+2.894(6.458)}_{-3.261(6.301)}$ & $292.3^{+27.49(54.81)}_{-27.65(54.42)}$ & $163.8^{+6.088(12.88)}_{-6.530(12.32)}$ & $17.37^{+6.290(12.38)}_{-6.114(12.46)}$\\
    $0.30$ & $-$ & $+0.167^{+0.198(0.394)}_{-0.196(0.397)}$ & $-4.342^{+1.571(3.614)}_{-1.948(3.440)}$ & $282.5^{+27.12(54.56)}_{-27.42(54.72)}$ & $164.4^{+6.223(12.77)}_{-6.583(12.80)}$ & $16.54^{+6.556(12.43)}_{-6.037(12.99)}$\\
    $0.45$ & $-$ & $-0.132^{+0.113(0.228)}_{-0.112(0.227)}$ & $-1.136^{+0.781(1.755)}_{-0.915(1.656)}$ & $283.3^{+26.06(53.61)}_{-26.75(51.99)}$ & $165.0^{+6.009(13.02)}_{-6.594(12.37)}$ & $17.06^{+6.375(12.48)}_{-6.128(12.51)}$\\
    $0.60$ & $-$ & $-0.163^{+0.081(0.168)}_{-0.085(0.166)}$ & $-0.894^{+0.480(1.042)}_{-0.522(0.999)}$ & $282.7^{+26.72(52.78)}_{-26.47(53.66)}$ & $164.7^{+6.057(12.69)}_{-6.505(12.44)}$ & $16.76^{+6.459(12.33)}_{-6.127(12.65)}$\\
    $0.75$ & $-$ & $-0.269^{+0.068(0.134)}_{-0.068(0.134)}$ & $-0.091^{+0.369(0.761)}_{-0.399(0.759)}$ & $284.9^{+26.50(53.26)}_{-26.81(53.09)}$ & $164.5^{+6.007(12.58)}_{-6.363(12.01)}$ & $17.03^{+6.476(12.38)}_{-6.051(12.64)}$\\
    \multicolumn{7}{c}{Pad\'e (1,2)}\\
    $0.15$ & $-$ & $+2.644^{+0.526(1.071)}_{-0.540(1.085)}$ & $-36.24^{+3.802(9.949)}_{-5.485(8.925)}$ & $292.5^{+27.58(55.62)}_{-27.93(55.27)}$ & $163.7^{+6.031(12.55)}_{-6.458(12.63)}$ & $17.32^{+6.405(12.22)}_{-6.082(12.66)}$\\
    $0.30$ & $-$ & $+0.087^{+0.163(0.325)}_{-0.163(0.332)}$ & $-3.168^{+1.042(2.250)}_{-1.150(2.183)}$ & $283.1^{+26.61(54.48)}_{-27.16(53.30)}$ & $164.3^{+6.185(12.75)}_{-6.558(12.79)}$ & $16.60^{+6.428(12.34)}_{-6.142(12.69)}$\\
    $0.45$ & $-$ & $-0.128^{+0.086(0.180)}_{-0.090(0.175)}$ & $-1.469^{+0.368(0.772)}_{-0.393(0.755)}$ & $283.4^{+26.80(53.11)}_{-26.56(53.42)}$ & $164.9^{+6.152(12.72)}_{-6.501(12.58)}$ & $16.93^{+6.384(12.53)}_{-6.153(12.64)}$\\
    $0.60$ & $-$ & $-0.127^{+0.060(0.119)}_{-0.059(0.120)}$ & $-1.504^{+0.156(0.326)}_{-0.169(0.325)}$ & $283.2^{+26.53(53.31)}_{-26.62(53.15)}$ & $164.8^{+6.112(12.79)}_{-6.528(12.62)}$ & $16.82^{+6.430(12.54)}_{-6.129(12.64)}$\\
    $0.75$ & $-$ & $-0.181^{+0.046(0.093)}_{-0.046(0.093)}$ & $-1.324^{+0.089(0.189)}_{-0.096(0.184)}$ & $284.0^{+26.81(53.17)}_{-26.34(53.30)}$ & $164.6^{+6.171(12.57)}_{-6.481(12.74)}$ & $17.00^{+6.373(12.31)}_{-6.047(12.61)}$\\
    \hline\hline
    \end{tabular}
\caption{Constraints on the Hubble constant $H_0$ when Pantheon+\&SH0ES sample is adopted and on the cosmographic parameters $(q_0,\, j_0)$ for all three SNe Ia catalogs and on the velocity amplitude $v_0$ in units of $\mathrm{km\,s^{-1}}$ and direction across the sky $(\text{ra},\ \text{dec})$ in degree units. Both parameters show $1$-$\sigma$($2$-$\sigma$) error bars for $z\leq z_{\rm HEL}$ adopting a third order Taylor expansion and $(1,\ 2)$ Pad\'e approximant, respectively for Pantheon+\&SH0ES, DES-SNY5 and Union3 data catalogs.}
    \label{tab:cosmoDIPOLE}
\end{table*}

\begin{table*}
    \centering
    \setlength{\tabcolsep}{3em}
    \renewcommand{\arraystretch}{1.2}

    \begin{tabular}{cccc}
        \hline\hline
        $z_{\rm HEL}$ & $H_0$ & $q_0$ & $j_0$ \\
        \hline
        \multicolumn{4}{c}{
            $v_0 = 369.82~\mathrm{km\,s^{-1}}$,\quad
            $\mathrm{ra} = 167.942^{\circ}$,\quad
            $\mathrm{dec} = -6.944^{\circ}$
        } \\
        \hline\hline

        \multicolumn{4}{c}{\textbf{Pantheon+\&SH0ES}} \\
        \hline

        \multicolumn{4}{c}{Taylor} \\
        $0.15$ & $71.69^{+1.205(2.486)}_{-1.239(2.414)}$ & $+0.486^{+0.705(1.397)}_{-0.717(1.356)}$ & $-10.98^{+8.106(21.57)}_{-12.45(24.67)}$ \\
        $0.30$ & $72.09^{+1.021(2.179)}_{-1.110(2.184)}$ & $+0.055^{+0.250(0.491)}_{-0.250(0.496)}$ & $-3.291^{+1.768(4.322)}_{-2.383(3.996)}$ \\
        $0.45$ & $72.53^{+1.078(2.265)}_{-1.173(2.173)}$ & $-0.218^{+0.153(0.306)}_{-0.149(0.308)}$ & $-1.007^{+0.939(2.242)}_{-1.212(2.181)}$ \\
        $0.60$ & $72.75^{+1.042(2.141)}_{-1.066(2.117)}$ & $-0.372^{+0.114(0.223)}_{-0.111(0.226)}$ & $+0.294^{+0.690(1.536)}_{-0.811(1.473)}$ \\
        $0.75$ & $72.71^{+1.082(2.098)}_{-1.071(2.177)}$ & $-0.395^{+0.092(0.178)}_{-0.089(0.181)}$ & $+0.439^{+0.488(1.089)}_{-0.576(1.037)}$ \\

        \multicolumn{4}{c}{Pad\'e $(1,2)$} \\
        $0.15$ & $71.65^{+1.175(2.405)}_{-1.217(2.392)}$ & $+0.396^{+0.624(1.343)}_{-0.715(1.314)}$ & $-9.031^{+8.553(18.76)}_{-9.866(17.45)}$ \\
        $0.30$ & $72.01^{+1.008(1.963)}_{-0.973(2.079)}$ & $+0.024^{+0.198(0.405)}_{-0.202(0.409)}$ & $-2.768^{+1.118(2.477)}_{-1.298(2.399)}$ \\
        $0.45$ & $72.55^{+1.086(2.201)}_{-1.086(2.154)}$ & $-0.208^{+0.112(0.221)}_{-0.111(0.223)}$ & $-1.509^{+0.403(0.860)}_{-0.451(0.836)}$ \\
        $0.60$ & $72.61^{+1.055(2.130)}_{-1.104(2.085)}$ & $-0.277^{+0.078(0.158)}_{-0.080(0.157)}$ & $-1.209^{+0.204(0.426)}_{-0.220(0.417)}$ \\
        $0.75$ & $72.52^{+1.028(2.110)}_{-1.075(2.095)}$ & $-0.247^{+0.059(0.118)}_{-0.059(0.118)}$ & $-1.332^{+0.110(0.226)}_{-0.115(0.221)}$ \\

        \hline
        \multicolumn{4}{c}{\textbf{DES-SNY5}} \\
        \hline

        \multicolumn{4}{c}{Taylor} \\
        $0.15$ & $-$ & $+1.643^{+1.273(2.412)}_{-1.161(2.499)}$ & $-13.98^{+8.133(30.68)}_{-15.53(25.28)}$ \\
        $0.30$ & $-$ & $+0.475^{+0.423(0.824)}_{-0.412(0.836)}$ & $-4.407^{+1.613(5.240)}_{-2.984(4.505)}$ \\
        $0.45$ & $-$ & $-0.035^{+0.183(0.355)}_{-0.176(0.352)}$ & $-1.116^{+0.906(2.199)}_{-1.232(2.072)}$ \\
        $0.60$ & $-$ & $-0.134^{+0.113(0.227)}_{-0.112(0.230)}$ & $-0.592^{+0.466(1.103)}_{-0.584(1.020)}$ \\
        $0.75$ & $-$ & $-0.118^{+0.092(0.178)}_{-0.089(0.184)}$ & $-0.883^{+0.294(0.689)}_{-0.370(0.655)}$ \\

        \multicolumn{4}{c}{Pad\'e $(1,2)$} \\
        $0.15$ & $-$ & $+1.705^{+1.123(2.413)}_{-1.236(2.288)}$ & $-16.67^{+10.47(26.78)}_{-15.36(24.65)}$ \\
        $0.30$ & $-$ & $+0.387^{+0.348(0.699)}_{-0.341(0.717)}$ & $-3.423^{+1.353(3.514)}_{-1.907(3.172)}$ \\
        $0.45$ & $-$ & $-0.019^{+0.138(0.266)}_{-0.138(0.268)}$ & $-1.439^{+0.394(0.882)}_{-0.468(0.835)}$ \\
        $0.60$ & $-$ & $-0.070^{+0.080(0.158)}_{-0.080(0.159)}$ & $-1.344^{+0.127(0.272)}_{-0.136(0.261)}$ \\
        $0.75$ & $-$ & $-0.043^{+0.058(0.123)}_{-0.062(0.125)}$ & $-1.522^{+0.064(0.134)}_{-0.070(0.135)}$ \\

        \hline
        \multicolumn{4}{c}{\textbf{Union3}} \\
        \hline

        \multicolumn{4}{c}{Taylor} \\
        $0.15$ & $-$ & $+2.735^{+0.494(0.958)}_{-0.472(0.948)}$ & $-32.94^{+2.732(6.257)}_{-3.171(6.077)}$ \\
        $0.30$ & $-$ & $+0.182^{+0.197(0.384)}_{-0.193(0.392)}$ & $-4.394^{+1.516(3.521)}_{-1.900(3.284)}$ \\
        $0.45$ & $-$ & $-0.134^{+0.111(0.223)}_{-0.111(0.221)}$ & $-1.055^{+0.795(1.713)}_{-0.890(1.679)}$ \\
        $0.60$ & $-$ & $-0.155^{+0.083(0.162)}_{-0.080(0.166)}$ & $-0.898^{+0.464(1.017)}_{-0.529(0.970)}$ \\
        $0.75$ & $-$ & $-0.259^{+0.067(0.131)}_{-0.066(0.132)}$ & $-0.108^{+0.354(0.763)}_{-0.393(0.738)}$ \\

        \multicolumn{4}{c}{Pad\'e $(1,2)$} \\
        $0.15$ & $-$ & $+2.671^{+0.533(1.084)}_{-0.545(1.061)}$ & $-36.07^{+4.201(9.788)}_{-5.316(9.354)}$ \\
        $0.30$ & $-$ & $+0.094^{+0.165(0.330)}_{-0.165(0.330)}$ & $-3.160^{+1.061(2.245)}_{-1.172(2.215)}$ \\
        $0.45$ & $-$ & $-0.124^{+0.087(0.173)}_{-0.087(0.175)}$ & $-1.437^{+0.358(0.754)}_{-0.394(0.747)}$ \\
        $0.60$ & $-$ & $-0.124^{+0.059(0.120)}_{-0.059(0.118)}$ & $-1.484^{+0.157(0.333)}_{-0.170(0.326)}$ \\
        $0.75$ & $-$ & $-0.173^{+0.045(0.091)}_{-0.046(0.091)}$ & $-1.314^{+0.091(0.192)}_{-0.096(0.184)}$ \\

        \hline\hline
    \end{tabular}

    \caption{Constraints on the Hubble constant $H_0$ when Pantheon+\&SH0ES sample is adopted and on the cosmographic parameters $(q_0,\, j_0)$ for all three SNe Ia catalogs with $1$-$\sigma$($2$-$\sigma$) error bars for $z\leq z_{\rm HEL}$ adopting a third order Taylor expansion and $(1, 2)$ Pad\'e approximant, respectively for Pantheon+\&SH0ES, DES-SNY5 and Union3 data catalogs. Here, the observer velocity determined by the CMB dipole and its direction are fixed \cite{Planck:2013kqc, Planck:2018nkj}.}
    \label{tab:cosmoCMBdipole}
\end{table*}

\onecolumngrid

\twocolumngrid

\onecolumngrid

\end{document}